\documentclass[letter]{elsarticle}
\usepackage{amsmath}
\usepackage{amssymb}
\usepackage{amsthm}
\usepackage{amsfonts}
\usepackage{listings}
\usepackage{enumerate}
\usepackage{latexsym}
\usepackage{bbold}

\usepackage{psfrag}

\usepackage{bm}
\usepackage{bbm}
\usepackage{hyperref}

\newcommand{\beq}{\begin{equation}}
\newcommand{\eneq}{\end{equation}}
\newcommand{\bra}[1]{\left\langle#1\right|}
\newcommand{\ket}[1]{\left|#1\right\rangle}

\renewcommand{\eqref}[1]{Eq.~(\ref{#1})}

\def\be{\begin{equation}}
\def\ee{\end{equation}}
\def\ba{\begin{eqnarray}}
\def\ea{\end{eqnarray}}

\def\mathid{\mathbbm{1}}

\input{epsf}

\begin{document}

\tolerance 10000

\newcommand{\vk}{{\bf k}}

\title{Fractional Quantum Hall Matrix Product States For Interacting Conformal Field Theories}

\author[p6]{B. Estienne}
\author[pct,ens]{N. Regnault}
\author[pct]{B. A. Bernevig}
\address[p6]{LPTHE, CNRS, UPMC Univ Paris 06 Bo\^ite 126, 4 Place Jussieu, F-75252 PARIS CEDEX 05}
\address[pct]{Department of Physics, Princeton University, Princeton, NJ 08544}
\address[ens]{Laboratoire Pierre Aigrain, ENS and CNRS, 24 Rue Lhomond, 75005 Paris, France}

\begin{abstract}
Using truncated conformal field theory (CFT), we present the formalism necessary to obtain exact matrix product state (MPS) representations for any fractional quantum hall model state which can be written as an expectation value of primary fields in some conformal field theory. The auxiliary bond space is the Hilbert space of the descendants of primary fields and matrix elements can be calculated using well-known CFT commutation relations. We obtain a site-independent MPS representation on the thin annulus and cylinder, while a site-dependent representation is possible on the disk and sphere geometries. We discuss the complications that arise due to the presence of null vectors in the theory's Hilbert space, and present an array of procedures which optimize the numerical calculation of matrix elements. By truncating the Hilbert space of the CFT, we then build an approximate MPS representation for Laughlin, Moore-Read, Read-Rezayi, Gaffnian, minimal $M(3,r+2)$ models, and a series of superconformal minimal models. We show how to obtain, in the MPS representation, the Orbital, Particle, and Real Space entanglement spectrum in both the finite and infinite systems.
\end{abstract}

%%\date{\today}
%%\pacs{03.67.Mn, 05.30.Pr, 73.43.-f}

\maketitle
\tableofcontents

\newpage

\section{Introduction}

Our theoretical understanding of many-body quantum states of matter faces the significant challenge of the exponential growth of their Hilbert space as the system size is increased. In most situations, the expansion of a many-body interacting state in a basis of many-body non-interacting states involves a large number of coefficients which render both numerical and analytical calculations impossible. Exact diagonalization techniques, whereby the ground-state (and possibly some excited states) of a Hamiltonian is numerically obtained by Lanczos methods, soon hit the exponential barrier for small number of particles. This is true even for states where the first quantized expression of the state is known, such as some model wave functions whose universal properties are thought to describe the physics of the Fractional Quantum Hall (FQH) effect. Even for simplest states, such as the model Laughlin states\cite{laughlin-PhysRevLett.50.1395} for which a first quantized expression in terms of electron coordinates in the Lowest Landau level (LLL) is available in the form of a power of a Jastrow factor, analytical calculations for a large number of particles are impossible (without assuming that the plasma is in the screening phase) and numerical calculations have to rely on Monte-Carlo methods beyond $N_e=10-20$ electrons. An immediate question is whether a more economical representation of quantum states exists, even if approximate, in which we can replace the exponentially exploding many-body Hilbert space by a much smaller, yet still faithful, representation.

An important improvement in our understanding of the answer to the previous question came with the advent of the density matrix renormalization group\cite{White-PhysRevLett.69.2863} (DMRG), or in its more modern exemplification, matrix products states\cite{fannes1992finitely,Perez-Garcia:2007:MPS:2011832.2011833} (MPS). If we try to simulate a gapped state of matter, the crucial observation is that the entanglement between two parts of this system scales as the dimension of the \emph{boundary} between these states \cite{Bombelli-PhysRevD.34.373,Srednicki-PhysRevLett.71.666,Calabrese-2004JSMTE..06..002C,Kitaev-PhysRevLett.96.110404,levin-06prl110405}. Therefore, in one spatial dimension, the entanglement between two subparts of a system, as measured by the entanglement entropy, is constant\cite{verstraete-PhysRevLett.94.140601}, while in two spatial dimensions, it is proportional to the perimeter of the cut, or, for a finite density states, with $\sqrt{N_e}$. If we think of the entanglement entropy as a measure of the information stored in the state,  then becomes clear that gapped states of matter contain less "information" than the growth of the particle Hilbert space suggests. 

A matrix product representation of a state makes this last point explicit. It allows for an efficient approximate (and sometimes exact)   depiction of a state. In particular, when expanded in a non-interacting basis $ \ket{m_1\ldots m_{N_s}}  $ of sites $i =1 \ldots N_s$ and occupation number (for fermions) $m_i=0,1$, a many-body state $\ket{\psi}$ has the expression
\beq
\ket{\psi} = \sum_{\{m_i\}} c_{m_1\ldots m_{N_s}} \ket{m_1\ldots m_{N_s}}  
\eneq 
and there are $2^{N_s}$ such coefficients $c_{m_1\ldots m_{N_s}} $. Any of these coefficients can also be written in the following way: $c_{m_1 \ldots m_{N_s}}=  \bra{\alpha'}B^{m_{N_s}}[N_s] \cdots B^{m_1}[1]\ket{\alpha} $ where $B^{m_j}[j]$ are a set of two (for $m_j=0,1$) site-dependent matrices (in a translational-invariant MPS, the matrix does not depend on the site $[j]$ - which can reside on a $D$-dimensional lattice). For most generic states, to express them exactly, one has not made any significant improvement: even though we now have only $2$ matrices per site, the size of these matrices - or their "auxiliary bond dimension" - still grows exponentially as roughly $2^{N_s}$. However, when we physically cut the state in two regions, it becomes clear that the auxiliary bond dimension of the MPS matrices $B^m$ also serves as the dimension of the reduced density matrix. Since we know the generic scaling of the entanglement between two regions of a gapped phase in $D$ dimensions is proportional to $D-1$ surface separating the regions, it is now clear a more efficient representation of the MPS matrix is possible: in $1$ spatial dimension, entanglement is constant to leading order, which means that an MPS representation of a state with finite bond dimension is possible; in $2$ spatial dimensions, the matrices $B$ can have a Hilbert space which grows exponentially with $L \propto \sqrt{N_e}$. Both imply the possibility of a large improvement over exact diagonalization studies.

For a generic ground-state of a local Hamiltonian, obtaining these matrices involves numerical computations through DMRG. In this paper, we will look at the case of topological states of matter in the FQH, for which an alternate analytic avenue to obtaining the MPS is possible\cite{zaletel-PhysRevB.86.245305,estienne-PhysRevB.87.161112}. A large part of our understanding of the FQH physics of strongly interacting electrons (with quenched kinetic energy) comes from model wave functions. These wave functions (most of which are expressed in the LLL) are analytic in the electron coordinates $z_j= x_j + i y_j$, have no variational parameter, and are supposed to describe the universal properties of the bulk of the FQH state that could be stabilized by some short-range realistic Hamiltonian. Some of the most famous examples are the Laughlin \cite{laughlin-PhysRevLett.50.1395}, Jain\cite{Jain:1989p294}, Moore-Read \cite{Moore1991362}, Read-Rezayi \cite{read-PhysRevB.59.8084} and Gaffnian \cite{simon2006} states, which are known to be exact zero modes of Haldane pseudopotential \cite{haldane-PhysRevLett.51.605} Hamiltonians. Beyond these cases, families of other wave functions such as Jack polynomials \cite{Bernevig-PhysRevB.77.184502,Bernevig-PhysRevLett.100.246802} and superconformal \cite{Estienne2010539} are also possible.

A preferred method for obtaining bulk FQH model wave functions is to express them as correlators of some primary fields  $V(z)$ \cite{Moore1991362,Fubini-1991MPLA....6..347F,Cristofano-1991PhLB..262...88C,Cristofano-1991MPLA....6.2985C,Cristofano-1991MPLA....6.1779C} in a Conformal Field Theory (CFT). In a way which will be made more precise later, given any CFT, the electron operator has to be a field which exhibits trivial monodromy with every other field. In the standard construction the CFT is made of two parts : a neutral sector, and a charged sector, described by a U$(1)$ CFT. In turns the electron operator is of the form $V(z) =  \psi(z) \otimes :\exp(i \sqrt{q} \varphi(z)) :$, where $\varphi$ is a $U(1)$ free boson field and $q$ is a rational number which is related to the filling of the FQH state $\nu = 1/q$.  Then a first-quantized many-body model FQH wave function for $N_e$ electrons in the LLL at filling $1/q$ is defined to be $\Psi(z_1 \ldots z_{N_e}) = \langle V(z_1) \ldots V(z_{N_e}) \rangle$ where the expectation value is taken in the ground-state of the CFT, with a neutralising background charge left implicit here. There are strong non-analytic arguments\cite{Read-PhysRevB.79.245304} that the CFT whose operator we use to form the bulk model wave function must be identical to the CFT of the edge of the FQH droplet. We stress that this is a remarkable conjecture, as the correlation functions on the gapless edge are power-law decaying, while those in the bulk of a gapped state must necessarily be exponentially decaying despite the fact that they are both derived from the same CFT.

 Some of the known FQH wave functions are supposed to describe the universal properties of experimentally observable plateaus at Abelian fillings (such as $1/3$) or non-Abelian (such as $2+ 1/2$ or $2+2/5$). For a small part of these wave functions (such as the Laughlin and Moore-Read series), an explicit first quantized expression is available, but for most of them even such an expression is unknown. Performing analytic calculations of universal properties (such as braiding, fractional statics) is impossible except for the specific case of Laughlin, for which a plasma analogy between he quantum wave function and a classical one-component plasma system has made calculations of fractional statistics \cite{Arovas-PhysRevLett.53.722} possible with \emph{the (established) assumption} that the plasma is in its screening phase. A more intricate plasma analogy established the non-Abelian braiding in the FQH Moore-Read state \cite{Bonderson-PhysRevB.83.075303}, again under the assumption that the (in this case two-component) plasma screens. Performing numerical calculations is also not easy - the sizes for which we can compute things range from 15-30 particles through either exact diagonalization or Jack polynomial techniques\cite{Bernevig-PhysRevLett.103.206801}. Hence obtaining braiding, topological entanglement entropy, and many other universal properties for most FQH states is currently not possible. Most of the bulk properties of FQH wave functions that we know of are actually not computed but \emph{assumed} to be identical to those of the CFT whose correlators give the first quantized expression of the bulk state. One of the other outstanding questions is what is the fate of a bulk wave function which can be expressed through the expectation value of operators in a non-unitary CFT. If the primary field $V(z)$ comes from a non-unitary CFT then, in most cases, the correlations on the edge are power-law \emph{enhancing}, signaling the fact that the non-unitary CFT cannot be used to describe the edge of a gapped state. Some arguments that such states (of which the prime representative is the Gaffnian wave function\cite{simon2006}) are bulk gapless have been given by Read \cite{Read-PhysRevB.79.245304}, but so far no analytical or numerical calculation has shown that due to the difficulties in dealing with the large Hilbert of these states.

Hence an efficient MPS description of FQH model states which would take us to further computational limits is desirable. From analyzing the entanglement spectra\cite{li2008} of these model states we can immediately infer that their MPS description will also be special: the entanglement spectrum exhibits far less states than the generic spectrum of a $2D$ wave function, a statement to their peculiarity. The crucial step towards this was taken by Zaletel and Mong \cite{zaletel-PhysRevB.86.245305} who were able to obtain the MPS matrices $B^0, B^1$ for the Laughlin ground-state and quasiholes and for the Moore-Read ground-state. Zaletel and Mong exploited two different things: the fact that the bulk wave functions can be written in terms of a CFT expectation value (which was briefly pointed out as being useful for MPS earlier by Sierra \cite{Cirac-PhysRevB.81.104431} and Dubail and Read \cite{dubail-PhysRevB.86.245310}), and the fact that both the Laughlin and Moore-Read states can be written in terms of free CFTs: the Laughlin state is the expectation value of the $U(1)$ boson free field operator while the Moore-Reads adds to that the product of a $p+i p$ superconductor. In their paper, Zaletel and Mong obtained several known quantities such as Laughlin quasiholes statistical angle, entanglement spectrum and topological entanglement entropy and showed the power of the method by being the first to reliably calculate the Laughlin and Moore-Read topological entanglement entropies. They have also emphasized on the role of the cylinder geometry to efficiently implement DMRG for FQH systems, already leading to some interesting results beyond model states\cite{Zaletel-PhysRevLett.110.236801,Varjas-PhysRevB.88.155314}.

In this paper we show the detailed steps to obtain an MPS representation of states described by interacting CFTs that were shortly described in Ref.~\cite{estienne-PhysRevB.87.161112}. These include the $\mathbb{Z}_3$ Read-Rezayi as well as the Gaffnian state. Starting from the conformal block description of FQH states, we re-derive in a more transparent way the Laughlin MPS already obtained in Ref.~\cite{zaletel-PhysRevB.86.245305}, and then move to the interacting CFTs with an underlying Virasoro algebra. Working in full generality, we define the infinite Hilbert space of the CFT as the MPS auxiliary bond Hilbert space and show how to compute overlaps of CFT basis states and matrix elements of the vertex operators. Due to the intricate commutation relations of the algebras in question, these need to be implemented numerically. We here concentrate on the matrix elements of the electron field - necessary for obtaining ground-state wave functions. We will present the formalism to obtain the MPS for quasiholes in a future paper\cite{futureqhpaper}. We comment on the intricacies of the problem, on null and negative norm vectors (the latter for non-unitary CFTs). We then particularize to paired Jack states (states which are Jack polynomials and they are described by minimal models $M(3,r+2)$ ) which include the Moore-Read and Gaffnian states. We leave the particular formulas Read-Rezayi state and an infinite series of states described by superconformal field theories in the appendices. We then show how to use the exact MPS obtained in order to perform calculations for FQH wave functions. This first involves truncating the auxiliary Hilbert space of the MPS matrices in a procedure that is equivalent to those employed in truncated CFT\cite{Yurov-1990} in order to obtain an approximation of the MPS matrices. Within the truncated Hilbert space, we present several methods to significantly speed up the computation of the FQH states by further truncating the Hilbert space besides to achieve the optimal representation of the MPS matrices. The truncation parameter is roughly similar to the maximum conformal scaling dimension allowed for a field in the theory. A nice property of the approximate wave functions obtained by truncating is that they have the same weights (as the exact FQH model states on some of the many-body non-interacting states and exactly zero weight on others. We then comment on the infinite MPS, and show how to obtain the orbital, particle and real space entanglement spectra, in both finite and infinite MPS. We end with several appendices giving examples of the generic theory presented in the main body of the paper, including re-deriving, from the MPS, the thin torus limit, squeezing properties, cylinder normalization, etc.

This paper serves to develop the theory of matrix product states for non-Abelian FQH states. Using this, we have numerically implemented the wave functions whose MPS we present in this paper, and have analyzed the differences between unitary and non-unitary states by computing several topological universal properties of these states such as topological entanglement property, and several non-universal properties such as scaling of the transfer matrix gap. We have also very accurately computed the braiding properties of non-Abelian FQH states. These results will be presented in future papers \cite{regnaultinprep, yanglewuinprep}.

%%%%%%%%%%%%%%%%%%%%%%%%%%%%%%%%%%%%%%%%%%%%%%%%%%%%%%%%%%%%%%%%%%%%%%%%%%%%%%%%%%%%%

\newpage

\section{Conformal Blocks as Exact Matrix Product States}
\label{CFT section}

Throughout this paper we will consider model FQH states whose wave functions are given as the conformal blocks of a chiral conformal field theory in two dimensions ($1+1$). Although in the rest of this paper we will concentrate on the MPS formulation of fermionic wave functions, in this section we treat both the fermonic and the bosonic case. 
The main example of FQH state is the $\nu = \frac{1}{q}$ Laughlin state and its quasihole states. On the plane, the Laughlin wave functions are described by the following wave functions
 \begin{align}
 \prod_{i<j}^{N_e} (z_i -z_j)^q P(z_1,\cdots,z_{N_e}) \label{Laughlin zero modes}
 \end{align}
where $z_j= x_j+ i y_j$ is the $j$'th electron position on the complex plane and $P$ is an arbitrary (homogeneous) symmetric polynomial of the coordinates of the $N_e$ electrons. The Laughlin "ground-state" is usually referred to the wave function above when $P(z_1,\cdots,z_{N_e})=1$ . There is an infinite set of wave functions in \eqref{Laughlin zero modes}. They should be more appropriately described as the zero energy modes of the $\nu = 1/q$ Laughlin droplet with an arbitrary edge excitation, or, equivalently with an arbitrary number of quasi holes in the bulk. The notion of zero energy mode implies an Hamiltonian, and in this case it is a well known $2$-body Haldane interaction that penalize two electrons having a relative angular momentum less than $q$ \cite{haldane-PhysRevLett.51.605}. Similar pseudopotentials\cite{Simon-PhysRevB.75.195306} exist for many other \cite{Moore1991362,read-PhysRevB.59.8084,simon2006,Bernevig-PhysRevB.77.184502} but not all \cite{Bernevig-PhysRevB.77.184502,Estienne2010539,Jackson-PhysRevB.88.075313} FQH states derived from CFT conformal blocks. In this paper we focus on wave functions with edge excitations. Pinned quasiholes in the bulk will be analyzed in a future paper \cite{futureqhpaper}

Any Laughlin wave functions \eqref{Laughlin zero modes} can be written as conformal block 
\begin{align}
\langle \alpha' | V(z_1) \cdots V(z_n) | \alpha \rangle \label{conformal block wf}
\end{align}
where $V(z)$ is the vertex operator $V(z) = : \exp \, ( i \sqrt{q} \varphi(z)): $, and the states $| \alpha \rangle, |\alpha' \rangle$ are \emph{in} and \emph{out} states in the Hilbert space of the CFT $\textrm{U}(1)_{\sqrt{q}}$, to be described in more details below. Note that the conformal block \eqref{conformal block wf} has to be computed on the plane to reproduce the particular wave function \eqref{Laughlin zero modes}, but with the appropriate conformal transformations, the correspondence holds in various geometries, such as the cylinder and the torus.

\subsection{Conformal Field Theory for Fractional Quantum Hall States}
\label{CFT states intro}

 The expression \eqref{conformal block wf} exposes a generic way of building FQH polynomial wave functions from an underlying CFT. The primary field $V(z)$ is called the electron operator. In the Laughlin example, it is a simple vertex operator $V(z) = :e ^{i \sqrt{q}\varphi (z)}: $ but in the generic situation however the underlying CFT is not just a free field theory. For an almost exhaustive set \footnotemark \footnotetext{An exception would be the Jain states\cite{Jain:1989p294}, even in their CFT formulation\cite{Bergholtz-PhysRevB.77.165325}.} of CFT trial FHQ states, the electron operator is of the form
\begin{align}
V(z) = \Psi(z) \, \otimes \, : \exp \left( i \frac{1}{\sqrt{\nu}} \varphi (z) \right): \label{generic electron operator}
\end{align}
where $\Psi(z)$ is a field in the so called neutral conformal field theory $\text{CFT}_n$. The neutral CFT has it own chiral algebra, which is generated by the field $\Psi(z)$. We refer to this algebra as the \emph{neutral algebra}. Examples of such neutral algebra are the free Majorana fermion (for the Moore-Read state), the $\mathbb{Z}_k$ parafermionic theory \cite{fateev1985currents,Furlan-1989,Dotsenko-2003NuPhB.664..477D,Dotsenko-2003NuPhB.656..259D,Dotsenko-2004PhLB..584..186D,Dotsenko-2004NuPhB.679..464D,Dotsenko-2005PhLB..611..189D,jacob2005,dotsenko2012parafermionic,dotsenko2012two} (for the Read-Rezayi states), and the $N=1$ superconformal algebra for the states presented in Appendix~\ref{superconformal}.  The neutral side of the CFT encodes the information relevant to the so-called \emph{clustering} of the trial wave-function. This is characterized by the (para)-fermionic nature of the field $\Psi$, which is manifest in the OPE :
\begin{align}
\underbrace{\Psi \times \Psi \times \cdots \Psi}_{k} = 1 \label{parafermionicOPE}
\end{align}
for some strictly positive integer $k$. It turns out that such an OPE restricts severely the possible conformal dimension $h_{\Psi}$ of the neutral field $\Psi$. A possible family of solutions is given by 
$h_{\Psi} = \frac{r}{2} \frac{k-1}{k}$, 
where $r$ is a positive integer\cite{Estienne2010539}. Other solutions are possible, but as far as we know they have never been considered in the context of the Fractional Quantum Hall Effect (FQHE). The filling fraction $\nu$ entering the electron field \eqref{generic electron operator} is of the form
\begin{align}
\nu^{-1} = \frac{r}{k} + m
\end{align}
where $m \in \mathbb{N}$ is even for bosons and odd for fermions. The two integers $(k,r)$ characterize the vanishing properties of the corresponding (bosonic) $m=0$ trial wave function\cite{Bernevig-PhysRevB.77.184502,Bernevig-PhysRevLett.100.246802,Estienne2010539}. We refer to this behavior as $(k,r)$-clustering of the trial wave-function, even in the case $m \neq 0$. 

The vertex operator lives in the free theory of a compact boson. The radius of compactification depends both on the clustering parameter $k$ and the filling fraction $\nu$ 
\begin{align}
R = k/\sqrt{\nu} = k \sqrt{q}
\end{align}
The bosonic part encodes all the information relevant to electric charge, such as the filling factor. The total CFT is therefore a tensor product\footnotemark \footnotetext{To be more rigorous the situation is not entirely that of a simple tensor product: restrictions related to selection rules apply. We will come back to this semi-simple tensor product when describing the $k=2$ Jack states.} $\text{CFT}_{n} \otimes \text{U}(1)_{R}$.

The electron field \eqref{generic electron operator} obviously couples the neutral and $\text{U}(1)$ part of the underlying CFT. In order to understand the topological phase of the corresponding Fractional Quantum Hall State (FQHS), it is natural to enlarge the chiral algebra to include the electron operator $V(z)$ and its conjugate field 
\begin{align}
V^*(z) = \Psi^*(z) \, \otimes \, : \exp \left( - i \frac{1}{\sqrt{\nu}} \varphi (z) \right): 
\end{align}
The modes of $V$ and $V^*$ span an enlarged algebra that includes both the U$(1)$ and the neutral algebra \footnotemark \footnotetext{because the U$(1)$ current appears in the OPE $V(z) V^*(0) = z^{-2h}\left( 1 + \sqrt{q} z \, j(z) + \cdots \right) $, as was pointed out in Ref.~\cite{Jackson-PhysRevB.88.075313}}. In the usual cases this algebra is a WZW algebra or a $N=2$ supersymmetric algebra, but in general it can be much richer. We refer to this algebra as \emph{electronic algebra}. 

The total conformal field theory underlying a FQHS has to be rational \emph{w.r.t} the electronic algebra, and it contains a finite number of highest weight states $| \Phi_a \rangle = \Phi_a(0) \ket{0}$. One can then decompose the total Hilbert space $\mathcal{H}$ according to topological sectors $\mathcal{H}_a$, \emph{i.e.} irreducible representations of this enlarged algebra. 
\begin{align}
\mathcal{H} = \bigoplus_a \mathcal{H}_a
\end{align}
The irreducible representation $\mathcal{H}_a$ is spanned by acting on a primary field (highest weight state) $\ket{\Phi_a}$ with the modes of $V(z)$ and $V^{*}(z)$, and it is relevant for computing bulk wave functions with a quasi-hole $\Phi_a$ located at the origin, and its conjugate $\Phi_{\bar{a}}$ at infinity :
\begin{align}
\langle \Phi_{\bar{a}}(\infty)  V(z_1) V(z_2)\cdots   V(z_{N_e}) \Phi_a(0) \rangle       = \bra{\Phi_a}  V(z_1) V(z_2)\cdots   V(z_{N_e}) \ket{\Phi_a}
\end{align}  
with an implicit neutralizing background charge. In the context of the topological field theory one talks about the sector with topological charge '$a$'.

The electronic algebra is the most natural algebraic description of the CFT, since the sectors $\mathcal{H}_a$ are in one-to-one correspondence with the topological sectors of the corresponding topological quantum field theory. In particular the electron field belongs to the vacuum sector $\mathcal{H}_0$. Indeed the electron field is no longer primary, it is a descendant of the identity operator, namely $| V\rangle = V_{-h} | 0\rangle$, where $h = q/2 + h_{\Psi}$ is the conformal weight of the electron field $V(z)$. Each topological sector $\mathcal{H}_a$ is characterized by its character (edge modes/ entanglement spectrum counting) and its quantum dimension $d_a$. In the MPS formalism this decomposition of the Hilbert space into topological sectors is not only transparent but numerically beneficial as it allow to reduce the size of of the MPS matrices $B^m$ (see section~\ref{B block structure}). 

The decomposition of the CFT into $\text{CFT}_{n} \otimes \text{U}_{R}$ comes in handy when writing down a site-independent MPS. Therefore it is advantageous to understand the topological sectors in terms of this decomposition into neutral and charge parts. Primary fields $\Phi_a$ are of the form $\Phi_a(z) = \phi(z) \otimes e^{i Q \varphi (z)}$. Since the field $\Phi_a$ is primary \emph{w.r.t.} the electronic algebra, the field $\phi$ is a primary field \emph{w.r.t.} the neutral algebra. This stems from the fact that the electronic algebra contains then neutral one. One can think of the primary field $\Phi_a(z)$ as an operator creating a quasi-hole of topological charge '$a$'. The U$(1)$ charge $Q$ encodes the electric charge of that quasi-hole, and is not a free parameter. For a $k$-clustered state we have
\begin{align}
\Phi_a(z) = \phi(z) \otimes \exp \left[ i \sqrt{\nu} \left( \frac{b_{\phi}}{k} + j \right) \varphi(z) \right]
\end{align}
where $0 \leq b_{\phi} \leq k-1$ and $j$ are integers. While $b_{\phi}$ is fixed by the requirement of trivial monodromy with the electron field, $j$ is a free parameter, describing the presence of $j$ elementary Abelian quasi-hole $\exp (i \sqrt{\nu} \varphi)$. In this language the topological charge $a = (\phi,j)$ is generically labeled by a neutral primary field $\phi$ and an integer $j$. The range of $j$ is bounded for the following reason. Given that $k$ electron can fuse to a pure U$(1)$ charge $:\exp (i  k \sqrt{q} \varphi):$, we have to identify $j = j+ kq$ since they describes the same topological sector. Therefore for a given neutral sector $\phi$, the integer $j$ can generically assume $kq$ values  $ j = 0,\cdots,kq-1$\footnotemark \footnotetext{To be more precise the identification depends on the neutral sector $\phi$, namely $j = j + k_{\phi} q$, where $k_{\phi}$ is the smallest integer such that $\underbrace{\Psi \times \Psi \times \cdots \Psi}_{k_{\phi}} \times \phi = \phi$. In most cases it is $k$ itself, but it can happen to be a divisor of $k$. For instance this happens for the Moore-Read state ($k=2$) : the $\sigma$ sector has $k_{\sigma}=1$.}.

\subsection{Site Dependent MPS}

A natural many-body basis for $N_e$ particles in the LLL is the occupation basis of angular momentum orbitals. Partitions $\lambda = (\lambda_1\ldots \lambda_{N_e})$  ( i.e. sequences of non increasing, strictly positive integers) provide a convenient way to label these states. Accordingly, arbitrary many-body wave functions for $N_e$ electrons can be expanded in monomials (resp. Slater determinants) for bosons (resp. for fermions ): 
\begin{align}
m_{\lambda} = \frac{1}{\sqrt{N_e! \prod_i m_i !}}\sum_{\sigma \in \mathfrak{S}_{N_e}} \epsilon(\sigma) \prod_{i=1}^{N_e} \, z_{\sigma(i)}^{\lambda_i} \label{unnormalizedbasis}
\end{align}
where $m_i = m_i(\lambda)$ is the number of occurrences of the integer $i$ in the partition $\lambda$. The sum is taken over the group of permutations of $N_e$ objects $\mathfrak{S}_{N_e}$, while $\epsilon(\sigma)$ is equal to $1$ for bosons, and to the signature $\text{sgn}(\sigma)$ of the permutation for fermions. Note that \eqref{unnormalizedbasis} defines an un-normalized basis: the expression contains no information about, amongst others, the geometry and the magnetic length (which we set to unity in this paper).

As pointed out in \cite{dubail-PhysRevB.86.245310,zaletel-PhysRevB.86.245305,estienne-PhysRevB.87.161112}, wave functions of the form \eqref{conformal block wf} take the form of matrix product states. If expanded in a polynomial set, upon inserting a complete basis of states we get:
\begin{eqnarray}
\langle  \alpha' | V(z_1) \cdots V(z_{N_e}) | \alpha \rangle &= \sum_{\alpha_1,\cdots,\alpha_{N_e-1}}& \langle \alpha' | V(z_1) | \alpha_1 \rangle \langle \alpha_1 | V(z_2) | \alpha_2 \rangle \nonumber\\
&&\cdots   \langle \alpha_{N_e-1} | V(z_{N_e}) | \alpha \rangle \nonumber\\   
&= \sum_{\lambda}& c_{\lambda} m_{\lambda}(z_1,\cdots,z_{N_e})
\end{eqnarray}

The coefficient $c_{\lambda}$ can then be obtained from the mode expansion of the electron operator
\begin{align}
V(z) = \sum_{n \in \mathbb{Z}} z^n V_{-n-h}\label{electronoperatorexpansion}
\end{align}
where $h$ is the conformal dimension of $V(z)$. As a consequence of the absence of singular term in the operator product expansion (OPE) $V(z_1)V(z_2)$, the electron modes $V_{-n-h}$ (anti)commute for (fermions)bosons. This ensures the (anti)symmetric nature of $m_{\lambda}$ upon expanding the fields $V(z_i)$ in the conformal block. Through this expansion the conformal block becomes a site (\emph{which here is an angular momentum orbital}) dependent MPS
\begin{eqnarray}
\frac{c_{\lambda}}{\sqrt{N_e!}} &  = & \frac{1}{\sqrt{\prod_i m_i !}} \langle \alpha' |  V_{-\lambda_{1}-h}  V_{-\lambda_2-h} \cdots V_{-\lambda_{N_e}-h} | \alpha\rangle\nonumber\\
&  = & \langle \alpha' |  A^{m_{N_{\Phi}}}[N_{\Phi}] \cdots A^{m_1}[1] A^{m_0}[0]| \alpha\rangle \label{orbital dependent MPS}
\end{eqnarray}
 with the matrices for orbital $j$ being
\begin{align}
 A^m[j] = \frac{1}{\sqrt{m!}}\left( V_{-j-h} \right)^m  \label{MPS A matrix}
\end{align}
The matrices $A^m[j]$ (anti-)commute, and the ordering in \eqref{orbital dependent MPS} is insignificant. For fermions the only allowed occupation number are $0$ and $1$ ($ A^m[j] =0$ for $m\geq 2$). The corresponding matrix elements are
\begin{eqnarray}
&\langle \alpha' | A^0[j] | \alpha \rangle  = \delta_{\alpha',\alpha},&\nonumber\\
&\langle \alpha' | A^1[j] | \alpha \rangle  =  \frac{1}{2\pi i} \oint \text{d}z \,z^{-j-1} \, \langle \alpha' | V(z) | \alpha \rangle = \delta_{\Delta_{\alpha'},\Delta_{\alpha}+h+j} \langle \alpha' | V(1) | \alpha \rangle&
\end{eqnarray}
where the integration contour is around the origin and $\Delta_{\alpha}$ is the conformal dimension of the state $\ket{\alpha}$. In the above, we have used $\bra{\alpha'} V(z) \ket{\alpha} = z^{\Delta_{\alpha'}- \Delta_{\alpha} - h} \bra{\alpha'}V(1) \ket{\alpha}$ which follows from scale transformations. For most geometries, the orbitals used in our monomial/Slater basis in \eqref{unnormalizedbasis} are un-normalized. The true normalized orbitals are $N_j z^j$, where
\begin{align}
N_j =  \frac{1}{\sqrt{2\pi}} \frac{1}{\sqrt{2^j j!}} & \qquad \text{on the plane} \label{plane normalization} \\
N_j = \sqrt{ \frac{N_{\Phi}+1}{4\pi R^2} \binom{N_{\Phi}}{j}}\left(\frac{1}{2R}\right)^j& \qquad \text{on the sphere}  \label{sphere normalization} \\
N_j =   \frac{1}{\sqrt{L \sqrt{\pi}}}\exp \left\{- \left(\frac{2 \pi}{L} \right)^2 \frac{j^2}{2}\right\}& \qquad \text{on the cylinder}  \label{cylinder normalization}
\end{align} and a site-dependent MPS for expansions in these orbitals can be easily obtained from the un-normalized basis by the transformation $A^m[j] \rightarrow A^m[j]/(N[j])^m$. Also notice that the un-normalized basis corresponds physically to the geometry of the thin annulus, which can be converted into an expression where every orbital is "normalized" by the same amount. In this limit, every angular momentum $j$ is the sum of a large, infinity, angular momentum $J\rightarrow \infty $ of the inner circle of the annulus and a deviation from it which represents the thickness of the annulus $\delta j$. In the limit of a thin annulus, the ratio in normalization between two \emph{many-body} states  $m_{\lambda}, m_\mu$  of the same total angular momentum $|\lambda| = \sum_i \lambda_i = |\mu|$, with all momenta of any particle $i$, $\lambda_i$ and $\mu_i$ in the thin annulus is just $\sqrt{\prod_i \lambda_i!/\mu_i!} \approx 1$.

\subsection{Site Independent MPS}
\label{Site Independent MPS}

In \eqref{MPS A matrix} the un-normalized matrices $A^m[j]$ depend on the orbital of angular momentum $j$, hence forming a site-dependent MPS. This convention is very convenient when working with first quantized wave function, but it is not so natural in the MPS framework; the origin of the site dependence is the neutralizing background charge at infinity. The first hint that a site-independent MPS is possible is to realize that we can shift the index $j$ from the $A^1$ matrix (denoting an occupied orbital) into the string of exactly $j$ unoccupied orbitals described by the $A^0$ matrices preceding the occupied orbital. This effectively amounts to spreading the background charge uniformly over the droplet \cite{zaletel-PhysRevB.86.245305}, irrespective of whether a site is occupied or not.  There are $N_\Phi= N_e/ \nu$ orbitals over which we need to spread an $N_e$ background charge. This amounts to inserting an electric charge $- \nu e$ (\emph{i.e.} a $\text{U}(1)$ charge $-\sqrt{\nu}$) between consecutive orbitals. At a more algebraic level, this means inserting an operator $e^{-i\sqrt{\nu}\varphi_0}$, where $\varphi_0$ is the zero mode of the bosonic field $\varphi(z)$ (see section~\ref{section free boson}). Since the electron field $V(z)$ of generic form \eqref{generic electron operator} carries a $\text{U}(1)$ charge $\sqrt{q} = 1/\sqrt{\nu}$, its modes enjoy the property 
\begin{align}
V_{-n-h} = e^{in\sqrt{\nu}\varphi_0} V_{-h} e^{-in\sqrt{\nu}\varphi_0} \label{charge - V mode interplay}
\end{align}
Upon expanding the wave function $\ket{\psi} = \sum_ m c_m \ket{m}$ onto the (un-normalized) occupation basis \eqref{unnormalizedbasis}, we can express the coefficient $c_{m}$ corresponding to the occupation $m = m(\lambda)$ as
\begin{align}
 c_{m} & = \left( \frac{N_e! }{ \prod_i m_i !} \right)^{1/2} \langle \alpha' |  V_{-\lambda_{N_e}-h} \cdots V_{-\lambda_2-h} V_{-\lambda_{1}-h} | \alpha\rangle  \\
  &  =   \left( \frac{N_e! }{ \prod_i m_i !} \right)^{1/2}  \langle \alpha' |  e^{i\frac{\lambda_1}{\sqrt{q}} \varphi_0} V_{-h} e^{i\frac{(\lambda_2-\lambda_1)}{\sqrt{q}}\varphi_0} V_{-h}  \cdots e^{i\frac{(\lambda_{N_e}-\lambda_{N_e-1})}{\sqrt{q}}\varphi_0}    V_{-h}  e^{-i\frac{\lambda_{N_e}}{\sqrt{q}}\varphi_0}  | \alpha \rangle 
\end{align}
On the disk we have $N_{\Phi} \geq \lambda_1 \geq \lambda_2  \geq \cdots \geq \lambda_{N_e} \geq 0$, and we can write 
\begin{align}
\frac{c_{m}}{ \sqrt{N_e! }} =  \langle \alpha_R |  B^{m_{N_{\Phi}}} \cdots B^{m_1} B^{m_0}| \alpha_L \rangle,  \label{orbital independant MPS} \\  \ket{\alpha_L} = \ket{\alpha}, \, \ket{\alpha_R}  =  e^{-i (N_{\Phi}+1)\sqrt{\nu} \varphi_0} | \alpha' \rangle
\end{align}
with $B^m =    e^{-i \sqrt{\nu}\varphi_0}  V_{-h}^m/\sqrt{m!}$. The neutralizing background charge in the bulk modifies the out state $| \alpha' \rangle \to \ket{\alpha_R} =  e^{-i (N_{\Phi}+1)\sqrt{\nu} \varphi_0} | \alpha' \rangle$ but leaves the in state $\ket{\alpha_L} = \ket{\alpha}$ invariant. 
Up to the orbital normalization $N_j$, we now have an orbital independent MPS representation of all conformal blocks of the form $\langle \alpha' | V(z_1) \cdots V(z_{N_e}) | \alpha \rangle$, for arbitrary in and out states $| \alpha \rangle$, $|\alpha' \rangle$.

Some comments about the form of the matrices $B^m$ are in order. An important point is the degree of arbitrariness in the choice of the matrices $B^m$. Any change of basis of the form
\begin{align}
B^m \to G B^m G^{-1}, \qquad \bra{\alpha_R} \to \bra{\alpha_R}G^{-1}, \qquad \ket{\alpha_L} \to G \ket{\alpha_L} \label{gauge transformation}
\end{align}
lives the coefficients $c_{m}$, and hence the wave function, invariant. As a consequence other choices of $B$ matrices are possible. In particular, the index $h$ of the mode of the electron operator $V_{-h}$ is rather conventional. While natural on the disk (since orbitals index are positive $\lambda_i \geq 0$), it becomes much more arbitrary and a matter of personal convenience on the cylinder (where $\lambda_i \in \mathbb{Z}$). In the latter case a more natural choice would be 
 \begin{align}
B^m =    e^{-i \sqrt{\nu}\varphi_0/2} \left( \frac{1}{\sqrt{m!}} V_{0}^m  \right) e^{-i \sqrt{\nu}\varphi_0/2} \label{MPS B matrix}
\end{align}
This is the choice we make in this paper, since we are going to work mainly on the cylinder. In particular with this choice the mirror symmetry of the cylinder is manifest in 
\begin{align}
B^{\dagger m} = C B^m C \label{B to Bdagger}
\end{align}
where $C$ stands for charge conjugation. In the following we adopt the convention of \eqref{MPS B matrix}.

Another important issue is the ordering of the MPS matrices $B^m$. The orbital-dependent matrices $A^m[j]$ \eqref{MPS A matrix} (anti)-commute, and there the ordering does not matter. However the orbital-independent ones $B^m$ in \eqref{MPS B matrix} do not commute, and the ordering is no longer conventional. It has to be the one of \eqref{orbital independant MPS}\footnotemark \footnotetext{The MPS representation with the reversed order $  B^{m_{0}} \cdots B^{m_1} \cdots B^{m_{N_{\Phi}}}$ is also possible. This amounts to take the hermitian conjugate of \eqref{orbital independant MPS}, and then using \eqref{B to Bdagger}. This trick is possible since the coefficients $c_{\lambda}$ can be chosen to be real, which is ensured by the behavior of the FQH problem under time reversal.}. 

Finally, the MPS representation is very convenient to describe FQH states in the grand canonical ensemble. A change of chemical potential boils down to a rescaling of the $B$ matrices of the form
\begin{align}
B^m \to e^{\mu m} B^m \label{chemical potential}
\end{align}

\subsection{MPS on the Cylinder}

We have so far developed the MPS formalism in the thin annulus limit, by taking the single particle orbitals to be $z^m$. In this geometry, we argued that the coefficient $c_{\lambda}$ of a many-body state with $N_e$ electrons occupying the orbitals of angular momenta $\lambda = (\lambda_1, \cdots , \lambda_{N_e})$ is 
\begin{align}
\frac{c_{m}}{ \sqrt{N_e! }}  =  \bra{\alpha_R}B^{m_{N_\Phi}} \ldots B^{m_1} \ket{\alpha_L} 
\end{align}
where we have chosen to label the occupation of the first orbital by $m_1$ rather than $m_0$ since we are working on the thin annulus (and later on the cylinder). We will stick to this convention in the rest of the paper.

 In different geometries, the MPS matrices still depend on the orbital index $j$ as $B^m \rightarrow B^m/N_j^m$, but only through the one-body normalization factor $N_j$. In general these normalization factors cannot be absorbed: in the case of the sphere for instance, where orbitals close to the pole are different form those close to the equator, and likewise on the plane in the symmetric gauge. On the other hand on the cylinder all orbitals are equivalent to each other, and it is possible to improve the MPS \eqref{MPS A matrix} to a site independent one on this geometry (Ref.~\cite{zaletel-PhysRevB.86.245305}). 

On the cylinder, the correctly normalized coefficient $d_{m}$ corresponding to the orbital occupation $m = m(\lambda)$ can be expressed in terms of the un-normalized $c_{m}$ : 
\begin{align}
d_{m} = c_m \, e^{\tau\sum_i \frac{\lambda_i^2}{2}}  (L\sqrt{\pi})^{N_e/2}, \qquad \tau = \left( \frac{2\pi}{L} \right)^2
\end{align}
where the range of occupied orbitals $\lambda_i$ depends on the position of the droplet on the cylinder. In the following we consider a droplet localized at an arbitrary position on the droplet, with occupied orbitals ranging from $n_L$ to $n_R$, for a total number of orbitals  $N_{\Phi} = n_R- n_L +1$. The left and right edges of the droplet are then located around $x_L = 2\pi n_L /L$ and $x_R = 2\pi n_R /L$.

It is possible to find a MPS representation of this coefficient, \emph{i.e.} to absorb the factor $\sum_i \lambda_i^2 $ by a redefinition of the matrices $B^m$ (the factor $\sqrt{L \sqrt{\pi}}$ can easily be accounted for through a change of the form  \eqref{chemical potential}). As was pointed out in (Ref.~\cite{zaletel-PhysRevB.86.245305}), the trick is is to insert a Hamiltonian $-\frac{2\pi}{L}L_0$ propagation on the cylinder (in the CFT sense) over the intra-orbital distance $\frac{2\pi}{L}$. A possible choice for site independent $B$ matrices on the cylinder is given  by
\begin{align}
C^m = (L\sqrt{\pi})^{m/2} \, e^{- \frac{\tau}{2} L_0 }B^m e^{- \frac{\tau}{2} L_0 } \label{normalizedmps}
\end{align}
Such a modification produces the desired coefficient $\frac{1}{2} \sum_i \lambda_i^2$ (see Appendix~\ref{Appendix appendix cylinder normalization})
\begin{align}
\bra{\alpha_R}C^{m_{N_\Phi}} \cdots C^{m_1} \ket{\alpha_L} = e^{  \tau \left(  \frac{1}{2} \sum_i \lambda_i^2 - E_{R} + E_L \right) }\,  (L\sqrt{\pi})^{N_e/2} \,  \bra{\alpha_R}B^{m_{N_\Phi}} \ldots B^{m_1} \ket{\alpha_L}  \label{matrixelementscylinder}
\end{align}
up to corrections $E_L$ and $E_R$
\begin{align}
E_L  & =   \left( n_L - \frac{1}{2} \right) \left(\Delta_{L} - \frac{\nu}{24} \right) +  \frac{n_L(n_L-1)}{2} \sqrt{\nu}Q_{L}   + \frac{\nu}{6} \left( n_L - \frac{1}{2} \right)^3 \\
E_R  & =   \left( n_R+ \frac{1}{2} \right) \left(\Delta_{R} - \frac{\nu}{24} \right) +  \frac{n_R(n_R+1)}{2} \sqrt{\nu}Q_{R}   + \frac{\nu}{6} \left( n_R + \frac{1}{2} \right)^3  \label{cylinder edge factors}
\end{align}
where $\Delta_R = ,\Delta_{\alpha_R}$  and $Q_R=Q_{\alpha_R}$ stand for the conformal dimension and $U(1)$ charge of the state $\ket{\alpha_R}$. These contributions come from the left and right edges of the droplet. They encode the scaling of (the norm of) the first quantized wave function as we change the edge structure or the size of the droplet. For instance the term proportional to $\Delta_{\alpha}$ is the cylinder analog of the scaling on the plane described in Eq. (3.16) in Ref.~\cite{dubail-PhysRevB.86.245310}, while the term proportional to $Q_{\alpha}$ describes the scaling upon changing the number of electrons (while keeping the background charge constant). For the particular case of the Laughlin state, these terms reproduce the change of energy of the plasma as we modify the system size. In particular the last term is simply the electrostatic self-energy of the background charge on the cylinder. 

On the conformal field theory side of the story, these edge contributions can be interpreted as the Hamiltonian propagation of the edge states $\ket{\alpha'}$ and $\ket{\alpha}$ in the presence of the uniform background charge $\sqrt{\nu} / 2\pi$. The usual Hamiltonian on the cylinder is $-\frac{2\pi}{L} L_0$ (up to a constant additive term proportional to the central charge), but upon adding the background charge the unitary operator encoding a translation $x \to x+ 2\pi l / L$ on the cylinder becomes
\begin{align}
\lim_{n \to \infty} \left(e^{- \tau \frac{ l}{n} L_0}e^{- i \frac{l}{n} \sqrt{\nu} \varphi_0} \right)^n  = e^{- \tau \left(l  L_0 + \frac{l^2}{2}\sqrt{\nu}a_0 + \frac{l^3}{6} \nu \right) } e^{-i l \sqrt{\nu}\varphi_0}  
\end{align} 
 To sum up, a first quantized FQH conformal block on the cylinder $\Psi (z_1, \cdots, z_{N_e})= \bra{\alpha'} V(z_1) \cdots V(z_{N_e}) \ket{\alpha}$ has a site independent MPS representation  
 \begin{align}
 \ket{\psi_{\alpha_R, \alpha_L}} =  e^{\tau(E_R - E_L) }\,  \sqrt{N_e !} \, \sum_{ \{ m_1, \cdots, m_{N_{\Phi}} \}} \bra{\alpha_R} C^{m_{N_{\Phi}}} \cdots C^{m_1} \ket{\alpha_L}  |m_1 ,\cdots, m_{N_{\Phi}} \rangle \label{MPS cylinder}
 \end{align}
 where the MPS in and out states are related to the CFT one through 
 \begin{align}
 \ket{\alpha_L} = e^{-i \sqrt{\nu}\left( n_L - \frac{1}{2} \right) \varphi_0} \ket{\alpha},\qquad  \ket{\alpha_R} = e^{-i \sqrt{\nu}\left( n_R + \frac{1}{2} \right) \varphi_0} \ket{\alpha'} \label{MPS cylinder in and out states}
\end{align}
Note that the position of the droplet on the cylinder is fully determined by the states $\ket{\alpha}$ and $\ket{\alpha'}$.

\newpage

\section{Laughlin State}

\subsection{The Free Boson} \label{section free boson}

We now explicitly show the ingredients used to build the auxiliary Hilbert space, a complete basis of states, norms and matrix elements necessary for an explicit expression of the MPS matrices in a conformal field theory. In all the cases presented, the MPS matrices are essentially the modes of the electron operator $V (z)$ of \eqref{electronoperatorexpansion}
\begin{align}
V_{-n-h} = \frac{1}{2\pi i} \oint \text{d}z \, z^{-n-1} V(z) 
\end{align}
and the auxiliary space is the full Hilbert space of the CFT. As for any field theory this space is of course infinite dimensional, but for a finite number of electrons, only a finite number of Hilbert space states are necessary to obtain the \emph{exact} wave function. In the thermodynamic limit however, an exact computation of the wave function decomposition in monomials/Slaters for an infinite number of electrons requires the whole infinite dimensional auxiliary space. However the CFT Hilbert space is graded by the conformal dimension, and this provides a natural cut-off. It turns out that the finite MPS approximation to the exact wave function obtained by truncating the Hilbert space should give extremely high overlaps with the exact trial wave function for very large numbers of electrons\cite{estienne-PhysRevB.87.161112} (based on extrapolations of the approximation).

The simplest fractional quantum Hall wave functions are the Laughlin states \cite{laughlin-PhysRevLett.50.1395}. It is then no accident that they have the simplest MPS representation \cite{zaletel-PhysRevB.86.245305}. We will use them as an example before moving to the more complicated non-Abelian states. We first spend some time reviewing the free boson. On one hand the free boson provides a pedagogical playground to introduce generic concepts of CFT. On the other hand, this CFT is extremely relevant to the FQH trial wave functions. Not only is the free boson the CFT underlying the Laughlin states, but it also plays an crucial role in the underlying CFT of any FHQ trial wave functions.

\subsection{Compact Boson : Hilbert Space and Grading}
\label{Compact boson radius}

 The chiral field $\varphi(z)$ is the holomorphic part of a free massless boson $\Phi(z,\bar{z})$ with action $\frac{1}{8\pi}\int d^2z \, \partial_{\mu} \Phi \partial^{\mu} \Phi$. Its mode expansion on the plane is
\begin{align}
\varphi(z) = \varphi_0 - i a_0 \log (z) +  i\sum_{n\neq 0} \frac{1}{n}a_n z^{-n}, \qquad [a_n,a_m] = n \delta_{n+m,0} \label{Heisenberg_algebra}
\end{align}
in the normalization  $\langle \varphi(z_1) \varphi(z_2) \rangle = - \ln (z_1-z_2)$. The commutation relation of the $a_n$ is called the Heisenberg algebra. 
The $\text{U}(1)$ symmetry implies the conservation of the current $J(z) = i \partial \varphi(z) = \sum_n a_n z^{-n-1} $. In particular the zero mode $a_0$ measures the $\text{U}(1)$ charge, and $\varphi_0$ is its canonical conjugate ($[\varphi_0,a_0]=i$). Within the normalization adopted, $\sqrt{\nu}a_0$ measures the electric charge (in units of the electron charge), $\nu$ being the filling fraction.
Primary fields are so-called vertex operators, \emph{i.e.} normal ordered exponential of the free boson $V_{\beta}(z) = :e^{i\beta \varphi(z)}:  $.

The Hilbert space of the CFT plays a central role in the MPS description of the corresponding FQH states. The auxiliary space of the matrix product state is nothing but the CFT Hilbert space. Therefore it is crucial to have the latter under control. In the following we describe the Hilbert space of the free boson.

For a compactified boson at radius $R$, the $\text{U}(1)$ charge is quantized as $Q \in  \mathbb{Z}/R$.  For a $k$-clustered QH state the compactification radius is $R = k/\sqrt{\nu}$. In particular we have $R = 1/\sqrt{\nu}$ for Laughlin states, $R = 2/\sqrt{\nu}$ for paired states (including $k=2$ Jack states and superconformal states) , and $R=3/\sqrt{\nu}$ for the $\mathbb{Z}_3$ Read-Rezayi state.

For the numerical implementation of the MPS it is natural to work with the integer 
\begin{align}
N = R Q\label{QNrelation}
\end{align}
rather than the $\text{U}(1)$ charge $Q$ itself. Therefore we adopt the integer index $N$ to index the $\text{U}(1)$ charge when writing explicitely the MPS matrix elements. In the various sections of this paper we will use $Q$ and $N$ interchangeably when no confusion can arise, depending on with is the most adapted.

To any vertex operator corresponds a state in the CFT Hilbert space $|N \rangle = V_{N/R}(0) |0\rangle = e^{i N/R \varphi_0} \ket{0} $. This state has a $\textrm{U}(1)$ charge $Q=N/R$, and it is a highest weight state with respect to the Heisenberg algebra 
\begin{align}
 a_0| N\rangle = \frac{N}{R} |N \rangle,  \qquad a_n | N \rangle= 0  \qquad n > 0.
 \end{align} 
 
The above states are called \emph{primary states} (with respect to the Heisenberg algebra). The rest of the CFT Hilbert space is obtained from these primary states by repeated action of the lowering (or creation) operators $a_{-n}$. Such states are called descendants.  Partitions ( i.e. sequences of non increasing, strictly positive integers) provide a convenient way to label these states. The length of a partition $\mu = (\mu_1, \cdots, \mu_{p})$ is $l(\mu)=p$, while the size of the partition is $|\mu| = \sum_{i} \mu_i$. For $\mu$ a partition with $ m_i = m_i(\mu)$ parts equal to $i$, the quantity $z_{\mu}$ is defined as 
$ z_{\mu} = (1^{m_1}2^{m_2}\cdots) m_1! m_2! \cdots = \prod_i i^{m_i} m_i !$. With these notations at hand, one can write down an orthonormal basis of the Hilbert space
\begin{align}
|N,\mu \rangle  \equiv   \frac{1}{\sqrt{z_{\mu}}} \prod_{i=1}^{l(\mu)} a_{-\mu_i} |N\rangle, =  \frac{1}{\sqrt{z_{\mu}}} \prod_{i=1}^{\infty} a_{-i}^{m_i} |N\rangle. \label{canonical_basis_U(1)}
\end{align}
A highest weight $\ket {N}$ together with its Heisenberg descendants $\ket{N,\mu}$ form an irreducible unitary representation of the Heisenberg algebra. We denote this vector space by
 \begin{align}
 h_{N} = \text{span} \{ a_{-n_1} a_{-n_2} \cdots | N \rangle, \qquad n_i \in \mathbb{N} \} \label{Heisenberg module}
 \end{align} 
Note that all descendant $\ket{N,\mu}$ have a $\text{U}(1)$ charge $N/R$, since acting with lowering operators does not modify the charge. Using the algebra \eqref{Heisenberg_algebra} it is straightforward to check that $\bra{N',\mu'} N,\mu \rangle =  \delta_{N',N} \delta_{\mu', \mu }$, the hermitian product being defined by $a_n^{\dagger}= a_{-n}$. The total Hilbert space of the conformal field theory U$(1)_R$ is
\begin{align}
\mathcal{H} = \oplus_{N \in \mathbb{Z}} h_{N}
\end{align}

As previously mentioned, a CFT Hilbert space is always graded by the so-called conformal dimension of its states. Conformal transformations are generated by the Virasoro operators $L_n$, which appear in the mode expansion of the stress-energy tensor 

\begin{align}
T(z) = \sum_{n \in \mathbb{Z}} z^{-n-2}L_n\label{stressenergytensor}
\end{align}

Amongst them, the zero mode $L_0$ plays a particularly important role. On the plane, it generates dilatations, which is why its eigenvalues are called scaling or conformal dimensions. They are related to the critical exponents of two dimensional statistical systems. On the other hand, conformal field theories also describe $1+1$ quantum critical systems, and in this interpretation $L_0$ is proportional to the Hamiltonian on the cylinder (or the torus). In the case of the free boson, the stress-energy tensor is $T = -\frac{1}{2} : \left(\partial \varphi \right)^2: $, and the Virasoro modes are quadratic in $a_n$. In particular the zero mode is
\begin{align}
L_0  = \frac{1}{2}a_0^2  + \sum_{m >0} a_{-m} a_m  \label{Virasoro U(1)}
\end{align}
and a descendant state $\ket{N,\mu}$ in \eqref{canonical_basis_U(1)} has conformal dimension $Q^2/2+ |\mu| = N^2/2R^2 + |\mu|$.

\subsection{Vertex Operators : Matrix Elements}

Besides the auxiliary space, an important ingredient for the MPS is the set of matrix elements. Having chosen an orthonormal basis of the CFT Hilbert space \eqref{canonical_basis_U(1)}, we now have to compute the matrix elements of vertex operators.
Due to the simplicity of the Heisenberg algebra \eqref{Heisenberg_algebra}, which is nothing but decoupled copies of harmonic oscillators, it is rather straightforward to get a closed form expression. One finds
\begin{align}
 \langle  N',\mu' | :e^{i\beta \varphi(z)}: |N,\mu \rangle =  z^{\beta N/R + |\mu'| - |\mu|} \, A^{(\beta)}_{\mu',\mu}\,  \delta_{N',N+R\beta}
\end{align}
where
\begin{align}
A^{(\beta)}_{\mu',\mu} & = \prod_{j= 1}^{\infty} \sum_{r =0}^{m_j'}   \sum_{s = 0}^{m_j} \, \delta_{m_j'+s,m_j+r} \, \frac{\left(-1\right)^s}{\sqrt{r! s!}} \,   \left(\frac{\beta}{\sqrt{j}}\right)^{r+s} \sqrt{\binom{m'_j}{s}\binom{m_j}{r}} \label{matrix A}
\end{align}
The matrix elements $A^{(\beta)}_{\mu',\mu}$ are real but not symmetric in this basis
\begin{align}
A^{(\beta) }_{\mu',\mu}= A^{(-\beta) }_{\mu,\mu'}
\end{align} Their real form is a large advantage for numerical studies, and justifies our pick of the normalization of the Heisenberg algebra. Matrix elements of vertex operators are used extensively for the MPS formalism of trial wave functions, not only to write down the $B$ matrices, but also for the quasi-hole matrix.

%%%%%%%%%%%%%%%%%%%%%%%%%%%%%%%%%%%%%%%%%%%%%%%%%%%%%%%%%

\subsection{Matrix Product State Representation of the Laughlin State}

\begin{figure}
 \centering
\includegraphics[width=0.7\linewidth]{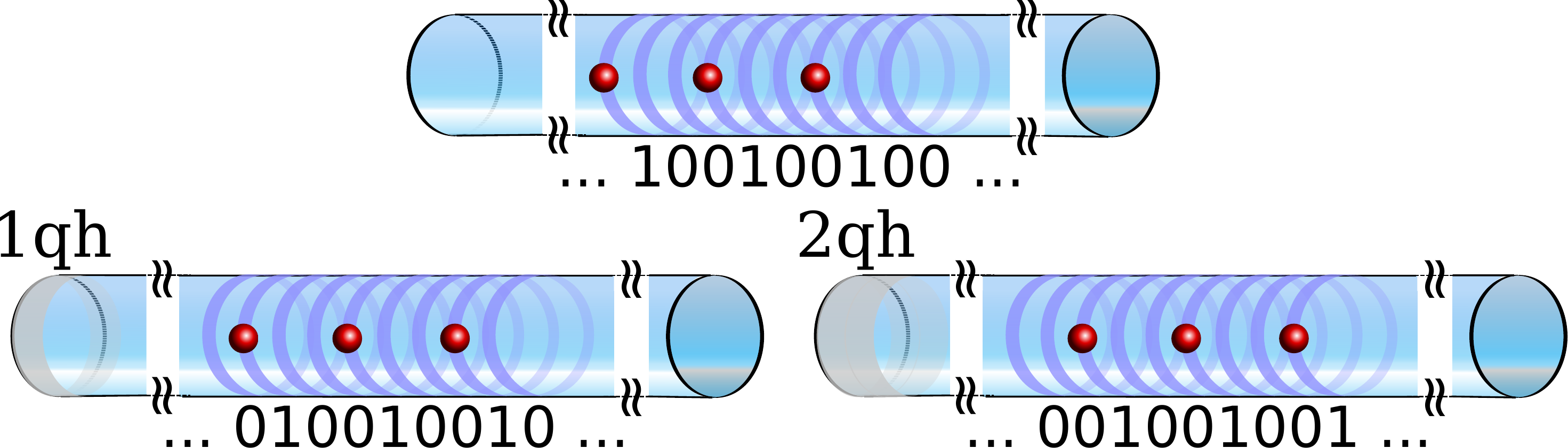}
 \caption{The three topological sectors for the Laughlin $\nu=1/3$ state corresponding to $0,1$ or $2$ quasiholes at one extremity of the infinite cylinder. These three topological sectors can also be deduced from the three nonequivalent root configurations on the infinite cylinder.}
 \label{threetopsectors}
\end{figure}

 The $\nu = \frac{1}{q}$ Laughlin wave function for $N_e$ electrons on the disk is simply obtained as the correlation function\cite{Moore1991362,Fubini-1991MPLA....6..347F,Cristofano-1991PhLB..262...88C,Cristofano-1991MPLA....6.2985C,Cristofano-1991MPLA....6.1779C} 
\begin{align}
\langle q N_e  | V(z_1) \cdots V(z_{N_e})|0 \rangle   = \prod_{i<j} (z_i -z_j)^q
\end{align}
where a $U(1)$ charge $N = -q N_e$ at infinity ensures charge neutrality. Edge states are obtained by changing the primary state $ \bra{qN_e}$ to a descendant $ \bra{qN_e, \mu}$. The corresponding electron operator is the vertex operator $V(z)=:\exp \, ( i \sqrt{q} \varphi(z)):$. Its electric charge is $1$, as it should. Quasi-holes on the other hand have an electric charge $1/q$, as can be seen from the corresponding vertex operator $:\exp \, ( i  \varphi(z) / \sqrt{q}):$. As discussed in section~\ref{CFT states intro}, the Hilbert space can be decomposed into topological sectors. For the $q$ Laughlin state, there are $q$ such sectors $\mathcal{H}_a = h_{a \text{ mod } q}$, defined as:
\begin{align}
 h_{a \text{ mod } q}  \equiv \bigoplus_{N = a \text{ mod } q} \,  h_{N} \label{Heisenberg modulo module}
\end{align}
These $q$ sectors can be thought as the $q$ Laughlin states with $0,1,...,q-1$ quasiholes at infinity, as depicted on Fig.~\ref{threetopsectors}. On the cylinder, a first-quantised wave functions in topological sector $a$ is generically of the form
\begin{align}
\langle  a + q N_e ,\mu' | V(z_1) \cdots V(z_{N_e})| a ,\mu \rangle \label{Laughlin cylinder a}
\end{align}
up to global translations along the cylinder (implemented by a shift of the U$(1)$ charge of the in and out states). The left and right edge excitations are encoded in the partitions $\mu$ and $\mu'$. The densest droplet is obtained for $\mu = \mu' = \emptyset$.

We are now ready to compute the MPS matrix elements for the Laughlin state. Plugging the explicit basis \eqref{canonical_basis_U(1)} in the generic formula of the $C^m$ matrix in \eqref{normalizedmps}, the $C^0$ and $C^1$ matrices read: 
\begin{align}
\langle N',\mu' | C^0 | N,\mu \rangle & = e^{-\tau (\delta' + \delta)/2}\delta_{N',N-1} \delta_{\mu',\mu} \\
\langle N',\mu'  | C^1 |N,\mu \rangle & = e^{-\tau (\delta' + \delta)/2} (L\sqrt{\pi})^{1/2}\, \delta_{N',N+q-1} \, \delta_{|\mu'|+N + \frac{q-1}{2},|\mu|} A_{\mu', \mu}^{(\sqrt{q})} \label{thinannulusmpslaughlin}
\end{align} 
where $N = R Q = \sqrt{q} Q \in \mathbb{Z}$, while $\delta = \frac{N^2}{2q} + |\mu|$ (resp. $\delta' = \frac{N'^2}{2q} + |\mu'|$) is the conformal dimension of $\ket{N,\mu}$ (resp. $\ket{N',\mu'}$). We have mentioned already than for a compact boson at radius $R = \sqrt{q}$, the $U(1)$ charge $Q$ is quantized as $\sqrt{q} Q \in {\mathbb{Z}}$. The expression of a Laughlin ground-state  \eqref{Laughlin cylinder a} with a topological charge (anyon) $a$ at infinity is 
\beq
\ket{\psi}_a =  \sum_{m_1, \ldots, m_{N_\Phi}} \bra{a} C^{m_{N_\Phi}} \ldots C^{m_1} \ket{a} \ket{m_1, \ldots, m_{N_\Phi}} \label{laughlinstateketandbra}
\eneq
where $N_{\phi} = qN_e$ is the flux on the cylinder. Descendants in place of the incoming (or outgoing) end states would produce states in the same topological sector, but with excitations localized on the left (or right) edge of the droplet. 

For finite $N_e$, only a known finite number of descendants are needed to obtain the Laughlin state exactly. This state exhibits by construction all the properties previously found for the Laughlin state, such as squeezing, which can be checked explicitly from the MPS construction (see Appendix~\ref{AppendixSqueezing}).

For a non truncated auxiliary space, the matrices above faithfully reproduce the Laughlin state on the cylinder (or the thin-annulus if we change $C^m$ to $B^m$). However, for efficient approximations of the Laughlin state, we now restrict the Hilbert space of the CFT. A very important parameter is the truncation parameter $|P_{\text{max}}|$ which gives the maximum size $|\mu|$ of the partition $\mu$ that we keep in our calculation. For $|P_{\text{max}}|=0$, we find that the state in \eqref{laughlinstateketandbra} contains only one Slater determinant, that of the occupation numbers $0^{q-1-a}10^{q-1}10^{q-1}\ldots 10^{q-1}10^a$, the root partition\cite{Bernevig-PhysRevB.77.184502,Bernevig-PhysRevLett.100.246802} (related to the thin torus\cite{bergholtz-PhysRevLett.94.026802,Bergholtz-2006-04-L04001}). For $|P_{\text{max}}|>0$, the state contains partitions squeezed from the root state by at most $|P_{\text{max}}|$ squeezings (see Appendix~\ref{AppendixSqueezing}). In the Appendix~\ref{AppendixSqueezing} we obtain the first few coefficients of the Laughlin state Slaters.

%%%%%%%%%%%%%%%%%%%%%%%%%%%%%%%%%%%%%%%%%%%%%%%%%%%%%%%%%%%%%%%%%%%%%%%%%%

\newpage

\section{Interacting CFTs}

In the MPS formulation for the Laughlin state we have used the fact that the CFT is a free theory. This led to the simple structure of the descendants \eqref{canonical_basis_U(1)}, to the closed form expression for the matrix elements \eqref{matrix A}, and to the fact that the Virasoro modes can be written in terms of decoupled harmonic oscillators as in \eqref{Virasoro U(1)}. The generic situation is much more complicated. The intricacies of building an MPS formulation of FQH states from CFT are similar for different states, and involve complications with interacting Hilbert spaces, null modes, and the computation of matrix elements. However, the details of the CFT descendants are different for different CFTs. We start by analyzing the simplest example possible, that of FQH states obtained as conformal blocks of Virasoro algebras. These include the Moore-Read\cite{Moore1991362} and Read-Rezayi states\cite{read-PhysRevB.59.8084}. For the Moore-Read state, a simpler representation is possible exploiting the representation of the state in terms of free Majorana fermions. We start off by reviewing some notions of CFT on the plane.

\subsection{Virasoro Algebra}

A conformal field theory is a field theory invariant under any conformal transformation, \emph{i.e.} local dilatation's/rotations. In two dimensions the conformal group is infinite dimensional, and conformal invariance yields as many symmetry constraints (Ward identities). This makes two dimensional CFTs exactly solvable, even in the presence of strong interactions. For a detailed introduction to CFT, we refer the reader to the book \cite{DSM}. Any CFT contains a field called the stress energy tensor $T(z)$. Its modes $L_n$ (see  \eqref{stressenergytensor}) are called the Virasoro modes. They are the generators of the conformal transformations and they obey the celebrated Virasoro algebra
\begin{align}
[L_n,L_m] = (n-m)L_{n+m} + \frac{c}{12}n(n^2-1)\delta_{n+m,0} \label{virasoroalgebra}
\end{align}
The quantity $c$ appearing in the r.h.s. is a real number called the central charge, and is the most important parameter of a given CFT. For instance a compactified boson $\text{U}(1)_R$ has central charge $c=1$, while a Majorana fermion has $c= 1/2$. As a rule of thumb, the larger the central charge, the richer the field content of the CFT.

\subsection{Structure of the Hilbert Space}\label{structureHilbertspace}

The CFT Hilbert space is spanned by the action of the Virasoro modes $L_{-n}$ on primary fields/states $ \ket{ \Phi_{\Delta} }$ characterized by
\begin{align}
 L_0  \ket{ \Phi_{\Delta} }= \Delta  \ket{ \Phi_{\Delta} }, \qquad  L_n  \ket{ \Phi_{\Delta} }= 0 \quad n>0 \label{virasoroprimaries}
\end{align}
 where the parameter $\Delta$ is the conformal dimension. The states obtained by acting with Virasoro modes on primary fields are called descendants, and are labeled by partitions $\theta = (\theta_1,\cdots, \theta_n)$
 \begin{eqnarray}
& \ket{ \Phi_{\Delta} , \theta } \equiv   \frac{1}{\sqrt{z_{\theta}}}  \prod_{i=1}^{l(\theta)} L_{-\theta_i} \ket{ \Phi_{\Delta}}  = \frac{1}{\sqrt{z_\theta}}  \prod_{i=1}^{\infty} L_{-i}^{m_i}   \ket{ \Phi_{\Delta}} ,&\nonumber\\
& L_0  \ket{ \Phi_{\Delta} , \theta }  = (\Delta+ |\theta|) \ \ket{ \Phi_{\Delta} , \theta }& \label{virasorodescendants}
 \end{eqnarray}
 with $ \theta _1 \geq  \theta _2 \geq \cdots \geq  \theta _n >0$ and $m_i$ is the number of parts equal to $i$ in the partition $\theta$. Note that the combinatorial prefactor $z_\theta = \prod_i i^{m_i} m_i!$ in \eqref{virasorodescendants} does not make the state normalised, but it prevents the appearance of huge terms in the overlaps between descendants. This is helpful for numerical stability. A highest weight $\ket {\Phi_{\Delta} }$ together with its Virasoro descendants form an irreducible representation of the Virasoro algebra \footnotemark \footnotetext{After removing the null vectors} . We denote this vector space by
 \begin{align}
 \mathcal{V}_{\Phi_{\Delta}} = \text{span} \{ L_{-n_1} L_{-n_2} \cdots | \Phi_{\Delta} \rangle, \qquad n_i \in \mathbb{N} \} \label{Virasoro module} 
 \end{align} 
 Using a hermitian scalar product defined by $(L_n)^{\dagger} = L_{-n}$, the Virasoro algebra \eqref{virasoroalgebra} and the highest weight condition \eqref{virasoroprimaries}, we can now compute any overlap between descendants $ \bra{ \Phi_{\Delta'} , \theta' } \Phi_{\Delta}, \theta \rangle $. An example of how to do this is given in Appendix~\ref{Appendix3}. 
\be
 \bra{ \Phi_{\Delta'} , \theta' } \Phi_{\Delta}, \theta \rangle \propto \delta_{\Delta', \Delta} \delta_{| \theta'|, | \theta|} \label{descendantsoverlap}
\eneq

A first difference with the U(1) case is that the descendants \eqref{virasorodescendants} are generically independent but not orthogonal - the $\delta$-function in \eqref{descendantsoverlap} 
applies to the total weight $| \theta| = \sum_i \theta_i$ of the two partitions and not to the partitions themselves. Hence, at each level $|\theta|$, we have to numerically build an orthonormal basis and compute its overlap Gram matrix. Indeed the MPS construction is obtained by inserting $\mathbb{1} = \sum_{\alpha} |\alpha \rangle \langle \alpha |$, which requires an orthonormal basis $|\alpha \rangle$. These calculations are done numerically (due to the complications of the Virasoro algebra the Gram matrix is not known analytically), but with exact number precision.

An additional complication is the presence of null vectors (states of vanishing norm). We exemplify such cases in Appendix~\ref{Appendix3}. For a generic conformal dimension $\Delta$, these states do not occur. However in a rational CFT (i.e. a CFT with a finite number of primary fields) there are generically null vectors. All the interesting FQH wave functions correspond to a rational CFT. Indeed a non-rational CFT would yield an unphysical FQH state with an infinite degeneracy on the torus. This is the case for instance with the Haffnian\cite{green-10thesis}. The number of non-independent states is known analytically from the CFT characters, and our numerical procedure for computing overlaps reproduces this counting.

Null vector are states perpendicular to all states (including themselves : they have zero norm). Such states decouple from the theory and have to be dropped out of the Hilbert space. For our purposes, upon reaching a level with a null vector before performing the Gram-Schmidt process, we need to detect and drop all null vectors at this level, i.e. all states with zero norm. For non-unitary CFTs (such as the one underlying the Gaffnian state \cite{simon2006}), another complication is that some descendants have a negative norm. These states have to be included, and the sign of their norm can be handled by an extra diagonal matrix D acting on the right of the MPS matrices :
$B^m \rightarrow  B^m D$.

There exists an alternate way of removing the null modes without actually finding them. At level $|\lambda|$, enumerate the Virasoro primaries of \eqref{virasorodescendants} and build the Gram matrix step by step by adding each descendant and computing its overlaps with the ones already in the list. At each step (for each addition of a descendant), compute the determinant of the Gram matrix. If it is non-zero, keep the state added to the Hilbert space. If the determinant of the Gram matrix becomes zero upon addition of the new descendant, discard the state as it means it is linearly dependent to the ones already there. Stop when the dimension of the Gram matrix has reached the analytically computable dimension of the non-null vector subspace. This procedure, starting from a primary state which is linearly independent, produces a complete non-orthogonal basis of states, and the Gram matrix can be diagonalized to obtain an orthonormal basis.

\subsection{Neutral Part of the Matrix Elements}\label{neutral_basis_section}

The matrix elements of the MPS depend on the actual field content of the CFT and on our choice for the electron operator of that CFT. The "electron operator" is a primary field of the following form
\begin{align}
V(z) & =  :e^{i\sqrt{q}\varphi (z)}: \qquad \text{for Laughlin}  \\
V(z) & =   \Psi(z)\, \otimes\,   :e^{i\sqrt{q}\varphi (z)}:  \qquad \text{more generally} \label{electron_operator}
\end{align}
where $\Psi(z)$ lives in the so-called neutral conformal field theory $\text{CFT}_n$, and $q = 1/\nu$ is the inverse filling fraction and need not be an integer anymore. In this case the underlying chiral CFT is a tensor product $\text{CFT}_n   \otimes  \text{U}(1)$. The CFT Hilbert space is of course a tensor product of the $\text{U}(1)$ Hilbert space and the neutral one. It is convenient to work in a factorized basis  $ |\alpha \rangle = | \alpha_n \rangle \otimes | \alpha_Q \rangle$, where $\ket{\alpha_n}$ is a state in $\text{CFT}_n$, and $|\alpha_Q \rangle$ is a state in $\text{U}(1)$. In this basis the 3-point function we want to compute factorizes as
\begin{eqnarray}
\langle \alpha_n';\alpha_Q'| V(1) | \alpha_n ; \alpha_Q \rangle &=& \langle \alpha_n' | \Psi(1) | \alpha_n \rangle \, \, \langle  \alpha_Q' | :e^{i \sqrt{q} \varphi(1)}: |\alpha_Q \rangle 
\end{eqnarray}
The U$(1)$ part $ \langle  \alpha_Q' | :e^{i \sqrt{q} \varphi(1)}: |\alpha_Q \rangle $ is a simple free boson calculation, and can be computed analytically as in the case of the Laughlin MPS.  The computation of the neutral part $ \langle \alpha_n' | \Psi(1) | \alpha_n \rangle $ on the other hand is more involved, and the tools depend on the details of the neutral CFT.

For the case of Virasoro descendants, it is convenient to work in the canonical "basis" \eqref{virasorodescendants} in order to compute matrix elements involving descendants. One has to keep in mind that it might be over-complete due to possible null vectors in the Verma module of $|\Phi_\Delta\rangle$ and $|\Phi_\Delta'\rangle$. Once the matrix elements are computed in this basis, they can be transformed to the orthonormal basis obtained by Gram-Schmidting the Gram matrix. In the descendant basis we need to compute all the matrix elements:
\begin{align}
 \bra{ \Phi_{\Delta'} , \theta' }  \Psi(1) \ket{ \Phi_{\Delta}, \theta } 
  \label{ME_Delta}
\end{align} 
This is a much larger number than the matrix elements of the Gram matrix as the conformal dimension is not conserved. Numerically, this is the most time-consuming part of the algorithm. Once these matrix elements are known, it is straightforward to compute the matrix in any orthonormal basis. We first need to be able to compute the level $0$ primary matrix element $\bra{ \Phi_{\Delta'} }  \Psi(1) \ket {\Phi_{\Delta}}$ but this is simply the OPE structure constant $D_{\Delta, \, \Delta'}$ in 
\begin{align}
\Psi(z) \Phi_{\Delta}(w) = D_{\Delta}^{\, \, \, \, \Delta'} \frac{\Phi_{\Delta'}(w)}{(z-w)^{h_{\Phi} + \Delta- \Delta'}} + \cdots
\end{align}
These are known in closed form for minimal models. As it will become clear these factors multiply all matrix elements as an overall prefactor. This factor is sector dependent and cannot be ignored. Once this factor is known, matrix elements between descendants can be addressed. They do not have an analytically closed formula but can be obtained numerically with exact accuracy by repeated action of the relation 
\begin{align}
\bra{\alpha'}  [ L_m -L_0 \, ,  \Psi(1) ] \ket{\alpha} = m h_{\Psi}\,  \bra{\alpha'}  \Psi(1)  \ket{\alpha}
\end{align}
A detailed method and some examples are given in Appendix~\ref{AppendixNeutralCFT}.

\subsection{MPS Matrix Elements}\label{secMPS Matrix Elements}

From the results derived in the prior sections, an over-complete "basis" of states in the CFT is given by
\begin{align}
\ket{ \Phi_\Delta, \theta \,; N, \mu}= \ket{ \Phi_\Delta, \theta} \otimes\ket{ N, \mu}\label{Vir x U(1) basis}
\end{align}
where $\ket{  \Phi_{\Delta}, \theta }$ is given by \eqref{virasorodescendants} and $ | N, \mu \rangle$ by \eqref{canonical_basis_U(1)}. The state $\ket{ \Phi_\Delta, \theta \,; N, \mu} $ has \emph{total descendant level} $P = |\theta| + |\mu|$ \footnotemark \footnotetext{For the $\mathbb{Z}_3$ Read-Rezayi state in the minimal model approach, there is an extra shift for the $W$ field such that the lowest $P$ value is 3 (see Appendix~\ref{RR}).}.  The quantity $P$ will be the truncation parameter for the MPS, i.e. we will build our Hilbert space only up to a total level $|P_{\text{max}}|$. 

The matrix elements of $C^0$ are
\begin{eqnarray}
&&\bra{ \Phi_{\Delta'}, \theta' \,; N', \mu'} C^0 \ket{ \Phi_\Delta, \theta \,; N, \mu} \nonumber\\ 
&=& e^{-\tau (\delta' + \delta)/2} \,\bra{ \Phi_{\Delta'}, \theta}
  \Phi_{\Delta}, \theta \rangle \; \delta_{\mu,\mu'} \, \delta_{N',N-k}  \label{B0 Virasoro}\nonumber\\ 
& =& e^{-\tau (\delta' + \delta)/2} \, \bra{ \Phi_{\Delta'}, \theta}
  \Phi_{\Delta}, \theta \rangle \; \delta_{\Delta',\Delta}\, \delta_{|\theta'|,|\theta|}\, \delta_{\mu,\mu'}  \,  \delta_{N',N-k} 
\end{eqnarray}
\begin{eqnarray}
  \addtolength{\fboxsep}{5pt}
&&\bra{ \Phi_{\Delta'}, \theta' \,; N', \mu'} C^1 \ket{ \Phi_\Delta, \theta \,; N, \mu}  \nonumber\\ 
& = &  e^{-\tau (\delta' + \delta)/2} (L\sqrt{\pi})^{1/2}\, \bra{ \Phi_{\Delta'} , \theta' }  \Psi(1) \ket{ \Phi_{\Delta}, \theta } \nonumber\\ 
&&\times \delta_{\Delta'+|\theta'|+|\mu'|+N/k+ \frac{q-1}{2},\Delta +|\theta|+|\mu|}
   \delta_{N',N+k(q-1)} \,   A_{\mu',\mu}^{(\sqrt{q})}    \label{B1 Virasoro}
\end{eqnarray}
where $\delta = \Delta+ \frac{N^2}{2k^2q} + |\theta| + |\mu|$ is the conformal dimension of $\ket{ \Phi_\Delta, \theta \,; N, \mu}$.

Once the matrices $C^0$ and $C^1$ have been computed in this "basis", the last step is to change to an orthonormal basis obtained as explained in section~\ref{structureHilbertspace}. For most applications the algebra of the neutral CFT is simply the Virasoro algebra, which we will show here. This will be the case for all $k=2$ Jack states (including Moore-Read and Gaffnian), but also the $\mathbb{Z}_3$ Read-Rezayi state (presented in Appendix~\ref{RR}). For a generic $k>2$ Jack states, one needs to deal with a CFT with an enlarged algebra (the so-called $\mathcal{W}_k$ algebra).  Although it is possible to work with such an extended algebra (see Appendix~\ref{RR}), it turns out to be numerically much more efficient to work only with the Virasoro one when available. This is the case for the $\mathbb{Z}_3$ Read-Rezayi states, which can be understood in terms of the Virasoro or the $\mathcal{W}_3$ algebra. There are cases however where a description in terms of the Virasoro algebra is not available (or to be more precise such a description would involve an infinite number of primary fields). One has then to deal with the extended algebra. This is the case for all unitary CFTs with a central charge $c \geq 1$. As an illustration of such an extended algebra, we give the MPS description of the superconformal minimal models in Appendix~\ref{superconformal}.

%%%%%%%%%%%%%%%%%%%%%%%%%%%%%%%%%%%%%%%%%%%%%%%%%%%%%%%%%%%%%%%%%%%%%%%%%

\newpage

\section{Paired Jack states}
\label{$k=2,r$ MPS}

The first and most important example of a neutral CFT are the Virasoro minimal models $M(p,p')$, which are labeled by two coprime integers $p$ and $p'$. For details about these CFT we refer the reader to Ref.~\cite{DSM}.  The minimal models contain a finite number of primary fields $\Phi_{(n|m)}$, with $1 \leq n \leq p'-1$ and $1\leq m \leq p-1$, and conformal dimension
\begin{align}
\Delta_{(n|m)} = \frac{(np-mp')^2-(p-p')^2}{4p p'}
\end{align}
This field content is not arbitrary and comes from the consistency of the theory on the torus (modular invariance). The same consistency relations are required for the corresponding FQH state. Indeed the manifold of ground states has to be invariant under modular transformations on the torus.
 
The $(k=2,r)$ Jack states - with filling fraction $ \nu = 2/(r+2m)$ - exist for any $r$ such that $3$ and $r-1$ are coprime. In the bosonic $m=0$ case these are wave functions that vanish as the $r$'th power of the difference of coordinates of $k+1$ particles coming together. For $r=2$ one recovers the Moore-Read state, while $r=3$ corresponds to the so-called Gaffnian. The corresponding neutral CFT is the minimal model $M(3,r+2)$, with electron operator 
\begin{align}
 V(z)=  \Psi(z) \, \otimes\,  :e^{i\sqrt{q}\varphi (z)}: , \qquad q = \nu^{-1} = r/2 + m
\end{align}
 It has conformal dimension $h= q/2+ h_{\Psi}$, and it involves the neutral primary $\Psi= \Phi_{(1|2)}$ with conformal dimension $h_{\Psi} = r/4$ and fusion rules
 \begin{align}
\Psi \times \Psi = 1 &, \qquad \Psi \times 1 = \Psi \label{Z2 fusion rules}
 \end{align}
from which the clustering parameter $k=2$ can be read off. 
 
\subsection{Hilbert Space of the Minimal Model $M(3,r+2)$}

In order to build the corresponding MPS we need to specify the CFT Hilbert space. In this paragraph we first give the list of all neutral primary fields (w.r.t. the neutral Virasoro algebra), and then their matching U$(1)$ charge. Finally we rearrange the total Hilbert space according to topological sectors. 

In the minimal model $M(3,r+2)$, primary fields come in pairs $\sigma_n = \Phi_{(n+1|1)}$ and $\varphi_n = \Phi_{(n+1|2)}$ with $n=0, \cdots , \lfloor r/2 \rfloor$. They have conformal dimension
\begin{align}
\Delta_{\sigma_n} = \frac{n}{4(2+r)}(3n+2-2r),\qquad \Delta_{\varphi_n} = \frac{(n-r)}{4(2+r)}(3n-2-r)
\end{align}
Any of these two fields can be obtained from the other by fusion with $\Psi$
\begin{align}
\Psi \times \sigma_n = \varphi_n &, \qquad \Psi \times \varphi_n = \sigma_n   \label{Z2 fusion rules qh}
\end{align} 
If we were to include $\Psi(z)$ in the neutral chiral algebra, $\ket{\varphi_n}$ would be a descendant of $\ket{\sigma_n}$. The $n=0$ fields are simply $\sigma_0 =1$ and $\varphi_0 = \Psi$. When $r$ is even there is a little subtlety : the fields $\sigma_{r/2}$ and $ \varphi_{r/2}$ have the same conformal dimension and are actually the same field (they have to be identified, $\sigma_{r/2} = \varphi_{r/2}$). All these primary fields must be taken into account in order to obtain the full ground-state degeneracy. This is the trademark of a non-Abelian quantum Hall state.

The bosonic part of the CFT underlying the $(k=2,r)$ Jack states in a compact boson at radius $R= 2\sqrt{q}$. However the full CFT is not just quite the tensor product $M(3,r+2) \otimes U(1)_{2\sqrt{q}}$. To ensure the polynomial nature of the trial wave function as we approach an electron, the electric charge of quasi holes is not free. For this particular case, the correct choice can be found in Ref.~\cite{Bernevig-2009JPhA...42x5206B}  
\begin{eqnarray}
&V_{\sigma_n}(z) = \sigma_n(z) \,  \otimes \,  :\exp \left(i \frac{n}{2}\sqrt{\nu} \varphi (z)\right):,& \nonumber\\
& V_{\varphi_n}(z) = \varphi_n(z) \,  \otimes \,  :\exp \left(i  \frac{r-n}{2}\sqrt{\nu}  \varphi (z)\right):\,.&
\end{eqnarray}
up to Abelian quasi-holes $:e^{im\sqrt{\nu}\varphi}:$. This leads to selection rules in the tensor product $M(3,r+2) \otimes U(1)_{2\sqrt{q}}$, and the allowed $U(1)$ charge in the non-Abelian topological sector $n$ is : 
\begin{itemize}
\item $\sigma_n$ and its descendants come with a $U(1)$ charge $N$ obeying $N =  n$ mod $2$
\item while for $\varphi_n$ and its descendants we have $N =  r-n$ mod $2$
\end{itemize}
The quasi-particle index $n$ ranges from $0$ to $\lfloor r/2 \rfloor$, and these states span the whole CFT Hilbert space.

\subsection{Topological Sectors, Ground State Degeneracy and Thin-Torus Limit} 

The total Hilbert space can be decomposed into $(r+1)q$ topological sectors. Using the notations of \eqref{Heisenberg modulo module} and \eqref{Virasoro module}, the topological sectors can be written in compact form
\begin{itemize}
\item for $n \neq r/2$ (i.e. $ 0 \leq n < r/2 $)
\begin{align}
\mathcal{H}_{\left(\sigma_n,j\right)} =  \left( M_{\sigma_n}^{(0)} \otimes  h_{n + 2 j \text{ mod } 4q} \right) \,     \oplus  \,\left( M_{\sigma_n}^{(1)}  \otimes \ h_{n + 2(j+q) \text{ mod }4q } \right) \label{topologicalsectorspairedstates1} 
\end{align}
where $j =0 , \cdots, 2q-1$, $M_{\sigma_n}^{(0)} = \mathcal{V}_{\sigma_n}$ and $M_{\sigma_n}^{(1)} = \mathcal{V}_{\varphi_n}$,
\item while when $r$ is even we have 
\begin{align}
\mathcal{H}_{\left(\sigma_{r/2},j\right)} =  M_{\sigma_{r/2}} \otimes  h_{\frac{r}{2} + 2j \text{ mod } 2q}  \label{topologicalsectorspairedstates2}
\end{align}
where $j =0 , \cdots, q-1$, $M_{\sigma_{r/2}} = \mathcal{V}_{\sigma_{r/2}}$.
\end{itemize}
It is straightforward to check that these sectors are stable under action of $V(z)$ and $V^*(z)$. Since the repeated action of the modes of $V$ and $V^*$ span both the Virasoro and Heisenberg modes, the topological sectors are nothing but the irreducible modules of the extended algebra generated by $V$ and $V^*$. 

Each topological sector $\mathcal{H}_{(\sigma_n,j)}$ yields a ground state on the cylinder, with the corresponding topological charge at infinity. In the particular case of the Jack states ground states can be characterized by a root partition obeying a generalized Pauli principle\cite{Bernevig-PhysRevB.77.184502,Bernevig-PhysRevLett.100.246802}. The sector $\mathcal{H}_{(\sigma_n,j)}$ corresponds to the root partition (see Appendix~\ref{AppendixThinTorus})
\begin{align}
\cdots 0^{r-n+m-1-j}10^{n+m-1}10^j \cdots
\end{align}

Let us illustrate this for the Moore-Read state ($r=2$) and the Gaffnian ($r=3$), in the bosonic $m=0$ case. In the MR case, there are $3$ ground states on the cylinder: $\cdots2020 \cdots$, $\cdots0202\cdots$ and $\cdots1111 \cdots$. Correspondingly there are three topological sectors
\begin{align} 
\mathcal{H}_{(1,j)} = \left(  M^{(0)}_{1} \otimes h_{2j \text{ mod } 4} \right) \, \oplus  \, \left( M^{(1)}_{1} \oplus h_{2(j+1)  \text{ mod } 4} \right), \qquad j = 0,1 
\end{align}
and 
\begin{align} 
\mathcal{H}_{(\sigma,0)} = \left(  M_{\sigma} \otimes h_{1 \text{ mod } 2} \right)\qquad \qquad (j=0) 
\end{align}
Meanwhile the Gaffnian has $6$ topological sectors
\begin{align} 
\mathcal{H}_{(1,j)} = \left(  M^{(0)}_{1} \otimes h_{2j \text{ mod } 6} \right) \, \oplus  \, \left( M^{(1)}_{1} \oplus h_{2j+3  \text{ mod } 6} \right), \qquad j = 0,1,2 
\end{align}
with root partitions pattern $200$, $020$ and $002$, and
 \begin{align} 
\mathcal{H}_{(\sigma,j)} = \left(  M^{(0)}_{\sigma} \otimes h_{1+2j \text{ mod } 6} \right) \, \oplus  \, \left( M^{(1)}_{\sigma} \oplus h_{2j+4  \text{ mod } 6} \right), \qquad j = 0,1,2 
\end{align}
corresponding to $110$, $011$ and $101$.

\newpage

\section{Performing Computations with the MPS}

Crucially, the implementation of the MPS allows us to compute physical properties without resorting to the original many-body Hilbert space. While this is standard procedure in the MPS community, we briefly repeat it here for inclusiveness. We will introduce the standard MPS transfer matrix (or $E$ matrix) with which  computations on the infinitely long cylinder such as entanglement spectrum will be conducted. We first present the block structure of the MPS matrices relative to the various topological sectors of the FQHS.  We then particularize our description to the $E$ matrix of FQH states, and discuss specific issues such as its block structure and the properties of its largest eigenvalue eigenvector(s). 

In this section the notation $B^m$ stands for a generic site independent MPS matrix. For our purposes it means either the thin-annulus matrix $B^m$  of \eqref{MPS B matrix}  or the cylinder matrix $C^m$ of \eqref{normalizedmps}.

\subsection{Block Structure of the $B$ Matrices}
\label{B block structure}

So far we have considered FQH trial wave functions with a total underlying CFT of the form $\text{CFT}_{n} \otimes \text{U}_{R}$. The electron operator 
\begin{align}
V(z) = \Psi(z) \, \otimes \, : \exp \left(  i \frac{1}{\sqrt{\nu}} \varphi (z) \right): 
\end{align}
spoils this factorization by coupling the neutral and the $\text{U}(1)$ parts of the CFT. As we argued previously, in order to understand the topological phase of the corresponding FQHS, it is natural to enlarge the chiral algebra to the one spanned by the electron operator $V(z)$ and its conjugate $V^*(z)$. This allows to rearrange the total Hilbert space according to topological sectors $\mathcal{H}_a$, \emph{i.e.} irreducible representations of this enlarged algebra.  
Such an irreducible representation is spanned by acting on a primary field (highest weight state) $\ket{\Phi_a}$ with the modes of $V(z)$ and $V^{*}(z)$. This sector is relevant for computing bulk wave functions with a quasi-hole $\Phi_a$ located at the origin, and its conjugate $\Phi_{\bar{a}}$ at infinity :
\begin{align}
\langle \Phi_{\bar{a}}(\infty)   V(z_1) V(z_2)\cdots   V(z_{N_e}) \Phi_a(0) \rangle       = \bra{\Phi_a}  V(z_1) V(z_2)\cdots   V(z_{N_e}) \ket{\Phi_a}
\end{align}  
with an implicit neutralizing background charge. 

 It is clear that the electron operator $V(z)$ cannot couple sectors with different topological charges, and the MPS matrices $A^m[j]$ of \eqref{MPS A matrix} - which do not include the neutralizing background charge - are block diagonal according to topological charge. The site independent matrices $B^m$ of \eqref{MPS B matrix} however do include a charge shift $Q \to Q - \sqrt{\nu}$ (\emph{i.e.} $N \to N - k$) accounting for the amount of background charge per orbital (see section~\ref{Site Independent MPS}). Note that the insertion of an Abelian quasi-hole $:e^{-i\sqrt{\nu}\varphi(w)}:$ would yield the same charge shift. Therefore the matrices $B^m$ change the Abelian part of the topological charge by one unit, and are not block diagonal in the topological sectors.

In order to understand how the decoupling between topological sectors $\mathcal{H}_a$ comes about, one must take a closer look at their decomposition onto $\text{CFT}_n \otimes U(1)$. As mentioned in section~\ref{CFT section}, the topological charge $a = (\phi,j)$ is characterized by a neutral sector $\phi$ and an Abelian charge $j$. At the level of the field $\Phi_a$ we have
\begin{align}
\Phi_a(z) = \phi(z) \otimes \exp \left[ i \sqrt{\nu} \left( \frac{b_{\phi}}{k} + j \right) \varphi(z) \right]
\end{align}  
For a given neutral sector $\phi$, the Abelian charge lives in $j=0,\cdots,kq-1$\footnotemark \footnotetext{As mentioned in section~\ref{CFT section} the identification can be $\phi$ dependent $j = j + k_{\phi} q$, where $k_{\phi}$ is a divisor of $k$. We present here the most common case $k_{\phi}=k$, the generalization to other cases being straightforward.}. The neutral primary field $\phi$ is a highest weight \emph{w.r.t.} the modes of $\Psi(z)$. Neutral descendants of $\phi$ can be classified according to their $\mathbb{Z}_k$ charge \footnotemark \footnotetext{Or $\mathbb{Z}_{k_{\phi}}$ charge.}. Let us denote by $M^{(n)}_{\phi} $ the space spanned by the neutral descendants of $\phi$ with $\mathbb{Z}_k$ charge $n$, \emph{i.e.} the states obtained from the neutral primary field $\phi$ by action of ($n$ mod $k$) modes of $\Psi(z)$. The neutral part of the electron field increases this charge by one unit $\Psi \times M^{(n)}_{\phi} \in M^{(n+1)}_{\phi}$.

With these notations at hand, the explicit structure of the sector with topological charge $a = (\phi,j)$ is 
\begin{align} 
\mathcal{H}_{(\phi,j)} = \bigoplus_{i=0}^{k-1} \left( M^{(i)}_{\phi} \otimes h_{k(j+iq) + b_{\phi} \text{ mod } k^2q} \right); \qquad q = \frac{r}{k}+m 
\end{align}
Let us illustrate this construction with the $\nu =1$ Moore-Read case. In the neutral sector of the identity ($\phi=1$), there are $2$ topological sectors:
\begin{align} 
\mathcal{H}_{(1,j)} = \left(  M^{(0)}_{1} \otimes h_{2j \text{ mod } 4} \right) \, \oplus  \, \left( M^{(1)}_{1} \oplus h_{2(j+1)  \text{ mod } 4} \right), \qquad j = 0,1 
\end{align}
where for the particular case of MR the sectors $ M^{(0)}_{1} = \mathcal{V}_1$ and $M^{(0)}_{1} = \mathcal{V}_{\Psi}$ (in the notations of \eqref{Virasoro module}) are the decomposition of the neutral vacuum sector according to electron parity.

\begin{figure}
 \centering
\includegraphics[scale=0.08]{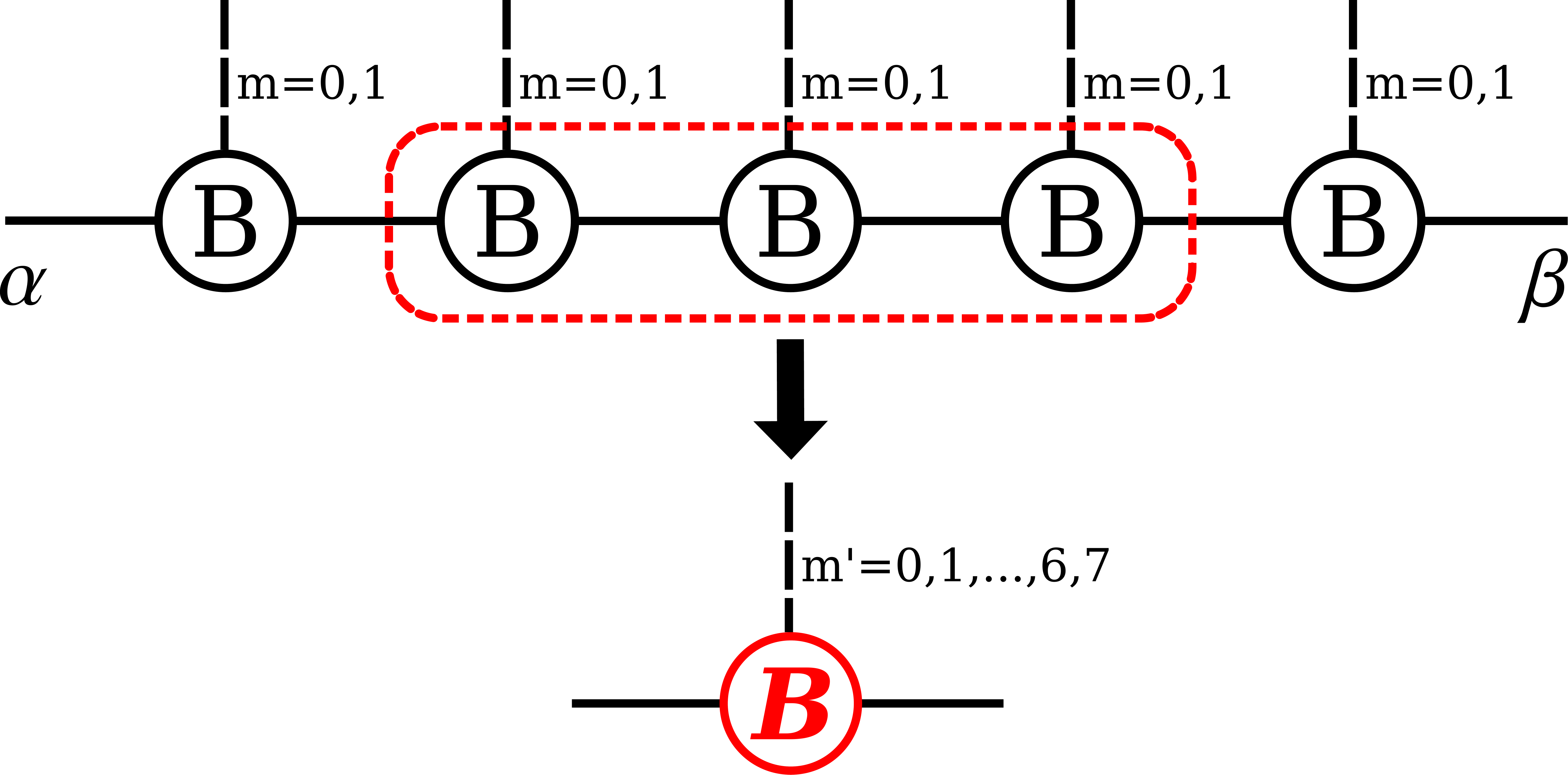}
 \caption{Going form a one-site MPS to a $kq$-site MPS. The bond dimension remains unchanged, but the new physical index $\{m\} = (m_1,\cdots ,m_{kq})$ runs over all possible occupations of $kq$ consecutive orbitals. This is the particular example of the $q=3$ Laughlin state ($k=1$). This state being fermonic we have $m \in \{ 0,1 \}$, while $m' \in \{ 000, \,001, \,010, \,011, \,100, \,101, \,110, \,111 \}  = \{0 , 1 \cdots ,7 \} $.}
 \label{kq site MPS}
\end{figure}

Going back to the generic case, we introduce $P^{(\phi,j)}$ the projector onto the topological sector $\mathcal{H}_{(\phi,j)}$. The matrices $B^m$ (and $C^m$) obey the following relation
\begin{align}
B^m P_{(\phi,j)} = P_{(\phi,j-1)} B^m \label{P B relation}
\end{align}
Since $P_{(\phi,j)} = P_{(\phi,j+kq)}$, it is clear that a product of $kq$ matrices $B^m$ is block diagonal in all topological sectors \footnotemark \footnotetext{In the most generic case we have $j = j + k_{\phi} q$. This means that we could in principle work with a $k_{\phi}q$-site MPS instead of a $kq$-site MPS if we restrict ourselves to the neutral sector $\phi$. However in order to be handle all neutral sectors at once, we have to adopt a $kq$-site MPS. This is fine because $k_{\phi}$ is a divisor of $k$.}. Therefore it is beneficial to work with a $kq$-site MPS rather than a one-site MPS. The relevant matrices for the $kq$-site MPS are
\begin{align}
{\bf B}^{\{m\}} = B^{m_{kq}}\cdots B^{m_2}B^{m_1} \label{kq block B matrix}
\end{align}
where the composite index $\{m\} = (m_1,\cdots ,m_{kq})$ runs over all the possible occupations of $kq$ consecutive orbitals. For fermions there would be $2^{kq}$ such matrices. Note that the auxiliary space is unchanged (see Fig.~\ref{kq site MPS}). The same property holds for the matrices $C^m$.

At a more algebraic level, the matrices ${\bf B}^{\{m\}}$ are generically products of modes of the electron operator $V(z)$ together with a U$(1)$ charge shift
\begin{align}
{\bf B}^{\{m\}} = e^{-ik\sqrt{q}\varphi_0} V^{m_{kq}}_{-h - kq+1} \cdots V^{m_2}_{-h-1} V^{m_1}_{-h} \label{algebraic B}
\end{align}
In particular ${\bf B }^0 $ is a pure charge shift $S = e^{-ik\sqrt{q}\varphi_0}$. The neutralizing background charge for $kq$ orbitals induces this charge shift $N \to N - k^2q$, and this leaves topological sectors invariant. As a consequence all product of $kq$ matrices $B^{m_i}$ - \emph{i.e.} all the matrices ${\bf B}^{\{m\}}$ - are block diagonal w.r.t. the topological charge $a$: 
\begin{align}
{\bf B}^{\{m\}} = \left( \begin{array}{cccc}  {\bf B}^{\{ m \} }_0 & 0& 0 & 0  \\ 0 & \ddots & 0 &0  \\ 0 & 0 & {\bf B}^{ \{m \}}_a & 0 \\ 0 & 0 & 0 & \ddots \end{array} \right) \label{B block}
\end{align}

\subsection{Transfer Matrix}
 \label{Transfer matrix}

Crucially, the implementation of the MPS now allows us to compute physical properties without resorting to the original many-body Hilbert space. While this is standard procedure in the MPS community, we briefly repeat it here as a motivation for introducing the MPS transfer matrix. This transfer matrix (or $E$ matrix) is an extremely efficient tool to  compute physical quantities such as entanglement spectrum on the infinite cylinder. Consider a MPS representation of a FQHE droplet with arbitrary edge excitations 
\begin{align}
\ket{\psi^{\alpha_R}_{\alpha_L}} = \sum_{\{ m_i \}}  \bra{\alpha_R}  B^{m_{N_{\Phi}}} \cdots B^{m_1} \ket{\alpha_L}  \, \, | m_1 \cdots m_{N_{\Phi}} \rangle
\end{align}
where $\bra{\alpha_R}$ and $\ket{\alpha_L}$ live in the auxiliary space $\mathcal{H}$ - which is nothing but the (possibly truncated) CFT Hilbert space. $m_i$ is the occupation of site $i$ and $B^{m}$ are the matrices for orbital occupation $m$. For the cylinder and for the conformal limit, they do not depend on the site $i$ but for the sphere and the infinite plane, they do depend on the site, hence we added the index $i$ on them. For the purpose of studying FQHS on the cylinder however, we can consider an orbital independent MPS.

\subsubsection{Computing Overlaps}

Changing the state $\ket{\alpha_R}$ (resp. $\ket{\alpha_L}$) modifies the right (left) edge excitation of the FQHE droplet. Overlaps between two right edge states can be computed as follow
\begin{align}
\langle\Psi^{\alpha_R'}_{\alpha_L} | \Psi_{\alpha_L}^{\alpha_R}\rangle & = \sum_{\{ m_i \}}     \bra{\alpha_R} B^{m_{N_{\Phi}}} \cdots B^{m_1} \ket{\alpha_L}   \bra{\alpha_R'} B^{m_{N_{\Phi}}} \cdots B^{m_1} \ket{\alpha_L}^*   \\
 &  = \sum_{\{ m_i \}}   \langle \alpha_R,\alpha_R' | \left( B^{m_{N_{\Phi}}} \otimes B^{m_{N_{\Phi}}\, *} \right) \cdots \left( B^{m_1}  \otimes B^{m_1\,*} \right) \ket{\alpha_L,\alpha_L}  \\
 & =  \langle \alpha_R,\alpha_R' | E^{N_{\Phi}} \ket{\alpha_L,\alpha_L} \label{Right overlap}
 \end{align}
where $\ket{\alpha_R,\alpha_R'}=\ket{\alpha_R} \otimes \ket{\alpha_R'}$, $B^{m\,*}$ stands for the complex conjugate of $B^m$, and the transfer matrix $E$ is defined as
\begin{align}
E =  \sum_{m}  B^{m}  \otimes  B^{m\, *}
\end{align}
Note that this operator is not hermitian, and the overlaps between two left edge states takes the form
\begin{align}
\langle\Psi^{\alpha_R}_{\alpha_L'}|\Psi^{\alpha_R}_{\alpha_L}\rangle =  \langle \alpha_L,\alpha_L' | E^{\dagger N_{\Phi} } \ket{\alpha_R,\alpha_R} \label{Left overlap}.
\end{align}

It is sometimes convenient to think of the transfer matrix as acting on the space of matrices rather that on the tensor product $\mathcal{H} \otimes \mathcal{H}$. These two descriptions are clearly equivalent, since the spaces of matrices $ X_{\alpha, \beta}$ in $\mathcal{H}$ is isomorphic to the two copies of the auxiliary space $\mathcal{H} \otimes \mathcal{H}$, through $\ket{\alpha , \beta} \to \ket{\alpha}\bra{\beta}$. The transfer matrix $E$ becomes a super-operator (a linear operator acting on matrices) $\mathcal{E}$
\begin{align}
\mathcal{E}(X) = \sum_m B^m X B^{m \dagger}, \qquad \mathcal{E}^{\dagger}(X) = \sum_m B^{m \dagger}  X B^m
\end{align}
while the hermitian product is now $\{X ,Y \} = \text{Tr}(X^{\dagger}Y)$.

The transfer matrix formalism becomes particularly powerful in the limit of an infinitely long cylinder ($N_{\Phi} \to \infty$). Assuming the transfer matrix is gapped, overlaps are dominated by its largest eigenvalue. The corresponding right (reps. left) eigenvector has a very simple interpretation : it is the overlap matrix between right (resp. left) edge modes on the infinite cylinder. 

 The transfer matrix is not hermitian : its eigenvalues are complex, and its eigenvectors are not orthogonal. Moreover, it is in general not diagonalizable (it has non trivial Jordan cells). However the super-operator $\mathcal{E}$ is a positive map, i.e. it preserves the subspace of positive matrices\footnotemark \footnotetext{A matrix $X$ is positive if $\forall \alpha, \, \bra{\alpha}X \ket{\alpha}$ is real and positive (this implies $X= X^{\dagger}$). Equivalently $X$ is positive iff $X = M^{\dagger}M$ for some $M$.}. Although the spectrum of the transfer matrix is not real, it is known that the largest eigenvalue (in modulus) is real positive, and the corresponding eigenvector can be chosen to be a positive matrix, \emph{i.e.} a hermitian matrix with positive eigenvalues. Note that this result is consistent with the interpretation of this eigenvector as an overlap matrix.

\subsubsection{Computing Expectation Values}

A similar language can be developed for correlation functions. Consider the expectation value of an arbitrary operator $O$ for an arbitrary FQH state $\ket{\psi} = | \psi_{\alpha_L}^{\alpha_R} \rangle$ 
\begin{eqnarray}
 \langle \psi | O | \psi \rangle   = \sum_{\{ m_i, m'_i \}}  &&   \bra{\alpha_R} B^{m_{N_{\Phi}}} \cdots B^{m_1} \ket{\alpha_L}   \bra{\alpha_R} B^{m'_{N_{\Phi}}} \cdots B^{m_1'} \ket{\alpha_L}^*     \nonumber\\
&&\langle m'_1 \cdots m'_{N_{\Phi}} | O | m_1 \cdots m_{N_{\Phi}} \rangle   \\
   =  \sum_{\{ m_i, m'_i \}} &&  \langle \alpha_R,\alpha_R | \left( B^{m_{N_{\Phi}}} \otimes B^{m'_{N_{\Phi}}\, *} \right) \cdots \left( B^{m_1}  \otimes B^{m'_1\,*} \right) \ket{\alpha_L,\alpha_L}   \nonumber\\
&&    \langle m'_1 \cdots m'_{N_{\Phi}} | O | m_1 \cdots m_{N_{\Phi}} \rangle
\end{eqnarray}
 Very often  $O = \bigotimes_i o_i$ is a (sum of) product of local observables $o_i$ acting on orbital number $i$. For instance the density of the droplet $O = \sum_j c^\dagger_j c_j$. The expectation value of such operators can be computed very efficiently. Indeed $\langle m'_0 \cdots m'_{N_{\Phi}} | O | m_0 \cdots m_{N_{\Phi}} \rangle = \prod_i \bra{m'_i} o_i \ket{m_i}$, and computing the expectation value boils down to an operator multiplication in auxiliary space
\begin{align}
 \langle \psi | \otimes_i o_i | \psi \rangle =     \langle \alpha_R,\alpha_R | O_{N_{\Phi}} \cdots  O_{1} \ket{\alpha_L,\alpha_L}   \label{eq tensor contraction}
\end{align}
where the $O_i$ are given by
\begin{align}
O_i = \sum_{m , m'}  \bra{m'} o_i \ket{m}  B^m \otimes  B^{m' \,*} 
\end{align}
They are linear operators acting in two copies of the auxiliary space $\mathcal{H} \otimes\bar{ \mathcal{H}}$. Even though the two copies are identical we denote by $\bar{\mathcal{H}}$ the second copy of the auxiliary space by analogy with the field theoretic interpretation of the MPS : we are now dealing with the full conformal field theory, containing both right ($\mathcal{H}$) and left ($\bar{\mathcal{H}}$) movers. After truncation we are dealing with an MPS with a finite the bond dimension $\chi$ (\emph{i.e.} finite dimensional auxiliary space), and the $O$ matrices are $\chi^2 \times \chi^2$ dimensional. Their matrix elements are:
\beq
(O_i)_{(\alpha'  , \beta') ; \, (\alpha , \beta)} = \bra{\alpha' , \beta '}O_i \ket{\alpha , \beta} = \sum_{m,m'} \bra{m'} o_i \ket{m}   B_{\alpha', \alpha}^{m} \,  B_{\beta',\beta}^{m'\,*} \label{Omatrices}
\eneq  
As for the transfer matrix they admit an equivalent description in terms of super-operators  
\begin{align}
\mathcal{O}_i(X) = \sum_{m, m'} \langle m' |o_i |m \rangle \, \, B^{m}XB^{m' \dagger} 
\end{align}
A good example would be the occupation number $n_j = c_j^{\dagger}c_j$ of orbital $j$, for which the corresponding MPS operator is
\beq
N = \sum_m m  B^{m}  \otimes  B^{m\, *}
\eneq
and the (un-normalized) occupation number of the orbital $j$ is:
\beq
n_{j} = \langle \alpha_R,\alpha_R | E^{N_{\Phi}-j} N E^{j-1}  \ket{\alpha_L,\alpha_L}   \label{Orbital occupation}
\eneq

\begin{figure}
 \centering
\includegraphics[scale=0.07]{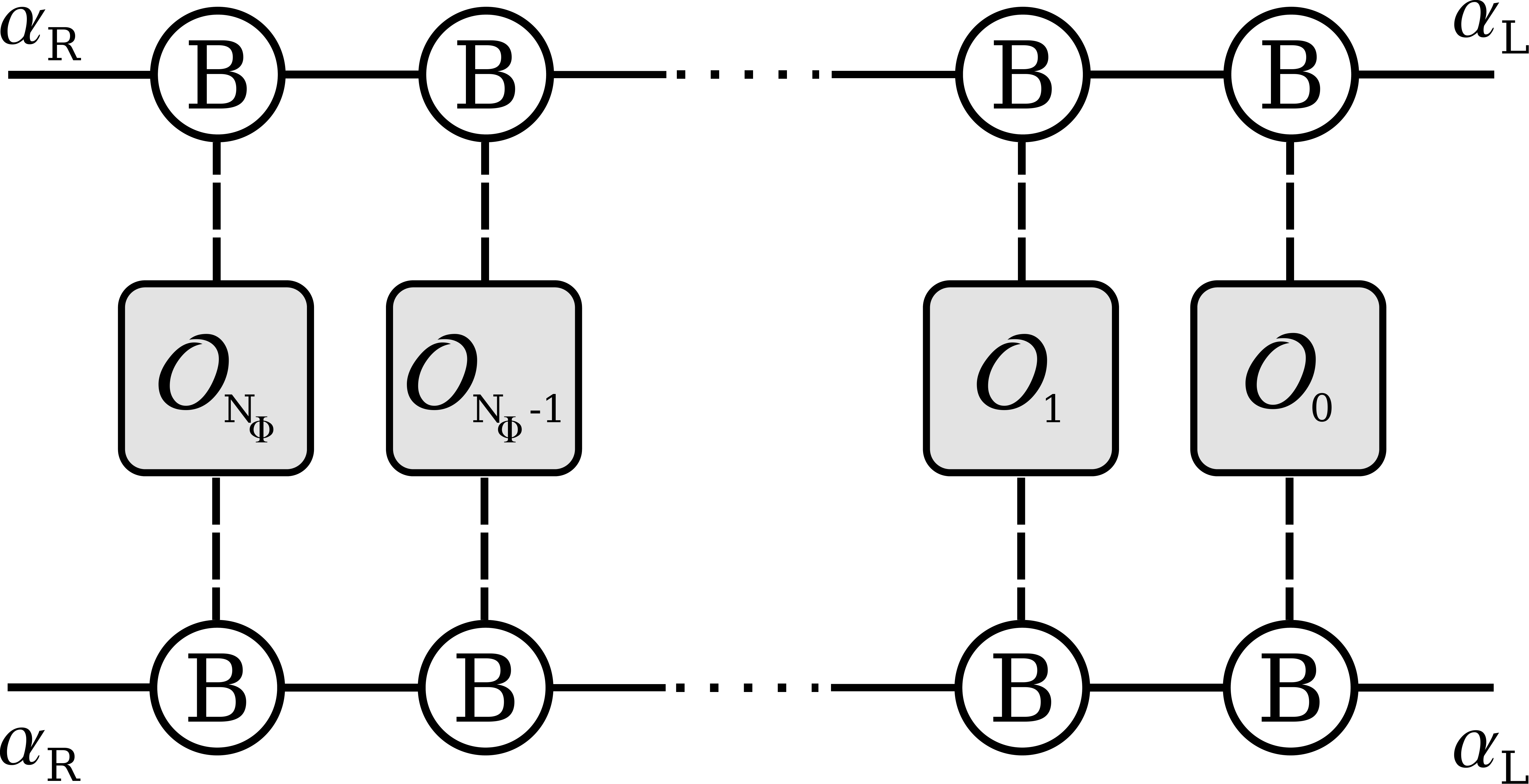}
 \caption{Tensor contraction corresponding to the computation of the expectation value of a product of local observables. The matrices $B^m$ are represented by circles, while the squares represent the observables $o_i$. Bonds correspond to index contraction (the vertical bonds are the occupation index $m$, and the horizontal bonds are the auxiliary space).}
  \label{tensor contraction}
\end{figure}

In MPS language the expectation value \eqref{eq tensor contraction} is given by the contraction of a tensor, as illustrated in Fig.~\ref{tensor contraction}.
Density-density correlations as well as any other expectation values can clearly be obtained in similar ways. Crucially, in the MPS literature it is well-known that the gap of the $E$ matrix bounds the decay length of any local correlation functions \cite{Fannes-1992}.

\subsection{Fixed Points and Canonical Form of the MPS} \label{canonical form}

As we argued the transfer matrix $E$ (or equivalently $\mathcal{E}$) plays an important role in understanding the infinite-MPS formulation of a state. As we consider a very long cylinder ($N_{\Phi} \to \infty$) the overlap between to right edge states, as given in Eq. \eqref{Right overlap}, is dominated by the largest eigenvalue of the transfer matrix. We will now delve deeper into the properties. Due to the block structure \eqref{B block} of the $kq$-site MPS, it is more convenient to work with the $kq$-site transfer matrix 
\begin{align}
\bm{\mathcal{E}}(X) = \sum_{\{m\}} {\bf B}^{ \{m\}} X {\bf B}^{ \{m\} \dagger}
\end{align} 
In order to be more specific we have to discuss about the different sectors of the transfer matrix. The most obvious quantum number of the transfer matrix comes from the different topological sectors of the FQH state. Since the map $\bm{\mathcal{E}}$ acts on two copies of the Hilbert space, it is block diagonal with respect to left and right topological charges both:
\begin{align}
\bm{\mathcal{E}} (P_a X ) = P_a \bm{\mathcal{E}} (X), \qquad \bm{\mathcal{E}} ( X P_b )  = \bm{\mathcal{E}} ( X ) P_b
\end{align}
and we can consider it's restriction to a particular sector $a \times b$, i.e. matrices $X$ such that $X = P_a X = X P_b$. Besides the left and right topological charge, another quantum number is the relative $\text{U}(1)$ (or electric) charge between left and right, as measured by $\mathcal{A}_0(X) = [a_0, X]$. 

In principle all topological sectors are required in order to compute generic correlation functions in the FQH droplet. However depending on the nature of the observables one can work with the restriction to a particular sector $a \times b$. For instance the electron propagator involves a particular non-diagonal sector, while quantities such as overlaps, density or entanglement spectra are purely concerned with diagonal sectors $a \times a$.  In this paper we are concerned with the latter, and we focus our attention on the diagonal sectors.  

The restriction of the transfer matrix to a diagonal sector $a \times a$ takes the form
\begin{align}
\bm{\mathcal{E}}_a (X)= \sum_{ \{m\}} {\bf B}^{\{m \}}_a X {\bf B}^{\{m \} \dagger}_a 
\end{align}
where ${\bf B}^m_a = P_a{\bf B}^m = {\bf B}^m P_a$ are the restriction of ${\bf B}^m$ to the topological sector $a$ as in \eqref{B block}. As was already mentioned, the transfer matrix $\bm{\mathcal{E}}_a$ is a positive map, i.e. it preserves the subspace of positive matrices. 

By an appropriate rescaling ${\bf B}^m_a \to e_a {\bf B}^{m}_a$, we can assume the spectral radius of the transfer matrix $\bm{\mathcal{E}}_a$ to be unity (assuming the transfer matrix is bounded, which is always true after truncation). By virtue of $\bm{\mathcal{E}}_a$ being a positive map, $1$ itself has to be an eigenvalue of $\bm{\mathcal{E}}_a $\cite{evans1978spectral} (and therefore for $\bm{\mathcal{E}}_a^{\dagger} $ as well) 
\begin{align}
\bm{\mathcal{E}}_a(R_a) =   R_a, \qquad \bm{\mathcal{E}}_a^{ \dagger} (L_a) =  L_a
\end{align}
In the case at hand - and for the particular choice of $B$ matrices \eqref{MPS B matrix} - the matrices $B^m$ and $B^{m \dagger}$ are related through a unitary transformation
\begin{align}
B^{m \dagger} = U B^m U^{\dagger} 
\end{align}
where $U = U^{\dagger} = C$ is the U$(1)$ charge conjugation. As a consequence the right and left eigenvectors are related through
\begin{align}
R_a = C L_a C
\end{align}
They are sometimes called fixed points in the MPS literature. It is now straightforward  - assuming the transfer matrix is gapped  - to compute the overlaps \eqref{Right overlap} and \eqref{Left overlap} in the limit of an infinitely long cylinder :
\begin{align}
\langle\psi^{\alpha_R}_{\alpha_L'}|\psi^{\alpha_R}_{\alpha_L}\rangle \propto \langle \alpha_L' | L_a |\alpha_L \rangle, \qquad \langle\psi^{\alpha_R'}_{\alpha_L}|\psi^{\alpha_R}_{\alpha_L}\rangle \propto \langle \alpha_R' | R_a |\alpha_R \rangle \label{overlap matrices and fixed points}
\end{align}
where $a$ is the particular topological sector in which the droplet $| \psi^{\alpha_R}_{\alpha_L} \rangle$ lives. Due to a generic theorem about positive maps, both left and right fixed points $L_a$ and $R_a$ can be chosen to be positive matrices. But this property is trivial in view of \eqref{overlap matrices and fixed points}, since the fixed points are nothing but overlap  matrices.  

From positivity some further properties can be derived. Since $\mathcal{A}_0$ commute with $\bm{\mathcal{E}}$, the fixed points are also eigenstates of $\mathcal{A}_0$. By positivity the corresponding eigenvalue has to be zero (it has to be real since $\mathcal{A}_0$ is hermitian, and also pure imaginary since $L_a$ and $R_a$ are hermitian). The left and right eigenvectors therefore commute with $a_0$. Moreover, as long as we are in the kernel of $\mathcal{A}_0$, the operator $\mathcal{L}_0(X) = [L_0,X]$ commute with $\bm{\mathcal{E}}$. Since $R_a$ and $L_a$ do belong to the kernel of $\mathcal{A}_0$, they are eigenstate of $\mathcal{L}_0$. And by positivity, they must commute with $L_0$. We end up with the following result : the left and right fixed points $L_a$ and $R_a$ are block diagonal \emph{w.r.t.} $U(1)$ charge and conformal dimension.

This generic property boils down to a trivial statement as soon as we realize that $R_a$ and $L_a$ are overlap matrices.  Indeed the U$(1)$ charge encodes the deviation of the electron number from the ground state, while the conformal dimension yields the deviation in total angular momentum. The symmetries of the FQH problem ensure that the overlap matrix is block diagonal \emph{w.r.t.} these two quantum numbers.

If there are more conserved quantities in the CFT which are compatible with the transfer matrix, this type of argument can be extended. For instance the CFT underlying the $\nu=1$ non-interacting case, which can be thought of a $q=1$ Laughlin state, is a free complex fermion. Its integrability is manifest in the set of commuting observables : the fermion occupation numbers (in momentum basis). It is rather straightforward to prove that the left and right fixed points commute with all these quantities, and therefore are diagonal in the fermion occupation basis (of the CFT). However so far no such structures have been found for interacting FQH states.

One can use the transformations \eqref{gauge transformation} to simplify the form of the left (or right) fixed points. Under a change of basis $G$ \begin{align}
{\bf B}^{ \{ m \}}  \to G  {\bf B}^{ \{ m \}} G^{-1}
\end{align}
the left and right fixed point transform according to
\begin{align}
R \to G R G^{\dagger},\qquad L \to G^{-1 \dagger } L G^{-1} 
\end{align}
It was proven  in Ref.~\cite{Prosen-0305-4470-39-22-L02} that there always exists such a change of basis after which the left and right fixed point are diagonal positive matrices:
\begin{align}
R_a = \Lambda_a^R = \text{diag} \{ \Lambda^R_{a,\beta}, \, \beta \in \mathcal{H}_a  \}, \qquad L_a = \Lambda_a^L = \text{diag} \{ \Lambda^L_{a,\beta}, \, \beta \in \mathcal{H}_a  \}
\end{align}
The algorithm to put the MPS in its canonical form is described in Refs.~\cite{Prosen-0305-4470-39-22-L02,Perez-Garcia:2007:MPS:2011832.2011833}. We repeat it here for the reader's convenience, and to enlighten the particularities of our case. The first step is to diagonalize the product $R_a L_a$ \footnotemark  \footnotetext{The product of two positive matrices is similar to a positive matrix, and therefore diagonalizable.}:
\begin{align}
R_a L_a = G_a D_a G_a^{-1}  
\end{align}
Since $R_a$ and $L_a$ are hermitian, we also have $ L_a R_a = G_a^{-1 \dagger} D_a G_a^{\dagger}$. We then absorb the change of basis $G_a$ in ${\bf B}^{ \{ m \}}  \to G_a^{-1}  {\bf B}^{ \{ m \}} G_a$. The new left and right eigenvectors obey $R_a L_a = L_a R_a =D_a$, and therefore they commute. As such they can be diagonalized simultaneously:
\begin{align}
R_a = U \Lambda^R_a U^{\dagger}, \qquad L_a = U \Lambda^L_a U^{\dagger}
\end{align} 
and the unitary change of basis $U$ can again be absorbed in ${\bf B}^{ \{ m \}}  \to U^{\dagger}  {\bf B}^{ \{ m \}} U$. An important point is that all the changes of basis involved preserve the $L_0$ and $a_0$ grading of the CFT Hilbert space (because both $R_a$ and $L_a$ commute with $a_0$ and $L_0$), and they do not mix topological sectors. This means that such a change of basis can be done while preserving all the good quantum numbers of the system : particle number, momentum and topological sector.

\newpage

\section{Truncation of the Hilbert Space}
\label{truncation section}

So far we have been dealing with an exact MPS representation of FQH trial wave-functions, at the cost of working with an infinite dimensional auxiliary space : the Hilbert space of the underlying 1+1 dimensional conformal field theory. In this section we describe the necessary truncation scheme allowing the numerical implementation of the MPS. We then show how to analytically optimize the MPS in order to reach the maximum sizes possible for the computation of certain physical quantities,  by presenting several analytical tricks through which we can massively improve the size of the numerical simulations reached. From now, we restrict to fermions for numerical purposes.

\subsection{Truncation Scheme and Matrix Dimensions}

The numerical implementation of the MPS and the calculation of physical quantities requires us to work with a finite dimensional auxiliary space. A key observation made in \cite{zaletel-PhysRevB.86.245305} is that the CFT provides a natural truncation parameter, namely the conformal dimension. It turns out that it is possible to describe trial wave functions with an extremely high overlap while working with a moderate bond dimension at least on the cylinder geometry \cite{estienne-PhysRevB.87.161112}. There are several tricks used to be able to increase the bond dimension used in numerical calculations. First, considering a $kq$-site MPS allows us to work with the restriction of the full ${\bf B}^{\{m\}}$ matrices to a given topological sector $\mathcal{H}_a$, namely  ${\bf B}^{\{ m\}}_a$. The number of topological sectors is the torus ground state degeneracy $N_{GS}$. In average this reduces the dimension of the $B$ matrices by a factor $N_{GS}$, thereby providing a $N_{GS}^2$ further reduction in the dimension of the transfer matrix. 

\begin{figure}
\centering
\includegraphics[width=0.75\linewidth]{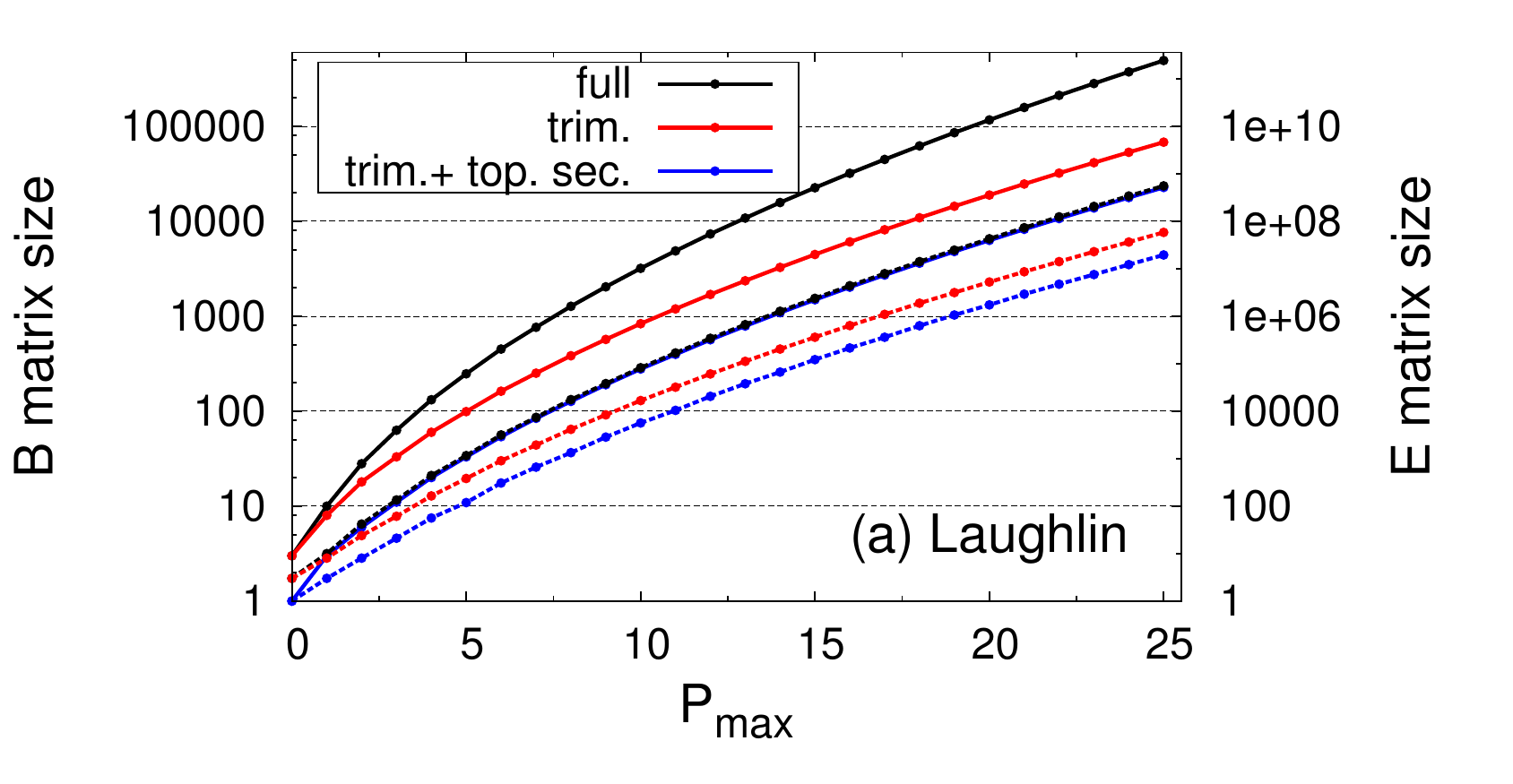}
\includegraphics[width=0.75\linewidth]{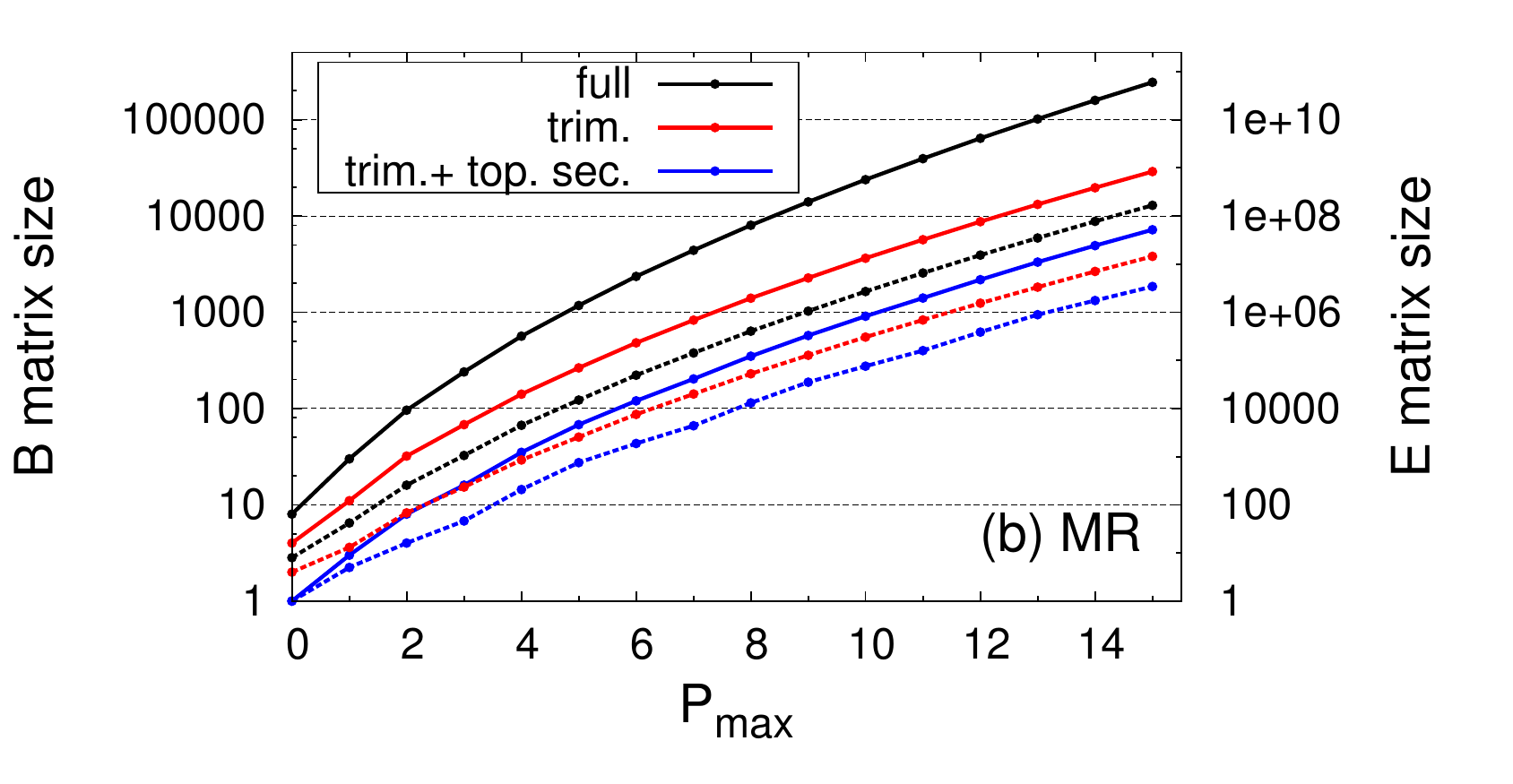}
\includegraphics[width=0.75\linewidth]{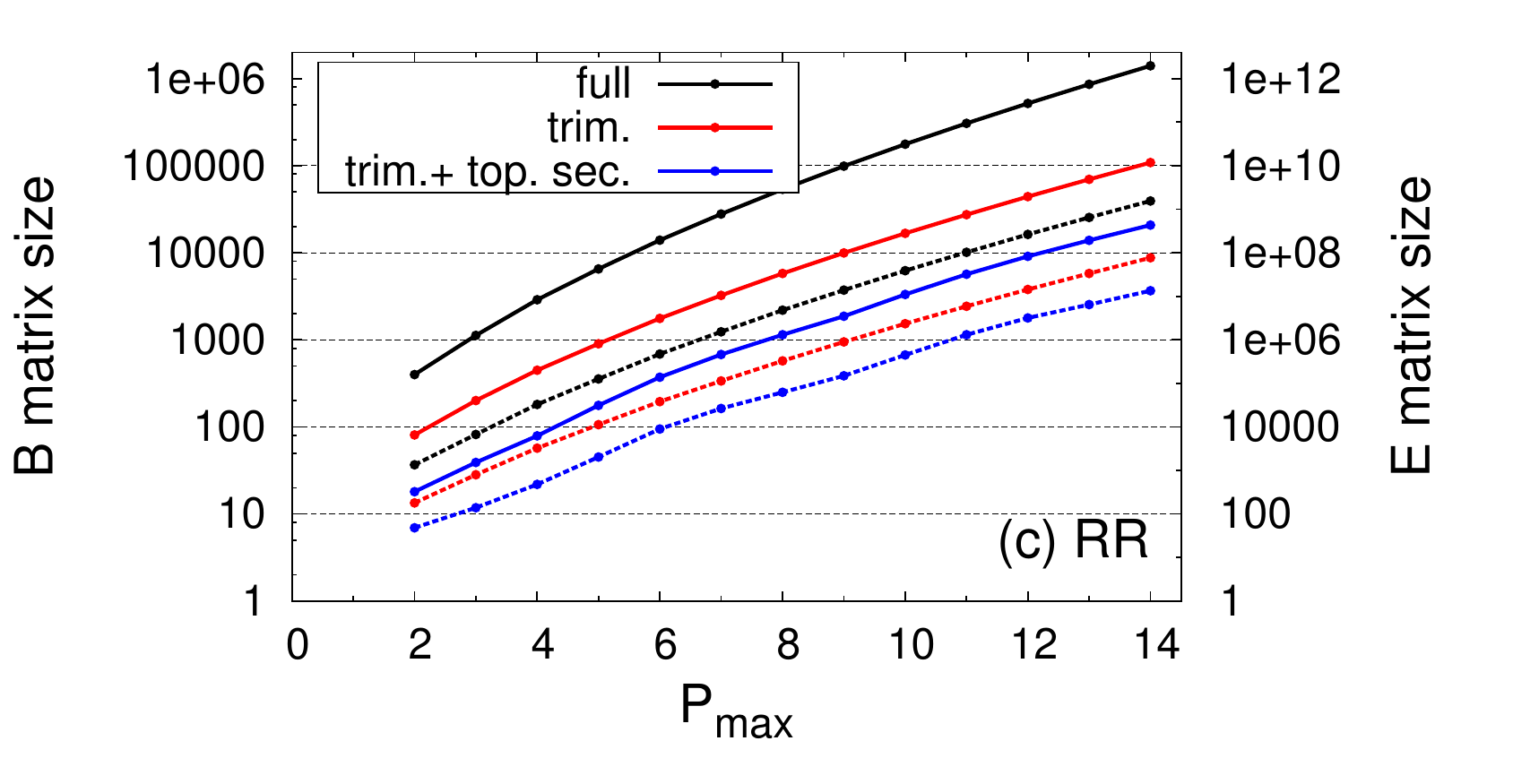}
\caption{Dimensions of the $B$ matrices and transfer matrices $E$ for $(a)$ the Laughlin $\nu=1/3$ state, $(b)$ the Moore-Read state (in the vacuum sector) and $(c)$ the $\mathbb{Z}_3$ Read-Rezayi state (in the vacuum sector). The solid lines denote the dimension of the $B$ matrices or the the $E$ matrices (which is the square of the $B$ matrix dimension) while the bold dashed lines denote the dimension of $E$ matrices in a given diagonal sector as defined in Section~\ref{canonical form}. The black color is used if the dimension is computed without any additional truncation scheme (full dimension). The red color corresponds to the case where the trimming procedure is applied and the blue color corresponds  to the case where one uses both the trimming procedure and the restriction to a given topological sector.}
 \label{bematrixsize}
\end{figure}

Since most of the MPS calculations require to compute the left and right largest eigenstates of the transfer matrix, it is important to use the smallest relevant part of the transfer matrix. As discussed in Section~\ref{canonical form}, we can restrict the transfer matrix to the subspace of $a \times a$ of states with the same left and right U$(1)$ charge and conformal dimension, when looking for the left and right fixed points. This is highly beneficial for numerics: depending on the state  and whether of not we rely on a $kq$-site MPS, it reduces the dimension of the space in which to search for the fixed point of the matrix from $D^2$ (if $D$ is the dimension of the $B$ matrix) to roughly $D^{1.6}-D^{1.7}$ according to numerical results. In Fig.~\ref{bematrixsize}, we give the dimensions of both the $B$ and transfer matrices for the Laughlin, Moore-Read and $\mathbb{Z}_3$ Read-Rezayi states. For these two latest cases, the data are limited to values of $P_{\text{max}}$ where the neutral part of the matrix elements can be numerically evaluated.

\subsection{Trimming}\label{trimming}

For many quantities such as entanglement spectrum\cite{li2008}, the momentum (which is identical to the total descendant level $P =|\theta| +| \mu|$ in \eqref{Vir x U(1) basis}) is a good quantum number. Thus in Refs.~\cite{zaletel-PhysRevB.86.245305} and~\cite{estienne-PhysRevB.87.161112} numerical calculations, the natural truncation parameter used is $P$, rather than the total conformal dimension $ P + Q^2/2 + \Delta$ where $\Delta$ is the conformal dimension of the neutral primary field at an MPS bond. In numerical calculations, one needs to implement the MPS auxiliary Hilbert space under which we will expand the MPS. Before truncation, this Hilbert space consists of a tensor product of $Q \otimes \Delta \otimes { \{ \mu , \theta, \cdots \}}$ where $\Delta$ describes the neutral sector of the CFT and the set of partitions encode the structure of the descendent as in \eqref{Vir x U(1) basis} or in \eqref{SV descendants} for instance. In particular the level of the descendent is $P = |\mu| + |\theta| + \cdots$. 

 Denote the maximum value of the the total descendant level kept $P_{\text{max}}$ (see Section~\ref{secMPS Matrix Elements}). In finite size, when we perform a truncation for values of $P< P_{\text{max}}$, it turns out that the values of $Q$ that can be reached are severely restricted by this truncation and the Hilbert space is no longer a tensor product  but contains a much smaller set of possible quantum numbers, which represents a much more efficient, "trimmed" basis in which the MPS can be expanded. We will give a generic prescription of how to trim the Hilbert space in the general case and then, in the Appendix~\ref{AppendixTrimming}, present examples of how to build this basis for Laughlin and Moore-Read cases and provide an analytic formula for the values of $Q$ possible at each $P$ for these two cases.

The general scheme for trimming the MPS Hilbert space of any state can be easily implemented numerically as follows. First establish the root partition \cite{Bernevig-PhysRevB.77.184502,Bernevig-PhysRevLett.100.246802} of the state by determining the only configuration which remains by setting $P_{\text{max}}=0$. The momentum and charge transfer at an MPS bond will be counted with respect to this root partition. Now set $P_{\text{max}}>0$. At any MPS site, which we now call "cut" in order to link it to the orbital entanglement spectrum cut that we will define in Section~\ref{OESsection}, we look at the auxiliary bond matrix elements or bond index $(N, P, \Delta)= (N, 0, \Delta)$ (note that we omit the explicit dependence in the partitions of the bond index and only quote the value of $P$): this is the bond of ground-state root partition (thin-torus) momentum ($P=0$) but with the $N, \Delta$ indices (the $U(1)$ charge $Q$ related to the integer index $N$ by \eqref{QNrelation} and $\Delta$ identifies the neutral sector) set to other values. It determines if $N>0$ or if $N<0$. If $N>0$, it means particles have been taken to the right of the "cut" but without changing the momentum (the $N<0$ is identical but with right changed to left). Depending on the filling factor $\nu$, roughly a number $(\Delta Q)/\sqrt{\nu}$ charges ($\Delta Q$ being the variation of the $U(1)$ charge $Q$) are transported across the cut.  This immediately means that the root partition of the configurations which, at a set bond cut (orbital) have quantum numbers $N, 0, \Delta$ is:
\beq
\text{[Root Partition]} B^0 \ldots B^0_{N,0,\Delta} |_{N,0,\Delta}B^1  \ldots B^1 \ldots B^1 B^0  \ldots B^0 \text{[Root Partition]} \label{trimmingroot}
\eneq  
It is easy to explain that this is the form of a root partition of configurations where $N$ particles have been taken across the cut: the most unsqueezed \cite{Bernevig-PhysRevB.77.184502,Bernevig-PhysRevLett.100.246802} configuration on the left of the cut is the one where the ground-state root partition is followed by as many unoccupied orbitals needed to transfer the particles to the right of the cut. This indeed gives momentum zero at the cut. The most unsqueezed partition to the right of the cut is the one that has the properties: 
\begin{itemize}
\item\textbf{1.} All the particles moved from the left of the cut have been placed as close to the cut as possible by fermionic statistics - this gives the string of $B^1$'s placed right by the cut. The length of this string is roughly $2*N$, as roughly the same number of particles from both the left of the cut and the right of the cut come together 
\item\textbf{2.} The particles close to the orbital cut are followed to the right by a string of unoccupied sites. This corresponds to sites depleted of roughly $N$ particles - the particles moved from the left to the right across the cut have to be compensated by $N$ particles moved to the left towards the cut from the root partition of the ground state on the right hand side of the cut. 
\item\textbf{3. } The string of unoccupied sites is followed by the ground-state root partition to the right.
\end{itemize}

\begin{figure}
\centering
\includegraphics[width=0.99\linewidth]{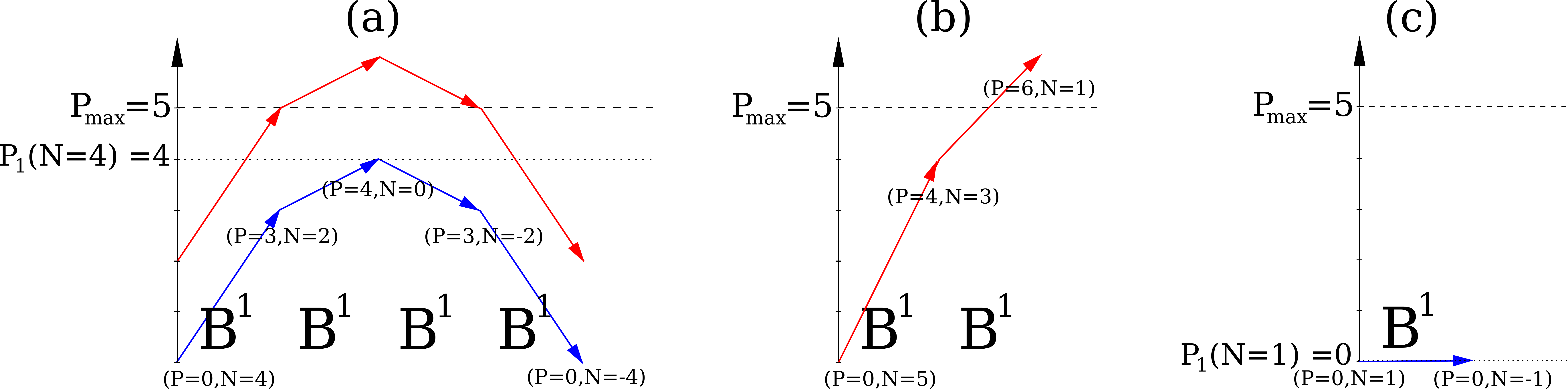}
\caption{Example of the evolution of the auxiliary bond index under a string of $B^1$ matrices. Here we consider the $\nu=1/3$ Laughlin state (so the $\Delta=0$ index is omitted), $P_{\rm max}=5$. $(a)$ We look at the evolution of one index $(P=0,N=4)$ when applying these matrices (according to \eqref{thinannulusmpslaughlin}). Following the blue path, we obtain that $P_1(N=4)=4$. Starting from $(P=2,N=4)$ instead of $(P=0,N=4)$ would give the red path. Since $P=2$ exceeds $P_{\rm max}-P_1(N=5)$, the string of $B^1$ generates indices beyond $P_{\rm max}$ which is forbidden by the truncation. $(b)$ If we start from $(P=0,N=5)$ and we apply $B^1$ twice, $P$ is larger than $P_{\rm max}$. Thus this value of $N$ has to be discarded. $(c)$ We now look at $(P=0,N=1)$. Applying a single $B^1$ does not change $P$, so $N=1$ should be kept for any value of $P$.}
\label{figtrimming}
\end{figure}

As every other element with indices $(N,0, \Delta)$ is squeezed from the partition \eqref{trimmingroot}, we now look at its properties. We can assume that the string of $B^1$'s to the right of the cut stretches out to infinity (we consider infinite MPS limit), and concentrate on this string. Knowing our MPS matrices, we can now compute the $P$ index at the bond of \emph{every} matrix $B^1$. This $P$ is well defined because the left matrix element momentum is $0$, and the $B^1$ matrix elements delta functions will uniquely determine the $P$ at each bond index in the partition $_{N,0,\Delta}B^1 B^1 B^1 \ldots B^1B^1B^1 \ldots $. Crucially, the $P$ will grow at first, reach a maximum value, then decrease (as shown on Fig.~\ref{figtrimming} for the Laughlin state). We know that it will decrease at some point because, the momentum has to go back to its ground state root partition value of $0$ at some point to the right of the cut. In other words, the momentum quantum number to both the left and right of the $B^1$ string is zero. Pick the maximum value of this momentum, call it $P_1(N, \Delta) $ where we explicitly denoted its dependence of $N, \Delta$. This can be trivially computed numerically as it involves only the delta function dependence of the $B^1$ matrices. 

Now try to increase the momentum of the auxiliary bond of charge index $N$ and primary field $\Delta$ where the cut was made and ask the question: can I increase this momentum all the way to the maximum value allowed by my truncation, $P_{\text{max}}$? The answer is no! Then the limit of the momentum that can be reached for the left bond index $N,\Delta$ and truncation $P_{\text{max}}$ is then $P_{\text{max}}- P_1(N, \Delta)$. A larger value of this momentum on the left bond would mean that the momentum in the string of $B^1$ matrices to the immediate right of the cut can go above $P_{\text{max}}$ which is impossible in our truncation. In this way, we can immediately associate with any $N, \Delta$, a maximum value possible for $P$ which is $P_{\text{max}}- P_1(N, \Delta)$. This severely restricts the auxiliary Hilbert space size. From a practical perspective, the trimming procedure greatly reduces (by one order of magnitude) the size of the $B^m$ and transfer matrices as shown in Fig.~\ref{bematrixsize} for the Read-Rezayi series.

\newpage

\section{Entanglement Spectra}

In this section we show how to compute the three main entanglement spectra, the orbital\cite{li2008,lauchli-PhysRevLett.104.156404}, particle\cite{Sterdyniak-PhysRevLett.106.100405} and real-space\cite{Sterdyniak-PhysRevB.85.125308,Dubail:2012p2980,Rodriguez-PhysRevLett.108.256806} entanglement spectra for a FQH system where an MPS description is available. Entanglement spectra have been extensively used \cite{li2008,Zozulya-PhysRevB.79.045409,Regnault-PhysRevLett.103.016801,Thomale-PhysRevLett.104.180502,Sterdyniak-NewJourPhys-13-10-105001} in the literature to distinguish topological phases in systems where the ground-state is obtained as a diagonalization of a Coulomb Hamiltonian. The three spectra reveal different information about the systems in which they are utilized, and for model states they have the property that the counting of their eigenvalues does not saturate the dimension of the Hilbert space. For realistic states in the same universality class as the model state, an entanglement gap appears between a set of levels which resembles those of the model state and the remainder, "spurious" set of remaining levels which saturate the Hilbert space dimension. The Orbital Entanglement Spectrum (OES) and the Real Space Entanglement Spectrum (RSES) reveal information about the edge theory of the FQH state. The orbital entanglement spectrum turns out to be the easiest to obtain, while the particle and real space spectrum require further manipulation of the state's MPS. We here show how to obtain all three spectra in both the finite and infinite system limit. While most of the derivation does not depend on the geometry, we focus on the cylinder. Indeed our formula involve the cylinder $C^m$ matrices instead of the generic $B^m$ matrices. Furthermore we show an interpolation between the real space entanglement spectrum and the orbital spectrum.

\subsection{Orbital Entanglement Spectrum}\label{OESsection}

We consider a FQH droplet $\ket{\psi} = \ket{\psi^{\alpha_R}_{\alpha_L}}$ in topological sector $a$,  \emph{i.e.} the auxiliary states $\ket{\alpha_L}$ and $\ket{\alpha_R}$ encoding the left and right edge excitations are in $\mathcal{H}_a$
\begin{align}
\ket{\psi} = \sum_{\{ m_i \}}  \bra{\alpha_R}  C^{m_{N_{\Phi}}} \cdots C^{m_1} \ket{\alpha_L}  \, \, | m_1 \cdots m_{N_{\Phi}} \rangle
\end{align}
Without any loss of generality we can assume the amount of orbitals to be a multiple of $kq$ (by padding the edges with empty orbitals if required). 

\begin{figure}
\centering
\includegraphics[width=0.35\linewidth]{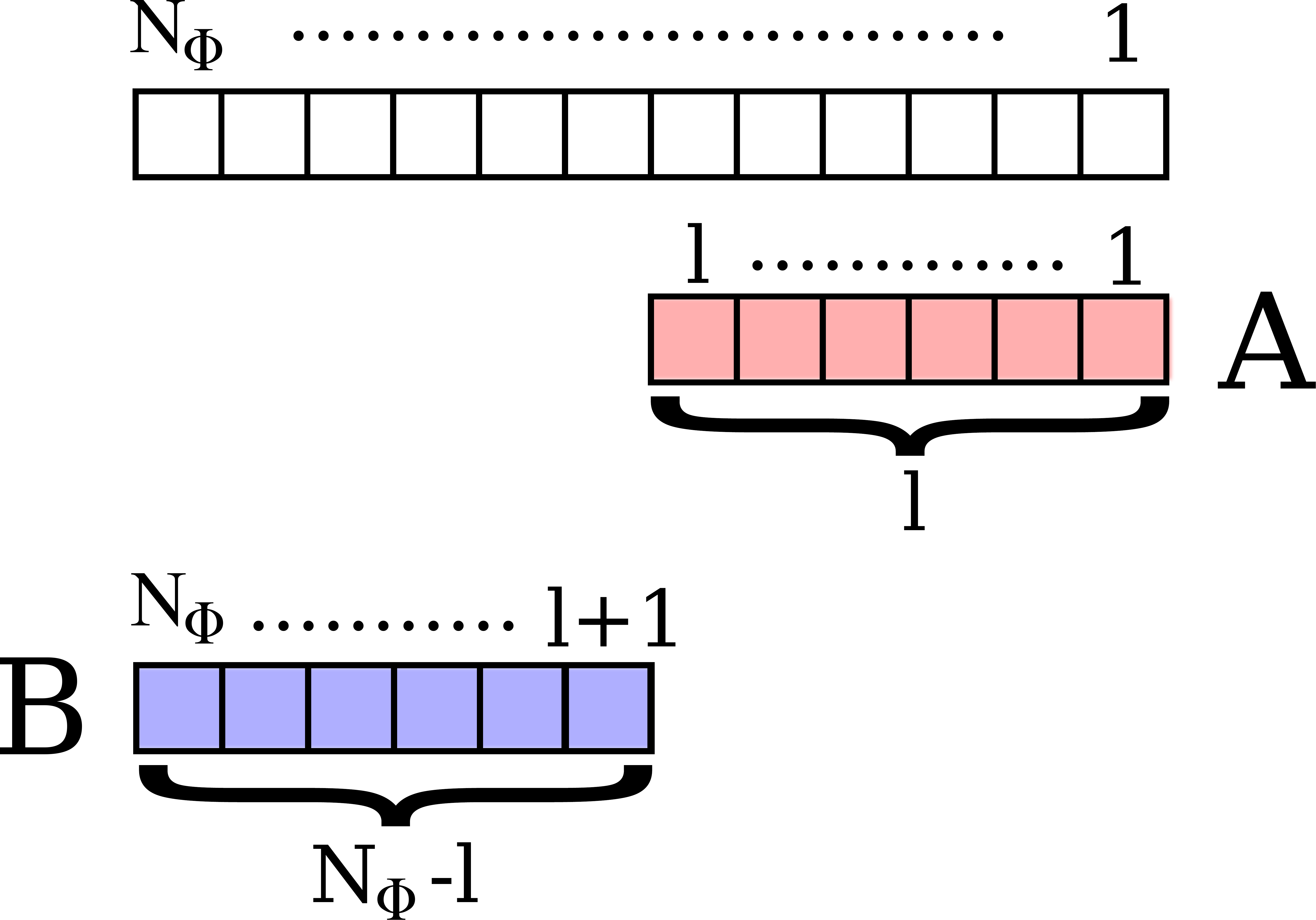}
\caption{Description of the orbital cut. We start with a system made of $N_{\Phi}$ orbitals. The part $A$ is built from the $l$ leftmost orbitals (light red) and the  part $B$ is built from the $N_{\Phi}-l$ rightmost orbitals (light blue).}
\label{oesfig}
\end{figure}

The partition we are interested in the the so-called orbital partition, in which we cut the system after the $l^{th}$ orbital (see Fig.~\ref{oesfig}). Although this is not mandatory, it is very convenient to choose for $l$ a multiple of $kq$ in order to benefit from the block structure of the transfer matrix, as explained in Section~\ref{trimming}. 
Part $A$ is made of the orbits located on the left of the cut $\{m_A\} = \{ m_1, m_2, \cdots, m_l\}$, while part $B$ is on the right $\{ m_B \} = \{m_{l+1}, \cdots, m_{N_{\Phi}} \}$.  The MPS representation of the state $\ket{\psi} $ is extremely convenient to obtain the Schmidt decomposition \emph{w.r.t.} the orbital partition.  After making explicit the dependence overt parts $A$ and $B$ in the MPS
\begin{align}
\ket{\psi^{\alpha_R}_{\alpha_L}} = \sum_{ \{ m_A\}, \{m_B\}} \bra{\alpha_R}  C^{\{ m_B\}} C^{\{ m_A\}} \ket{\alpha_L}    \ket{ \{ m_A\}, \{ m_B\}} \label{SchmidtDecomp1}
\end{align}
where $C^{\{m_A\}}$  stands for the product of matrices $C^{m_l} \cdots C^{m_1}$, it is straightforward to obtain the Schmidt decomposition by inserting the resolution of the identity $\mathid=  \sum_{\beta} \ket{\beta} \bra{\beta}$ 
\begin{align}
\ket{\psi^{\alpha_R}_{\alpha_L}} = \sum_{\beta}\, \sum_{ \{ m_A\}, \{m_B\}} \, \bra{\alpha_R}  C^{\{ m_B\}}  \ket{\beta}\, \bra{\beta} C^{\{ m_A\}} \ket{\alpha_L}   \ket{ \{ m_A\}, \{ m_B\}}  \label{SchmidtDecomp2}
\end{align}

Since we have chosen $l$ to be a multiple of $kq$, the only $\ket{\beta}$ that contribute are those in the same topological sector as $\ket{\alpha_L}$. This means that we can restrict the sum to the topological sector $a$, and insert  $\mathid_a =  \sum_{\beta \in \mathcal{H}_a} \ket{\beta} \bra{\beta}$ instead of $\mathid$. We end up with a decomposition of the state $\ket{\psi}$ as
\begin{align}
\ket{\psi} = \sum_{ \beta \in \mathcal{H}_a}   \ket{ \psi^A_{\beta}} \otimes \ket{ \phi^B_{\beta}}
\end{align} 
where 
\begin{align}
 \ket{ \psi^A_{\beta}} =  \sum_{ \{ m_A\}} \bra{\beta} C^{\{ m_A\}} \ket{\alpha_L}   \ket{ \{ m_A\}} , \qquad  \ket{\phi^B_{\beta}} = \sum_{ \{m_B\}} \bra{\alpha_R}  C^{\{ m_B\}}  \ket{\beta}   \ket{\{ m_B\}}
\end{align}
Naturally the state $|\psi^A_{\beta}\rangle $ can be interpreted as a FQH droplet living in part $A$, whose right edge excitation is encoded in $\beta$. As such overlaps of states in part $A$ can be computed using the transfer matrix. Likewise for states in part $B$.
\begin{align}
\bra{ \psi^A_{\beta'}}  \psi^A_{\beta} \rangle =   \bra{\beta' ,\beta}   E^{l} \ket{\alpha_L, \alpha_L}   , \qquad  \bra{ \phi^B_{\beta'}}  \phi^B_{\beta} \rangle =   \bra{\beta' ,\beta}   E^{ \dagger (N_{\Phi}-l )} \ket{\alpha_R , \alpha_R}   
\end{align}
In thermodynamic limit, obtained by sending to infinity the number of orbitals in part $A$ and $B$, respectively $l$ and $N_{\Phi}-l$,  the overlaps are dominated by the largest eigenvalue of the transfer matrix in the diagonal sector $a \times a$ - assuming the transfer matrix is gapped in this sector. They boil down to the right and left fixed points of the transfer matrix\footnotemark  \footnotetext{We have assumed implicitly that the eigenvalue $1$ belongs to a trivial Jordan block (remember that the transfer matrix need not be diagonalizable). If this is not the case, a rescaling is required to go to the thermodynamic limit.  }, in the corresponding topological sector $a$:
 \begin{align}
\bra{ \psi^A_{\beta'}}  \psi^A_{\beta} \rangle  \to  \left( R_a \right)_{\beta' ,\beta}    , \qquad  \bra{ \phi^B_{\beta'}}  \phi^B_{\beta} \rangle \to  \left(L_a\right)_{\beta' ,\beta}    
\end{align}
up to a global, \emph{i.e.} independent of $\beta$,  multiplicative prefactor that depends on the left (right) edge excitations $\ket{\alpha_L}$  (resp. $\ket{\alpha_R}$) of the whole droplet. 

This is where the canonical form explained in section \ref{canonical form} simplifies the discussion drastically. In the canonical gauge the left and right fixed points are (positive) diagonal matrices $\Lambda^L_a$ and $\Lambda^R_a$, and the states  $|\psi^A_{\beta}\rangle $ (as well as $|\phi^B_{\beta}\rangle $) are mutually orthogonal
\begin{align}
\bra{ \psi^A_{\beta'}}  \psi^A_{\beta} \rangle \propto  \Lambda^R_{a,\beta} \delta_{\beta' ,\beta}     , \qquad  \bra{ \phi^B_{\beta'}}  \phi^B_{\beta} \rangle \propto \Lambda^L_{a,\beta}\delta_{\beta' ,\beta}
\end{align}    
The reduced density matrix for part $A$ on an infinite cylinder is then 
\begin{align}
\rho_A = \text{Tr}_B(\rho) \propto \sum_{\beta \in \mathcal{H}_a}  \Lambda^L_{a,\beta} \Lambda^R_{a,\beta}  \ket{\tilde{\psi}^A_{\beta}}\bra{\tilde{\psi}^A_{\beta}}
\end{align}
 where the sum is over all $\beta \in \mathcal{H}_a$ such that $ \Lambda^L_{a,\beta} \Lambda^R_{a,\beta}  \neq 0$, and the states $ | \tilde{\psi}^A_{\beta} \rangle = ( \Lambda^R_{\alpha} )^{-1/2}   | \psi^A_{\beta} \rangle $ are orthonormal. 

\begin{figure}
\centering
\includegraphics[width=0.32\linewidth]{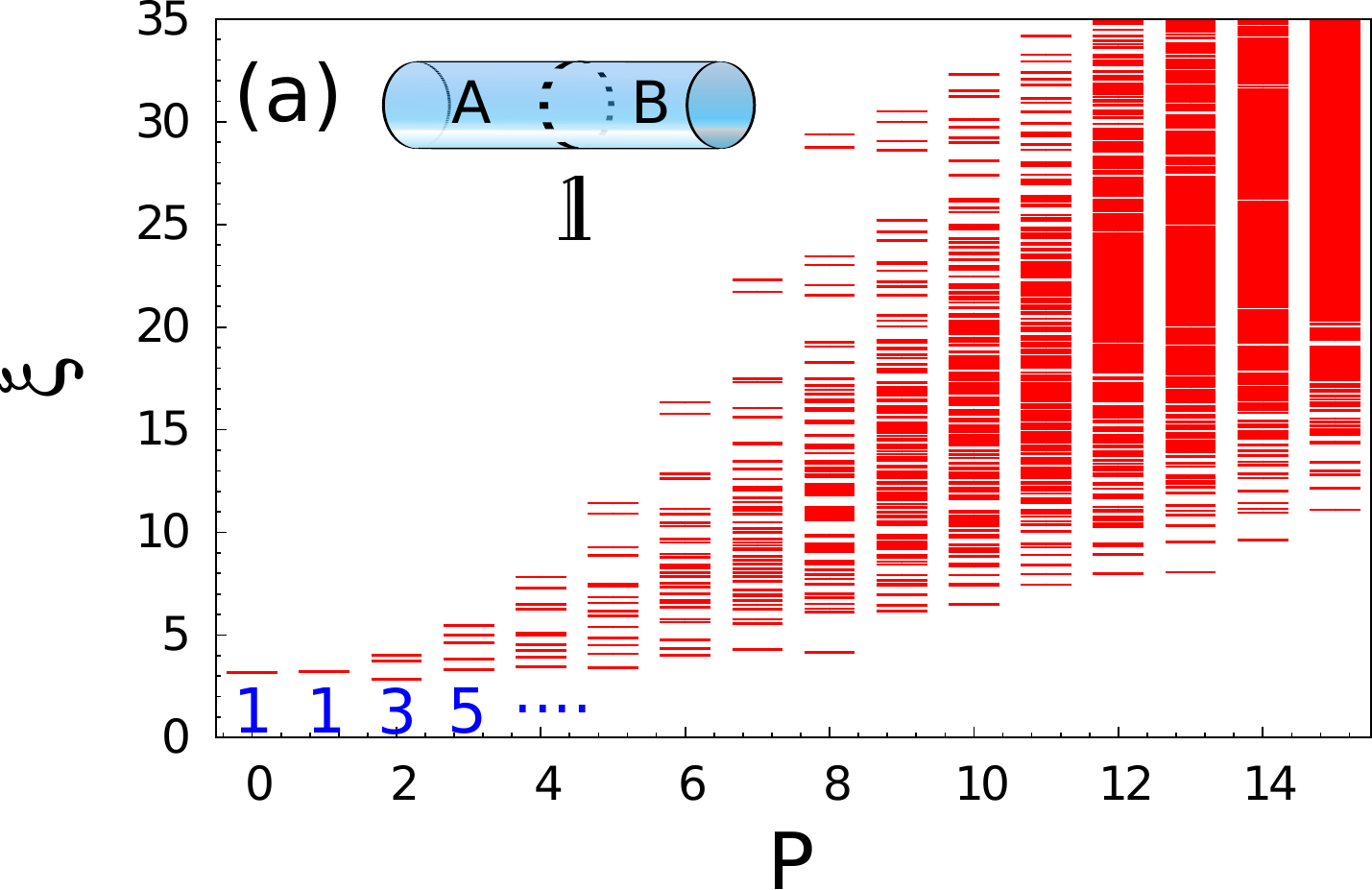}
\includegraphics[width=0.32\linewidth]{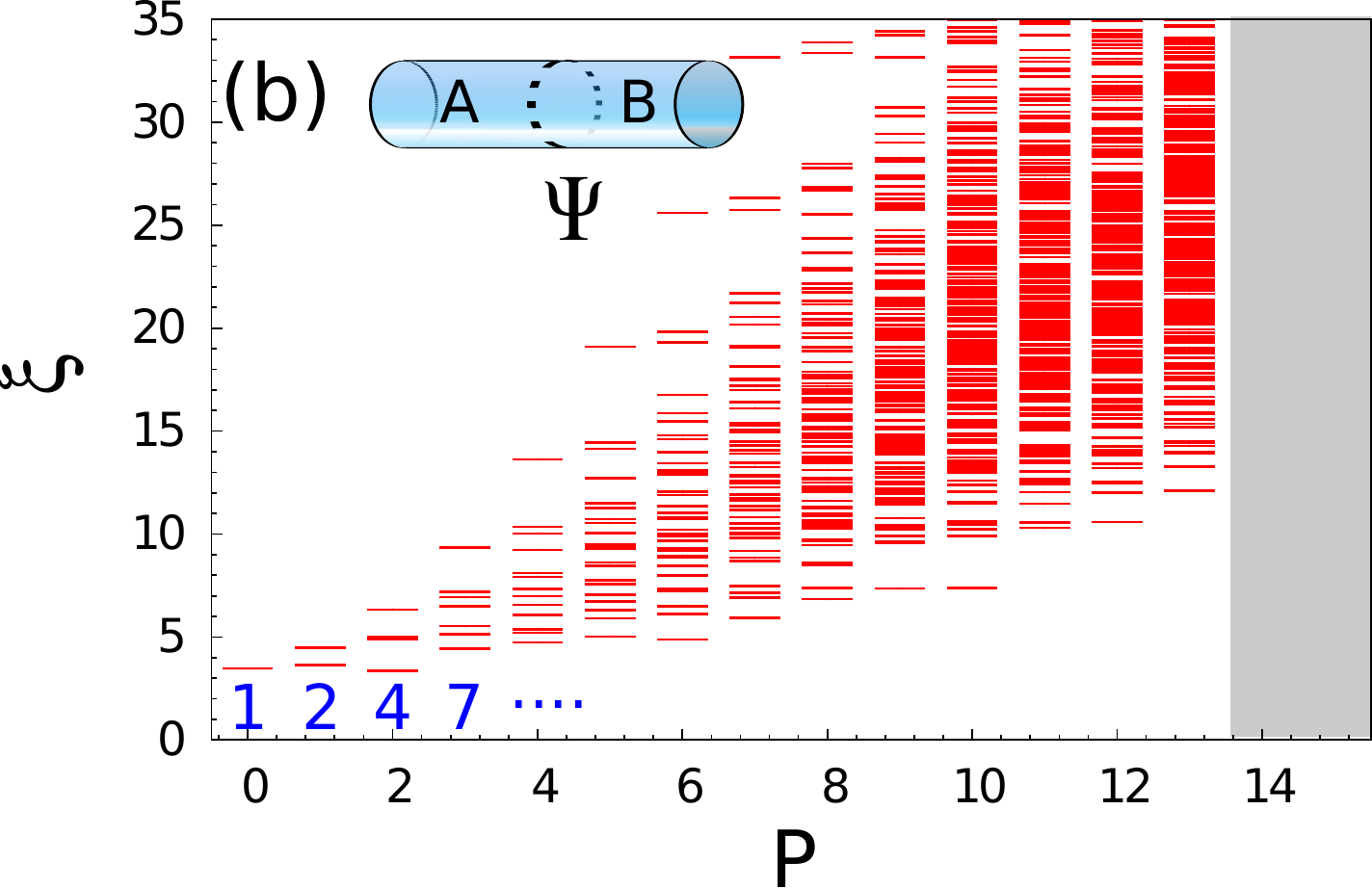}
\includegraphics[width=0.32\linewidth]{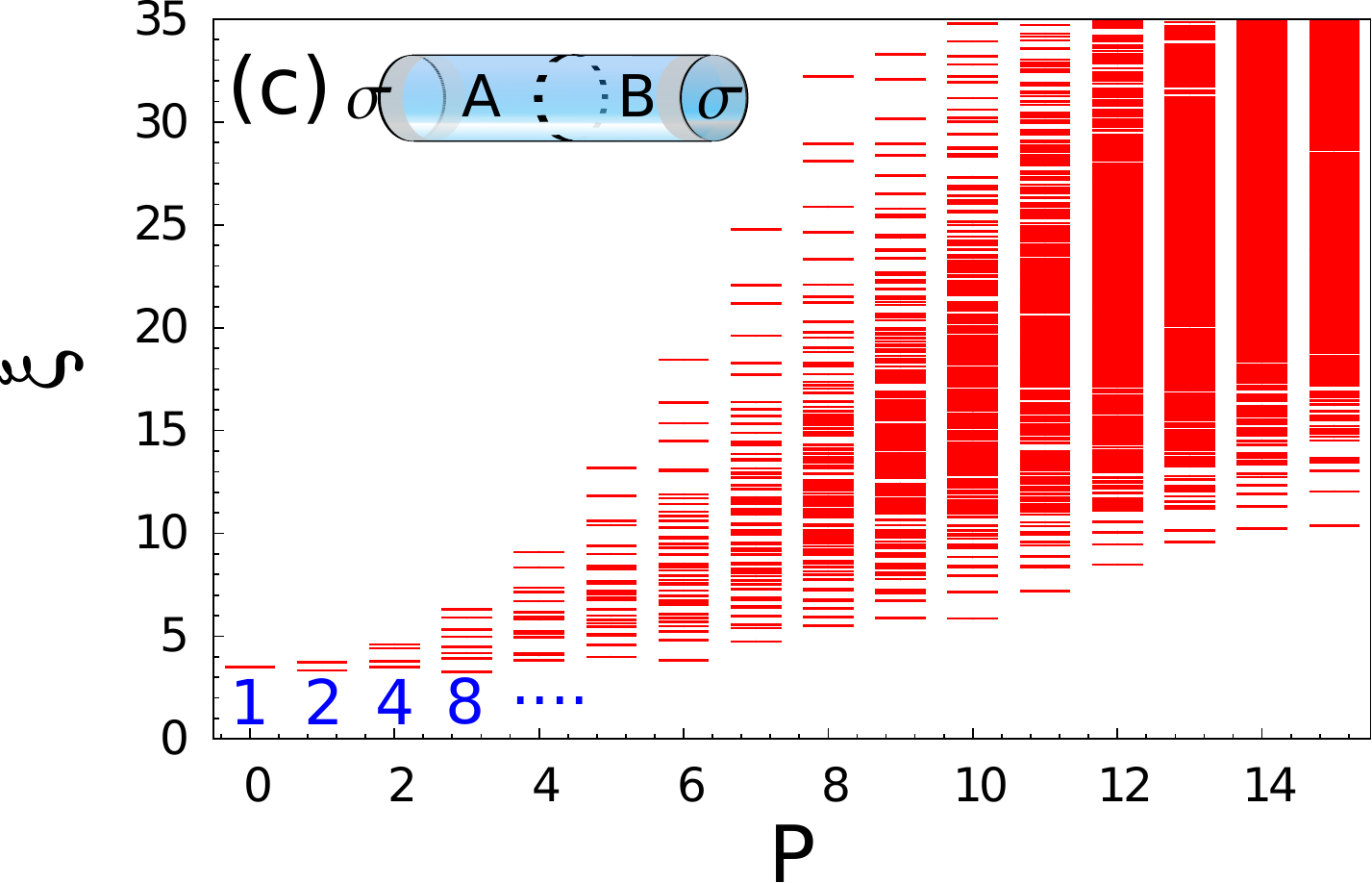}
\caption{OES for the Moore-Read state on a infinite cylinder of perimeter $L=25l_b$ and $P_{\text{max}}=15$. $(a)$ OES in the identity sector, $(b)$ OES in the $\Psi$ sector and $(c)$ OES with one quasihole at each end. the $\xi$'s are related to the eigenvalues $\lambda$ of $\rho_A$ through $\xi=-\log(\lambda)$. In each cases, we have fixed the charge sector $Q$ and the counting of states per momentum sector (in blue) matches the one of the descendants per level. Note that if one considers a particular topological sector, a global $P$-truncation might result in different accessible $P$ sectors for the identity and $\Psi$ sectors: Here the $P=14$ and $P=15$ are missing in the $\Psi$ sector (gray zone).}
 \label{oesmr}
\end{figure}

This form of  reduced density matrix is particularly simple because it is already diagonal, and one can read of immediately the orbital entanglement spectrum.
\begin{align}
\rho_A = \text{Tr}_B(\rho) =\sum_{\beta \in \mathcal{H}_a}  e^{-\xi_{\beta}} \ket{\tilde{\psi}^A_{\beta}}\bra{\tilde{\psi}^A_{\beta}}, \qquad  e^{-\xi_{\beta}}  = \frac{ \Lambda^L_{a,\beta} \Lambda^R_{a,\beta} }{\text{Tr} ( \Lambda^L_{a} \Lambda^R_{a} ) }
\end{align}
The spectrum of the reduced density matrix in topological sector $a$ is nothing but the (normalized) spectrum of $\Lambda^R_a \Lambda^L_a$, or more generally the spectrum of the product of left and right fixed point $R_aL_a$ (or equivalently $L_aR_a$). This spectrum can be plotted versus the quantum numbers relative to part A : momentum and particle number, because both left and right fixed points are block diagonal \emph{w.r.t.} U$(1)$ charge and conformal dimension, as was argued in section~\ref{Transfer matrix}. In Fig.~\ref{oesmr} we give an explicit example of OES for the Moore-Read state, showing how it enables to read out the truncated characters of the CFT by selecting the proper topological sectors.

\subsection{Particle Entanglement Spectrum}

We  now move to  the next entanglement spectrum and try to obtain the particle entanglement spectrum (PES) on the cylinder. For simplicity we consider a FQH droplet centered around $x=0$ on the cylinder, with no edge modes, the generalization to a droplet with edge excitations being straightforward. The corresponding first quantized wave function is the following cylinder conformal block 
\begin{align}
\psi(z_1,\ldots, z_{N_e}) =  \bra{ N_e \sqrt{q}/2} V (z_{1}) \ldots  V (z_{N_e}) \ket{-N_e \sqrt{q}/2} \label{cylinder GS first quantized}
\end{align}
where we have implicitly assumed $N_e$ to be a multiple of $k$. Likewise we assume in the following that $N_A$ and $N_B$ to be multiple of $k$. While this is not essential, it simplifies the discussion. In terms of occupation basis, this state is confined to the range of orbitals
\begin{align}
- q \frac{N_e}{2} +h \leq \lambda_i \leq  q \frac{N_e}{2} -h
\end{align}
with a number of flux $N_{\Phi} = q N_e - 2h$. Because of the MPS block structure described in section~\ref{B block structure}, it is more convenient and numerically efficient to work with a number of matrices $C^m$ which is a multiple of $kq$. To do see we choose the orbitals range to be $n_L \leq \lambda_i \leq n_R$, with 
\begin{align}
n_R = -n_L =  \frac{qN_e-1}{2}
\end{align}

We can obtain the particle density matrix of the state by integrating out $N_B = N_e - N_A$ particles, from $z_{N_e- N_A +1}$ up to $z_{N_e}$. The Hilbert space of the density matrix is the states of $N_A$ particles. In second quantized language this corresponds to keeping the full Hilbert space of the system, (i.e. a QH system with $q (N_A + N_B)$ orbitals), and letting the $N_A$ particles take the phase space through all the $N_\Phi$ number of fluxes. This procedure is easily performed at the level of conformal blocks :
 \begin{align}
 \psi(z_1, \ldots, z_{N_e})    = \sum_{\beta}   \psi^B_{\beta} (z_{N_e},  \ldots, z_{N_a+1}) \, \psi^A_{\beta} (z_{N_A}, \ldots , z_{1}) \label{PESschmidt}
 \end{align}
by simply inserting $\sum_{\beta} \ket{\beta} \bra{\beta}$ in the conformal block
 \begin{eqnarray}
&&\bra{ N_e \sqrt{q}/2}V (z_{N_e}) \ldots  V (z_1) \ket{-N_e \sqrt{q}/2}  \label{PESfirstquantizedwave function}\\
&=& \sum_{\beta}   \bra{ N_e \sqrt{q}/2} V (z_{N_e})  \ldots  V( z_{N_a+1}) \ket{\beta} \bra{\beta} V (z_{N_A}) \ldots  V (z_{1}) \ket{-N_e \sqrt{q}/2}\nonumber  
 \end{eqnarray}
How is this different from the OES? This is of course the same decomposition that we have used in the orbital entanglement spectrum, but the internal sum runs over a different part of Hilbert space. Indeed the U$(1)$ charge of $\beta$ is frozen to the value $Q_{\beta} = \sqrt{q} (N_A - N_B)/2$ by charge conservation\footnotemark \footnotetext{The constraint of having $N_A$ electrons before the cut freezes more than just the U$(1)$ charge of the state $\ket{\beta}$. It also freezes its neutral sector.  For instance when $N_A$ is a multiple of $k$, the neutral part has to be in the modulus of the identity $\ket{0}$. The generic constraint is that $\ket{\beta}$ has to appear in the OPE of $N_A$ primary fields $V$ acting on $\ket{\alpha_L}$.}. Therefore momentum is the only quantum number available for the PES, as we have lost particle number.  
Another difference is that the $N_A$ particles of part $A$ can be located anywhere on the initial droplet. In principle we have to keep all $q N_e$ sites for the description of the $A$ part.

The MPS representation  \eqref{MPS cylinder}-\eqref{MPS cylinder in and out states} can be used for all the conformal blocks involved in the decomposition  \eqref{PESfirstquantizedwave function}. The most naive way to do this is to keep the total number of orbitals $N_{\Phi} = q(N_A + N_B)$ for both $A$ and $B$ parts. For part $A$ we have
 \begin{align}
\ket{ \psi_{\beta}^A} = \sqrt{N_A!} \, e^{\tau (E_R(\beta)- E_L(0)) } \sum_{m_1, \cdots, m_{N_{\phi}}} \langle{\beta_R}|C^{m_{N_\Phi}} \cdots C^{m_1} \ket{0} \, \ket{m_1 \cdots m_{N_\Phi}} \label{naivePESpartA}
 \end{align}
 where $|\beta_R \rangle = e^{-i \sqrt{q}N_e \varphi_0/2} \ket{\beta}$ has a U$(1)$ charge $Q_{\beta_R} = -\sqrt{q}N_B$, since it is shifted by the background charge of the whole droplet. This means that effectively the state above describes a FQH state of $N_A$ particle with a giant quasihole (corresponding to the missing part $B$). Similarly for part $B$ the MPS representation is
  \begin{align}
\ket{ \psi_{\beta}^B} = \sqrt{N_B!} \, e^{ \tau ( E_R(0)- \tau  E_L(\beta))} \sum_{m_0, \cdots, m_{N_{\phi}}} \langle 0 |C^{m_{N_\Phi}} \cdots C^{m_1} \ket{\beta_L} \, \ket{m_1 \cdots m_{N_\Phi}} \label{naivePESpartB}
 \end{align}
where $|\beta_L \rangle = e^{i \sqrt{q}N_e \varphi_0/2} \ket{\beta}$ has a U$(1)$ charge $Q_{\beta_L} = \sqrt{q}N_A$. We then have - in principle - a MPS representation of the decomposition  \eqref{PESschmidt} with respect to the particle cut $N_e = N_A + N_B$
 \beq
 \ket{\psi} =\sum_P  \ket{ \psi_{\beta}^B} \otimes \ket{ \psi_{\beta}^A}
 \eneq 
where the overlaps of the states $ |\psi_{\beta}^B\rangle$  (and  $|\psi_{\beta}^A\rangle$) for different $\beta$ have yet to be computed.
 
In practice however the numerical implementation of the MPS states $|\psi_{\beta}^A\rangle$ and $|\psi_{\beta}^B\rangle$ as described in  \eqref{naivePESpartA} and  \eqref{naivePESpartB} hits a severe problem. The U$(1)$ charge of the bond indices $\ket{\beta_R}$ and $| \beta_L \rangle$, which encode how many electrons are missing in parts $A$ and $B$, scale with the number of electrons. As we increase the system size they eventually will lie outside of the truncation parameter regime. It would hence seem that the spectrum is not computable numerically.

 \subsubsection{Low Momentum Sectors of the PES}

The way around this issue is to implement a more efficient MPS representation of these states. We illustrate the method for part $A$, the extension to part $B$ being straightforward. 
The key ingredient is that states in part $A$ with a low momentum are mostly localized on the left of the cylinder. They have a large number of unoccupied sites on the right side of the cylinder, and therefore there is not need to keep all the $qN_e$ orbitals in their MPS description. 
In \eqref{PESschmidt} the state $\ket{\beta}$ ranges over all states of the CFT with a U$(1)$ charge $Q_{\beta} = \sqrt{q}(N_A - N_B)/2$. They are of course graded by the conformal dimension $\Delta_{\beta}$, which encode the momentum of part $A$
\begin{align}
L_z^A = \Delta_{\beta} - \frac{q}{8}N_e^2
\end{align}  
Since we have assumed $N_A$ to be a multiple of $k$ for simplicity, the smallest possible conformal dimension for $\ket{\beta}$ is obtained for the U$(1)$ primary $\ket{\beta_0} = \ket{\sqrt{q}(N_A- N_B)/2}$. This correspond to a very simple state for part $A$. All $N_A$ particles are located as far to the left of the total droplet as possible, and $\ket{ \psi_{\beta_0}}$ is the (densest) droplet of $N_A$ particles localized over the first $qN_A$ orbitals. A descendant at level $P_{\beta}$, i.e. a state $\ket{\beta}$ with a conformal dimension $\Delta_{\beta} = \Delta_{\beta_0} + P_{\beta}$, describes a state in part $A$ with an extra momentum $\delta L_z^A = P_{\beta}$. Electrons in $| \psi_{\beta} \rangle$ can occupy at most the first $qN_A +P_{\beta}$ orbitals. This means that orbitals beyond $qN_A + P_{\beta}$ are necessarily unoccupied. As a consequence we can rewrite the MPS of a states $| \psi^A_{\beta} \rangle$ as
\begin{align}
 | \psi^A_{\beta} \rangle \propto \sum_{m_1, \cdots, m_{qN_A +P}} \langle{\beta_R}| C^{m_{qN_A+P}} \cdots C^{m_1} \ket{0} \, \ket{m_1 \cdots m_{qN_A +P} 0 \cdots 0} \label{PES MPS A}
\end{align}
$|\beta_R \rangle = e^{-i (q(N_A-N_B)/2 +P) \sqrt{\nu} \varphi_0} \ket{\beta}$ has a U$(1)$ charge $Q_{\beta_R} = -\sqrt{\nu}P_{\beta}$ (with $P = P_{\beta}$). The gain in this representation is twofold : first there are fewer orbitals, and accordingly a much small Hilbert space to consider for part $A$ - though the number of orbital one needs to consider increases with $\Delta_\beta$. Second, and most important point, is that the U$(1)$ charge of $| \beta_R \rangle$ is now much smaller and independent of the number of particles in part $A$ and $B$. Not only is this now accessible within the truncation scheme, but we can even take the thermodynamic limit. 

\begin{figure}
\centering
\includegraphics[width=0.4\linewidth]{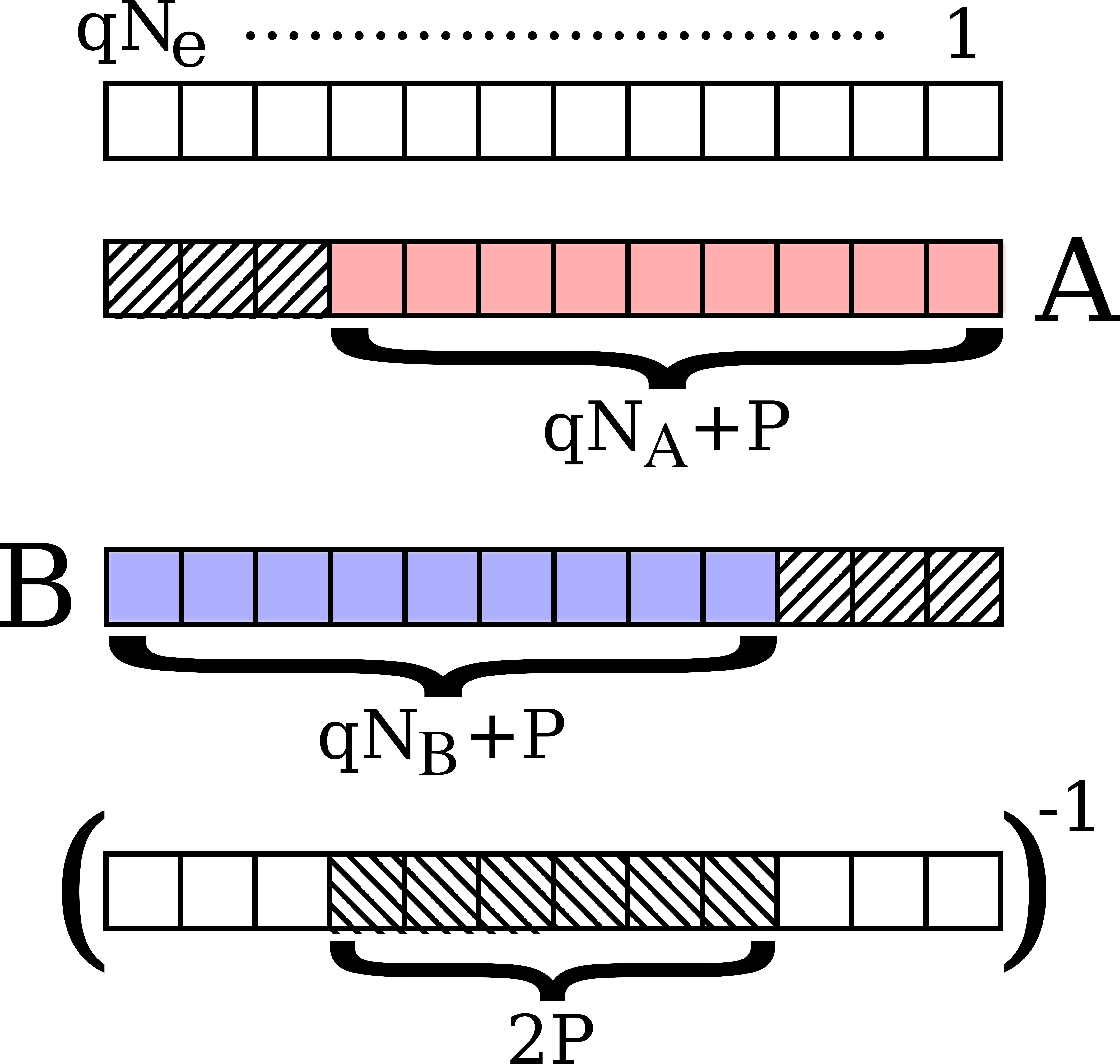}
\caption{Description of the particle cut when focusing on the low momentum sectors. Starting from the full system and fixing the momentum sector $P$ and the number of particles in $A$ and $B$, we only have to consider the $qN_A + P$ leftmost orbitals for $A$ (light red) and the $qN_B + P$ rightmost orbitals for $B$ (light blue). The empty orbitals (regions with stripes) for each part lead to a factor that is proportional (up to a global term that only depends on the total number of orbitals) to the invert of the one of the intersect region (the $2P$ orbitals displayed at the bottom of the figure).}
\label{pesfig}
\end{figure}

Performing the same trick for $|\psi^B_\beta\rangle$, and keeping track of all the prefactors, we get
 \begin{align}
 \ket{\psi} &= \sum_{\beta, Q_{\beta} \text{ frozen}} \sqrt{N_A! \, N_B!} e^{\tau (E_R(0) - E_L(0))} \left(\bra{\beta_R} \left( C^0 \right)^{2P} \ket{\beta_L} \right)^{-1} \nonumber\\
 &\left( \sum_{ \{m_A \}} \langle{\beta_R}| C^{\{ m_A\}} \ket{0} \, \ket{m_A} \right)  \otimes \left( \sum_{ \{m_B \}} \langle{0}| C^{\{ m_B\}} \ket{\beta_L} \, \ket{m_B} \right) \label{PES MPS}
 \end{align}
where for part A the occupation $\{ m_A \} = \{ m_1,\cdots, m_{qN_A + P_{\beta}} \}$ ranges over the first $qN_A +P$ orbitals, while for part $B$ the occupation $ \{ m_B \} = \{ m_{qN_A -P+1},\cdots, m_{qN_e} \}$ ranges over the last $qN_B +P$ orbitals. The intermediate edge states
\begin{eqnarray}
&|\beta_R \rangle = e^{-i (q(N_A-N_B)/2 +P) \sqrt{\nu} \varphi_0} \ket{\beta},&\nonumber\\
&|\beta_L \rangle = e^{-i (q(N_A-N_B)/2 -P ) \sqrt{\nu} \varphi_0} \ket{\beta}&
\end{eqnarray}
now have much smaller U$(1)$ charges ($\mp \sqrt{\nu} P$), and are manageable within the truncation scheme. Moreover they are independent of both $N_A$ and $N_B$. Fig. \ref{pesfig} gives a schematic description of the procedure.

There is a price to be paid for this method: as we increase the number of particles (for instance when moving to the infinitely long cylinder), we have to restrict ourselves to the low momentum sector of the PES. In term of the auxiliary space, this amount to keep $P_{\beta} = \Delta_{\beta} - Q_{\beta}^2/2$ of order $O(1)$ as the number of particle goes to infinity, which boils down to the truncation explained in section~\ref{truncation section}. The effect of the truncation is to confine the particle of part $A$ to the left of the droplet. While this is a poor approximation for the PES, for which particles of part $A$ have no reason not to spread over the whole droplet, it becomes a very good one for the RSES because of incompressibility.

\subsubsection{Infinite PES}

The density matrix can now be obtained by using the same techniques we had for the OES. In particular it is straightforward to use the transfer matrix formalism to reach the thermodynamic limit $N_A, N_B \to \infty$ and work on the infinitely long cylinder. 

For instance the overlaps of the states in part $A$, of the form 
\begin{align}
 \ket{\phi^A_{\beta}} = \sum_{ \{m_A \}} \langle{\beta_R}| C^{\{ m_A\}} \ket{0} \, \ket{m_A}
 \end{align}
are well behaved because the edge state $\beta_R$ does not depend on the number of particles. The transfer matrix formalism yields
 \begin{align}
 \langle \phi^A_{\beta'}\ket{\phi^A_{\beta}} = \delta_{\beta,\beta'} \Lambda^R_{\beta_R}, \qquad  \langle \phi^B_{\beta'}\ket{\phi^B_{\beta}} = \delta_{\beta,\beta'} \Lambda^L_{\beta_L}
   \end{align}
On the infinite cylinder the reduced density matrix becomes (up to a global multiplicative constant)
\begin{align}
\rho_{N_A} = \text{Tr}_{N_B} (\rho) \propto \sum_{\beta}  \frac{\Lambda_{\beta_L}^L \Lambda_{\beta_R}^R }{\bra{\beta_R}\left( C^0 \right)^{2P} \ket{\beta_L} ^2}     \ket{\psi^A_{\beta}} \bra{\psi^A_{\beta}}  
\end{align}
where  the states $|\psi^A_{\beta}\rangle  =  \left(\Lambda^{R}_{\beta_R} \right)^{-1/2} |\phi^A_{\beta}\rangle$ are orthonormal and the sum over $\beta$ is restricted to a specific U$(1)$ charge and neutral sector. 

As a side-remark, notice that the infinite PES does no longer depend on the particle number of the state. This is natural because in the infinite limit, we are removing a finite fraction of an infinite number of particles from the infinite number of particles. Hence the iPES cannot depend on the number of particles.

\subsection{Real Space Entanglement Spectrum}\label{RSES}

\begin{figure}
\centering
\includegraphics[width=0.8\linewidth]{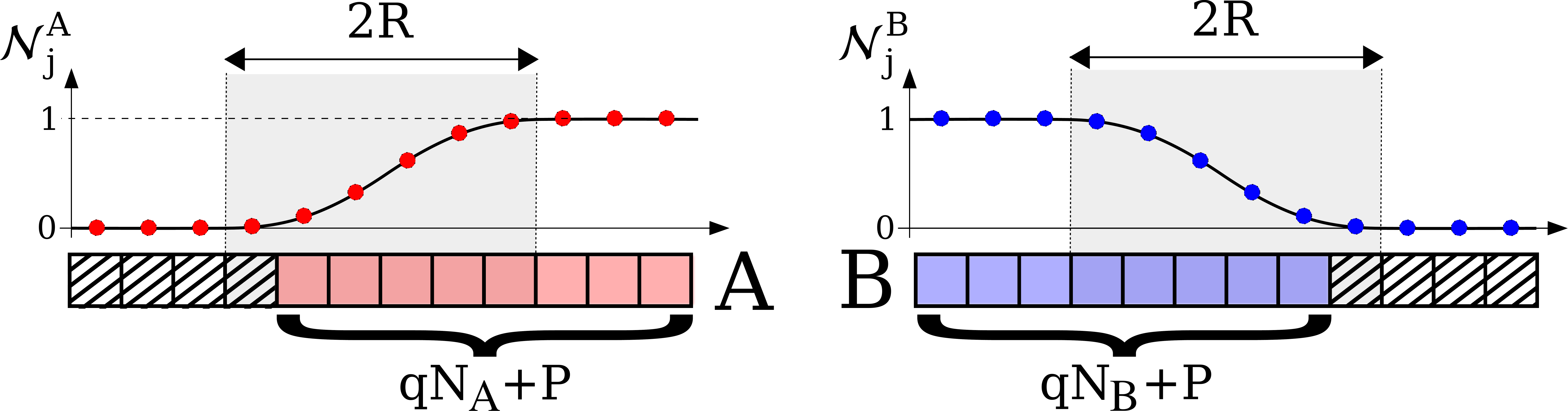}
\caption{Description of the real space cut. The procedure is similar to the one of the particle cut in Fig.~\ref{pesfig}, including the factor coming from the empty orbitals and the number of orbitals for each part. In the gray shaded region, one has to take into account that the orbitals have to be weighted with either $\mathcal{N}^A_j$ (for the part $A$) or  $\mathcal{N}^B_j$ (for the part $B$), leading to site dependent $C^m$ matrices in this region. Away from this finite size region, the weights are either 0 or 1 and the calculation is identical to the PES.}
\label{rsesfig}
\end{figure}

Upon defining the regions $A$ and $B$ by a real space cut, one obtains the so-called real space entanglement spectrum. The relation between the PES and the RSES is well established\cite{Sterdyniak-PhysRevB.85.125308,Dubail:2012p2980,Rodriguez-PhysRevLett.108.256806}. It is convenient to choose a spatial cut which does not destroy the momentum quantum number. On the cylinder this amounts to a cut along the perimeter, at some position $x_0$ along the $x$-axis and define
\beq
\text{Region A : } x < x_0, \qquad \text{Region B : } x > x_0
\eneq
 Each particle can be either in part $A$ or in part $B$, and the electron operator can be decomposed as the sum of an operator with support in $A$ and an operator with support in $B$
\begin{align}
V(z) = V^A(z) + V^B(z)
\end{align}
Let us consider the ground state on the cylinder in first quantized form Eq. \eqref{cylinder GS first quantized}. Plugging the decomposition of $V(z)$ into this conformal block 
we get
\begin{align}
\langle V(z_1) \cdots V(z_{N_e}) \rangle = \sum_{\{ \eta_i \} \in \{ A, B\}} \langle V^{\eta_1}(z_1) \cdots V^{\eta_{N_e}}(z_{N_e}) \rangle
\end{align} 
The number $N_A$ of particles in part $A$ is a good quantum number for this cut.  In each term in the r.h.s we can reorder the variable $z_1,\cdots, z_{N_e}$ so that all the particles in part $A$ are on the left. Let us denote $i_1,\cdots, i_{N_A}$ the particles in part $A$, and $j_1,\cdots, j_{N_B}$ the ones in part $B$.
\begin{align}
\langle V(z_1) \cdots V(z_{N_e}) \rangle =  \sum_{N_a=0}^{N_e}  \sum_{ i_1 \leq  \cdots \leq i_{N_A}}  \epsilon^{\sum_p i_p -p} \langle V(z_{i_1}^{A}) \cdots V(z_{i_{N_A}}^{A}) V(z^B_{j_1}) \cdots V(z^B_{j_{N_B}})\rangle
\end{align} 
where $\epsilon = +1$ for bosons and $-1$ for fermions. Each term in the r.h.s. can now be decomposed as a PES.
\begin{eqnarray}
&&\langle V(z_1) \cdots V(z_{N_e}) \rangle\\
& =&  \sum_{N_a=0}^{N_e}  \sum_{ i_1 \leq  \cdots \leq i_{N_A}}  \epsilon^{\sum_p i_p -p} \sum_{\beta }\langle V(z_{i_1}^{A}) \cdots V(z_{i_{N_A}}^{A}) \ket{\beta} \bra{\beta}V(z^B_{j_1}) \cdots V(z^B_{j_{N_B}})\rangle\nonumber
\end{eqnarray} 
Upon going to the occupation basis, the index $i_1,\cdots, i_{N_A}$ of  the particles in part $A$ does not matter any longer, up to a sign that while cancel $\epsilon^{\sum_p i_p -p}$. As far as the second quantisation is concerned, we can replace the previous decomposition by
\begin{align}
 \sum_{N_a=0}^{N_e} \binom{N_e}{N_A}\sum_{\beta }\langle V(z_{1}^{A}) \cdots V(z_{N_A}^{A}) \ket{\beta} \bra{\beta}V(z^B_{N_A+1}) \cdots V(z^B_{N_e})\rangle
\end{align} 
The r.h.s. - in a given $N_A$ sector- looks very much like a PES. As far as the first quantized wave function is concerned, this is exactly the PES. There is however an important difference when moving to the second quantization. Since the $N_A$ particles are localized in the real space region $A$, while the remaining $N_B$ particles are bound to the region $B$, the one-body normalization have to be modified. For this purpose, let us introduce  
\begin{align}
\mathcal{N}^A_j=\sqrt{\int_{\text{region A}}  dx dy \,    | \phi_{j}(x,y)|^2 }, \qquad \mathcal{N}^B_j = \sqrt{1 - (N^{A}_j)^2}
\end{align} 
where
\begin{align}
\phi_j(x,y) = \frac{1}{\sqrt{L\sqrt{\pi}}}e^{2 \pi i j y /L} e^{- \frac{1}{2}(x - \tau_j)^2 }, \qquad \tau_j = 2 \pi j /L \label{cylinder one body wf}
\end{align} 
The real space spectrum is obtained by taking the same MPS decomposition as for the PES matrix (see Fig.~\ref{rsesfig}), except that the $C^m$ matrices in the $A$ side on orbital $j$ become site-dependent (due to the broken translational invariance) 
\begin{align}
C^m \to C_A^m[j] = \left(\mathcal{N}^A_j \right)^m C^m \label{RSES Cm}
\end{align}
and likewise in region $B$ we have $C^m \to C_B^m[j] = \left(\mathcal{N}^B_j \right)^m C^m$. We end up with at the following decomposition 
 \begin{eqnarray}
 \ket{\psi} & =  e^{\tau (E_R(0) - E_L(0))} \sqrt{N_e!} \sum_{N_A=0}^{N_e}  \binom{N_e}{N_A}^{1/2} \ket{\psi_{N_A, N_B}} 
\end{eqnarray}
 \begin{align}
 & \ket{\psi_{N_A, N_B}}   = \sum_{\beta, Q_{\beta} \text{ frozen}} \left(\bra{\beta_R} \left( C^0 \right)^{2P} \ket{\beta_L} \right)^{-1}  \nonumber\\ 
 & \left( \sum_{ \{m_A \}} \langle{\beta_R}| \prod_j C_A^{m_j}[j] \ket{0} \, \ket{m_A} \right) \otimes \left( \sum_{ \{m_B \}} \langle{0}| \prod_j C_B^{m_j}[j] \ket{\beta_L} \, \ket{m_B} \right)\label{RSES MPS}
 \end{align}
 where $\beta$ has is U$(1)$ charge (and neutral sector) fixed by $N_A$, as was the case for the PES.
 
With such a site dependent MPS representation, it seems very difficult to go to the infinitely long cylinder limit for the real space entanglement spectrum. Indeed  the MPS matrices $C^m_A[j]$ and $C^m_B[j]$ depend on the position of the orbital $j$ relative to the real space cut. However in practice, since Landau level orbitals are localized in real space along the $x$ axis of the cylinder, orbitals very far from the cut either in the traced out side or in the remaining side, have norm either $1$ or $0$, as long as their Gaussian spread does not significantly intersect the cut.

In order to simplify the discussion, we consider a cut at $x_0=0$. 
\beq
\text{Region A : } x < 0, \qquad \text{Region B : } x > 0
\eneq
The single particle orbital $j$ as given by \eqref{cylinder one body wf} is localized in $x$ by the Gaussian factor around $\tau_j = \frac{2\pi j}{L}$, decaying within a few magnetic length  (we work in units such that $l_B=1$). Essentially the single particle orbital vanishes as soon as $|x - \tau_j|$ is larger than a few magnetic length. Three cases can be considered :
\begin{itemize}
\item if $\tau_j \gg 1$  the orbital $j$ is entirely located in part $B$, and we can approximate $\mathcal{N}^B_j \simeq 1$ (and $\mathcal{N}^A_j \simeq 0$). For such orbitals we can effectively work with $C^m_A = 0$ and $C^m_B = C^m$.
\item likewise if $\tau_j$ is much smaller than $-1$  the orbital $j$ is entirely located in part $A$, and we have $\mathcal{N}^A_j \simeq 1$. There we have $C^m_A = C^m$ and $C^m_B = 0$.
\item the orbitals in between, with support in both part $A$ and $B$, for which the matrices $C^m$ have to be site dependent as in \eqref{RSES Cm}. 
\end{itemize}
To implement this idea, we have to introduce a cutoff to control which orbitals are in part $B$, which are in part $A$, and which are in the in-between region.  We now define the \emph{orbital} length $R$ of a region over which the orbital overlap depends on the distance from the cut. The logic will be that, within this orbital distance away from the cut, the norm of the orbitals is dependent on the distance from the cut, but away from it the orbital norm becomes $1$ or $0$ depending whether the orbital is on the right or left side of the cut. In other words:
\beq
\mathcal{N}_j^A =1 \;\;\; \text{if}\;\;\; j < -R;\;\;\;\;\;\;\;\;\; \mathcal{N}_j^A = 0 \;\;\;  \text{if}  \;\;\;  j >R
\eneq
while in-between for $-R<j<R$ we keep the exact overlap $\mathcal{N}_j^A$.
\begin{align}
\mathcal{N}_j^A = \sqrt{\frac{1}{\pi}\int_{x < 0} dx \,  e^{- (x - \tau_j)^2 } }
\end{align}
The parameter $R$ provides an interpolation between the real space ($R \gg 1$) and the orbital spectrum ($R=0$).  To get a good approximation of the RSES it is sufficient to chose
the orbital region $R$ such that the distance between the cut and the R$^{th}$ orbital is larger than a few magnetic length, i.e.
\begin{align}
R > \frac{L}{2\pi}
\end{align}
In practice, we will want to take $R\ge |P_{\text{max}}|$.

Using this trick it is straightforward to compute the RSES on an infinitely long cylinder. Indeed in the MPS representation \eqref{RSES MPS}, anything at a distance greater than $\tau_R$ away from the real space cut has a translational invariant MPS representation (using $C^m$), while only a finite region of the cylinder - the orbitals ranging from $-R$ to $R$ - requires a site dependent representation. This is illustrated in Fig. \ref{rsesfig}.  This means that we can use the transfer matrix formalism to make the translation invariant regions, in which the overlaps are $1$ or $0$, arbitrarily large. We do not detail the method, as it is the same as for the OES and the PES.

\section{Conclusions}

In this manuscript we have detailed all the general steps necessary to obtain an MPS representation for any FQH wave function that can be written as an expectation value of CFT vertex operators. We also particularized and presented the MPS for many examples such as the minimal models, Read-Rezayi states, and superconformal minimal models. We have then truncated the Hilbert space so that numerical computations can be possible and presented ways to trim the Hilbert space further and hence optimize the MPS representation of the FQH states. We have also showed how to use the MPS to compute properties of states such as the orbital, particle, and real entanglement spectrum in both finite and infinite limits. In the current paper we have focused only on the FQH ground-state wave functions, but the formalism put in place can deal with both quasihole and quasielectron excitations, although the latter ones are more difficult to obtain. The MPS for many FQH ground-states has been implemented numerically and will be useful in understanding the difference between FQH states described by unitary versus non-unitary CFTs as well as obtaining the braiding of non-Abelian quasiholes that has so far escaped a first principle calculation \cite{regnaultinprep, yanglewuinprep}. Jain's composite fermion states\cite{Jain:1989p294}, as well as other, more complicated, spinful FQH states \cite{Ardonne-PhysRevLett.82.5096} may also have MPS representations which can be obtained with added effort. 

In future papers \cite{regnaultinprep, yanglewuinprep} we will analyze the physical properties of many FQH states, including the non-unitary Gaffnian state, compute their entanglement entropy, quasihole braiding, and place limits on the correlation lengths of local observables in these states.  

\section{Acknowledgements}
We thank thank M. Zaletel, R. Mong, J. Dubail, N. Read, Y.-L. Wu, A. Sterdyniak for discussions and especially Z. Papic for earlier collaboration on this work. BAB and NR were supported by NSF CAREER DMR-095242, ONR-N00014-11-1-0635, MURI-130- 6082, Packard Foundation, and Keck grant.

\newpage 

\appendix

\section{Cylinder Normalization for the MPS}
\label{Appendix appendix cylinder normalization}

We start from the $L= \infty$ MPS (the "conformal limit"), and we want to compute the change to the MPS coefficient 
\begin{align}
c_m = \langle \alpha' | B^{m_{J -1}} \cdots B^{m_1}B^{m_0} | \alpha \rangle
\end{align}
coming from the replacement of the $B^m$ matrix with its cylinder normalized\footnotemark \footnotetext{This is not the gauge chosen in \eqref{normalizedmps}, but the results given in \eqref{cylinder edge factors} follow from the calculation presented here.} $C^m$: 
\begin{align}
B^m \to C^m = \exp \left( - \left( \frac{2\pi}{L} \right)^2 L_0 \right) B^m
\end{align}
where the matrix $B^m$ are
\begin{align}
B^m =    e^{-i \sqrt{\nu}\varphi_0/2} \left( \frac{1}{\sqrt{m!}} V_{0}^m  \right)   e^{-i\sqrt{\nu}\varphi_0/2}
\end{align}
Let us introduce some useful notations :
\begin{align}
\ket{\alpha_j} =  C^{m_{j-1}} \ket{\alpha_{j-1}}, \qquad \ket{\alpha_0} = \ket{\alpha} 
\end{align} 
The state $\ket{\alpha_j}$ above has charge and dimension
\begin{align}
\sqrt{\nu}Q_j & = \sqrt{\nu}Q_{j-1} + m_{j-1}  - \nu \\
\Delta_j & = \Delta_{j-1} - \frac{1}{2} m_{j-1} + \frac{\nu}{2} - \sqrt{\nu}Q_{j-1}
\end{align} 
with the solutions:
\begin{align}
\sqrt{\nu}Q_j & = \sqrt{\nu}Q_0- \nu j + \sum_{k=0}^{j-1}m_k \\
\Delta_j &   =  \Delta_0  - \frac{j^2}{2}\nu - j\sqrt{\nu}Q_j +  \frac{1}{2} \sum_{k=0}^{j-1}m_k  + \sum_{k=0}^{j-1}k m_k \label{conformaldimensionsolution1}
\end{align}
A good check for these formulas is the non-interacting case. The $\nu=1$ filled band is obtained by taking all $m_i =1$, and to start with $Q_0= \Delta_0 =0$. One gets the expected $Q_j = \Delta_j =0$ for all $j$. 

To obtain the change in the state $\ket{\alpha_j}$ coming from the $B^m \rightarrow C^m$ replacement is a sum over all the $\Delta_k$ - eigenvalues of $L_0$ up to site $j$.  Introducing the intermediate notation $I_p = I_p(j) = \sum_{k=0}^{j-1} k^p m_k$, we have 
\begin{align}
\sum_{k=1}^j \sqrt{\nu} Q_k &  = j \sqrt{\nu}Q_j + \nu \frac{j(j-1)}{2} -  I_1
\end{align}
and likewise
\begin{align}
\sum_{k=1}^j \Delta_k & = j\Delta_0 + \frac{j(j+1)(2j+1)}{12}\nu  - \frac{j(j+1)}{2}\sqrt{\nu}Q_0- \frac{j^2}{2}I_0 -  \frac{1}{2}I_2  + j  I_1
\end{align}
Notice that $I_2$ is the usual Gaussian factor of the orbital normalization on the cylinder. Hence, we  found the difference between the coefficients as normalized on the cylinder $c_m(L)$ and in the conformal limit $c_m(\infty)$ as $c_m(L) = c_m(\infty) e^{-  \left( \frac{2\pi}{L} \right)^2 A }$ with $A = \sum_{k=1}^j \Delta_k $

%%%%%%%%%%%%%%%%%%%%%%%%%%%%%%%%%%%%%%%%%%%%%%%%%%%%%%%%%%%%%%%%%%%%%%%%%%%%%%%%%%%%%%%%

\newpage 
\section{Gram Matrix, Orthogonality,  and Null Vectors}\label{Appendix3}

We present an example of the computation of the Gram matrix at level $2$ where we also encounter a null vector. States with different descendant level are trivially orthogonal but in a given level different states have generically a non vanishing overlap. For instance at level $2$ there are two states $| \Phi_\Delta, (1,1) \rangle = L_{-1}^2 | \Phi_\Delta \rangle$ and $| \Phi_\Delta, (2) \rangle = L_{-2} | \Phi_\Delta \rangle$. Their overlap can be computed as follow
\begin{eqnarray}
\langle \Phi_\Delta, (1,1) | \Phi_\Delta, (2) \rangle &= \langle \Phi_\Delta | L_1^2 L_{-2} | \Phi_\Delta \rangle = & \langle \Phi_\Delta | L_1 [ L_1,L_{-2}] | \Phi_\Delta \rangle\nonumber\\
& = 3  \langle \Phi_\Delta | L_1  L_{-1} | \Phi_\Delta \rangle = &6 \langle \Phi_\Delta | L_0 | \Phi_\Delta \rangle = 6 \Delta
\end{eqnarray}
where we have used $L_{1} |\Phi_\Delta \rangle =0$ ($|\Phi_\Delta \rangle$ is a primary field/state of conformal dimension $\Delta$) and the Virasoro algebra $[L_{1},L_{-2}] = 3 L_{-1}$. This procedure can be used to compute any overlap between descendants. At level $2$ for instance we have
\begin{align}
\langle \Phi_\Delta, (2) | \Phi_\Delta, (2) \rangle  =  4 \Delta + \frac{c}{2}, \;\;\;\; \langle \Phi_\Delta, (1,1) | \Phi_\Delta, (1,1) \rangle =  4\Delta(2\Delta+1)
\end{align} From this we can build two orthogonal states
\begin{align}
\ket{\alpha_1} =  | \Phi_\Delta, (1,1) \rangle \quad \text{and} \quad  \ket{\alpha_2} = | \Phi_\Delta, (1,1) \rangle - \frac{2}{3}(2\Delta+1)  | \Phi_\Delta, (2) \rangle
\end{align}
and the last step is to normalize them. Their norm is
\begin{align}
\bra{\alpha_1} \alpha_1 \rangle = 4 \Delta(2\Delta+1), \qquad \bra{\alpha_2} \alpha_2 \rangle =  \frac{2}{9}(2\Delta+1)\left[2\Delta(8\Delta-5) + c(2\Delta+1) \right]
\end{align}
Depending on the values of the central charge $c$ of the CFT, and of the conformal dimension $\Delta$ of the primary under consideration $\ket{\Phi_\Delta}$, three things can happen :
\begin{itemize}
\item the norm of the descendant is (strictly) positive, and one can normalize the state in the usual way 
\item the norm is (strictly) negative (this can only happen in a non-unitary CFT)
\item the norm vanishes
\end{itemize}
In the latter case, one talks of { null states} or { null vectors} (this can never happen for primary fields). Such states decouple from the theory (it is orthogonal to all states, including itself), and they can be discarded from the theory. The simplest such null-state appears in the module of the identity field $\Phi_{(1|1)} = \mathid$, correspond to the vacuum state usually denoted by $|0 \rangle$, of conformal dimension $0$ ($L_0 | 0 \rangle = 0$). The level-$1$ descendant  $L_{-1}|0 \rangle$ has zero norm $\langle 0 |  L_1 L_{-1}|0 \rangle = 2 \langle 0 |  L_0 |0 \rangle = 0$. Another example is given by any primary field $\Phi$ whose conformal dimension obeys $2\Delta(5-8\Delta) = c (2\Delta+1)$. In this case the level $2$ descendant
\begin{align}
 \left(L_{-1}^2 - \frac{2(2\Delta+1)}{3}L_{-2}\right) |\Phi_\Delta \rangle,
 \end{align}
has zero norm. 
For a given CFT, the central charge is fixed. Null vectors can only occur amongst descendants of primary fields with very specific conformal dimensions $\Delta$. When null vectors occur, we call the  level ''made up of degenerate fields". Naively one would thing that this being a set of zero measure, it will be at best an extremely rare scenario, but this cannot be so as then the central charge of the CFT (which governs the growth of the CFT Hilbert space) would be the same as that of a $U(1)$ boson. It turns out that minimal models are made up entirely of degenerate fields. For instance the Ising CFT has central charge $c=\frac{1}{2}$, and contains $3$ primary fields, all of them degenerate.
\begin{itemize}
\item $\mathid$, with conformal weight $0$, is degenerate at level $1$,
\item $\Psi$, with conformal weight $1/2$, is degenerate at level $2$,
\item $\sigma$, with conformal weight $1/16$, is also degenerate at level $2$.
\end{itemize}
The number of null vectors in a degenerate level is analytically known. We use it to test that our numerical procedure gives the correct number of null vectors.

%%%%%%%%%%%%%%%%%%%%%%%%%%%%%%%%%%%%%%%%%%%%%%%%%%%%%%%%%%%%%%%%%%%%%%%%%%%%%%%%%%%%%%%

\newpage

\section{On Matrix Elements For Neutral CFT} \label{AppendixNeutralCFT}

We now give several examples of the methods exposed to compute matrix elements of the CFT primary fields. 

\subsection{The method}

The key ingredient to compute the matrix elements of a primary field in the basis of (Virasoro) descendants is the following relation, valid for all $m \in \mathbb{Z}$, $\Phi^{(h)}$ a primary field with conformal dimension $h$, 
\begin{align}
\langle\alpha' | L_{m}  \Phi^{(h)}(1) |\alpha \rangle = ( m h + \Delta_{\alpha'}-\Delta_{\alpha})\langle \alpha' |   \Phi^{(h)}(1)(1) |\alpha \rangle + \langle \alpha' |  \Phi^{(h)}(1) L_m|\alpha \rangle \label{FR_Virasoro}
\end{align}
where $\ket{\alpha}, \ket{\alpha'}$ are arbitrary Virasoro descendant states, namely $| \alpha' \rangle = \ket{\Phi_{\Delta'},\theta'}, | \alpha \rangle = \ket{\Phi_{\Delta},\theta}$ in the notations of equation \eqref{virasorodescendants} (we are going to drop the prefactors $z_{\theta},z_{\theta'}$ as they play no role in the evaluation of the matrix elements). Their conformal dimension is $\Delta_{\alpha} = \Delta + |\theta|$ and $\Delta_{\alpha'} = \Delta' + |\theta'|$. Assuming we have computed all matrix elements of the form \eqref{ME_Delta} up to total level $|\theta| + |\theta'| = N$, we can compute any matrix element
\begin{align}
\langle \Phi_{\Delta'} ,\theta' |  \Phi^{(h)}(1) |\Phi_{\Delta}, \theta \rangle = \langle \Phi_{\Delta'} |L_{\theta'_n} \cdots L_{\theta'_2}L_{\theta'_1}  \Phi^{(h)}(1) |\Phi_{\Delta}, \theta \rangle
\end{align}
 at total level $|\theta| + |\theta'| = N+1$ as follow
\begin{itemize}
\item commute the last mode $L_{\theta'_1}$ with the field $\Phi^{(h)}(1)$ using \eqref{FR_Virasoro}:
\begin{eqnarray}
&&\langle \Phi_{\Delta'} ,\theta' |  \Phi^{(h)}(1) |\Phi_{\Delta}, \theta \rangle \nonumber\\
& =& \left[ \theta'_1 (h-1) + \Delta' + |\theta'| - \Delta - |\theta| \right] \langle \Phi_{\Delta'} |L_{\theta'_n} \cdots L_{\theta'_2}  \Phi^{(h)}(1) |\Phi_{\Delta}, \theta \rangle \nonumber \\
& & + \langle \Phi_{\Delta'}  |L_{\theta'_n} \cdots L_{\theta'_2}  \Phi^{(h)}(1) L_{\theta'_1}|\Phi_{\Delta}, \theta \rangle \label{recursion_temp}
\end{eqnarray}
\item use the Virasoro algebra to decompose the level $|\theta| - \theta'_1$ descendant state $L_{\theta'_1}|\Phi_{\Delta}, \theta \rangle$ in the canonical basis \eqref{virasorodescendants} (if $|\theta| - \theta'_1 < 0$ this state vanishes):
\begin{align}
 L_{\theta'_1}|\Phi_{\Delta}, \theta \rangle &=   L_{\theta'_1} L_{-\theta_1} \cdots L_{-\theta_m}|\Phi_{\Delta} \rangle  = \left[ L_{\theta'_1},  L_{-\theta_1} \cdots L_{-\theta_m} \right]|\Phi_{\Delta} \rangle & \nonumber\\
&=  \sum_{\theta'' , |\theta''| = |\theta| - \theta'_1} D^{(h)}_{\theta'_1,\theta,\theta''} | \Phi_{\Delta} ,\theta'' \rangle
\end{align}
\item now all terms in the r.h.s. of \eqref{recursion_temp} are matrix elements of the form \eqref{virasorodescendants} at total level smaller than $N$, and they all have been computed previously. Therefore we have the new matrix element.
\end{itemize}

\subsection{Examples}

We give several explicit examples of how to compute the matrix elements in the neutral CFT.  An easy example to start with is
\begin{align}
\frac{\langle \Phi_{\Delta'}, (m)|\Phi^{(h)}(1) |\Phi_{\Delta}  \rangle}{ \langle\Phi_{\Delta'} |  \Phi^{(h)}(1) | \Phi_{\Delta} \rangle}  = \frac{ \langle \Phi_{\Delta'}|L_{m}  \Phi^{(h)}(1) |\Phi_{\Delta} \rangle}{\langle \Phi_{\Delta'} |  \Phi^{(h)}(1) | \Phi_{\Delta} \rangle} =  ( m h + \Delta'-\Delta)
\end{align} 

For a more complicated example, consider, for $m \geq n \geq 0$:
\begin{align}
&\frac{\langle \Phi_{\Delta'}, (m)| \Phi^{(h)}(1) |\Phi_{\Delta}, (n) \rangle}{ \langle\Phi_{\Delta'} |  \Phi^{(h)}(1) | \Phi_{\Delta} \rangle} = \frac{\langle \Phi_{\Delta'} |L_{m}  \Phi^{(h)}(1) L_{-n}|\Phi_{\Delta} \rangle}{ \langle\Phi_{\Delta'} |  \Phi^{(h)}(1) | \Phi_{\Delta} \rangle} & \nonumber\\
 = & (  m h + \Delta'-\Delta-n)\frac{\langle \Phi_{\Delta'} |  \Phi^{(h)}(1) L_{-n}|\Phi_{\Delta} \rangle}{ \langle\Phi_{\Delta'} |   \Phi^{(h)}(1) | \Phi_{\Delta} \rangle} + \frac{\langle \Phi_{\Delta'} |  \Phi^{(h)}(1)(1) L_m L_{-n}|\Phi_{\Delta} \rangle}{ \langle\Phi_{\Delta'} |   \Phi^{(h)}(1) | \Phi_{\Delta} \rangle} &\nonumber\\
 = & ( m h + \Delta'-\Delta-n)( n h + \Delta-\Delta') + \frac{\langle \Phi_{\Delta'} |  \Phi^{(h)}(1) [ L_m,  L_{-n} ]|\Phi_{\Delta} \rangle }{ \langle\Phi_{\Delta'} |  \Phi^{(h)}(1) | \Phi_{\Delta} \rangle}&
\end{align}
since $L_m | \Phi_{\Delta} \rangle= 0$ by virtue of $|\Phi_{\Delta} \rangle$ being primary. We can now use the Virasoro algebra to compute the last term. Finally we get for $m \geq n$:
\begin{align}
\frac{\langle \Phi_{\Delta'}, (m)|\Phi^{(h)}(1) |\Phi_{\Delta},(n) \rangle}{\langle \Phi_{\Delta'} |  \Phi^{(h)}(1) |\Phi_{\Delta} \rangle } & =   [& ( m h + \Delta'-\Delta-n)( n h + \Delta-\Delta') \nonumber\\
&& + \delta_{n,m} \left(2n \Delta + \frac{c}{12}n(n^2-1) \right) ]
\end{align} 
We now present the matrix element up to level 2 (with the convention $\langle \Phi_{\Delta'} |\Phi^{(h)}(1) |\Phi_{\Delta} \rangle =1$):
\begin{eqnarray}
\langle \Phi_{\Delta'},(1) |\Phi^{(h)}(1) |\Phi_{\Delta} \rangle & = &         (  h + \Delta'-\Delta)                \\
\langle \Phi_{\Delta'},(2) |\Phi^{(h)}(1) |\Phi_{\Delta} \rangle & =  &          ( 2 h + \Delta'-\Delta)                \\
\langle \Phi_{\Delta'},(1,1) |\Phi^{(h)}(1) |\Phi_{\Delta} \rangle & =  &         (  h + \Delta' + 1 -\Delta)   (  h + \Delta'-\Delta)                             \\
\langle \Phi_{\Delta'},(1) |\Phi^{(h)}(1) |\Phi_{\Delta},(1) \rangle & = &         ( h + \Delta' -\Delta -1 )   (  h - \Delta' + \Delta) + 2\Delta                       \\
\langle \Phi_{\Delta'},(2) |\Phi^{(h)}(1) |\Phi_{\Delta},(1) \rangle & = &            (  2h+ \Delta' -\Delta -1 )   (  h- \Delta' + \Delta)                   \\
\langle \Phi_{\Delta'},(1,1) |\Phi^{(h)}(1) |\Phi_{\Delta},(1) \rangle & =&      ( h + \Delta' -\Delta ) \nonumber\\
&&  \times [ (  h + \Delta' -\Delta -1 )   (  h- \Delta' + \Delta) + 4\Delta ]             \\
\langle \Phi_{\Delta'},(2) |\Phi^{(h)}(1) |\Phi_{\Delta},(2) \rangle & = &     ( 2 h + \Delta' -\Delta -2 )   ( 2  h  - \Delta' + \Delta)\nonumber \\
&& + 4\Delta  + \frac{c}{2}                       \\
\langle \Phi_{\Delta'},(1,1) |\Phi^{(h)}(1) |\Phi_{\Delta},(2) \rangle & = &    (2h + \Delta-\Delta'-2)   (  h  + \Delta' + 1 -\Delta)   ( h + \Delta'-\Delta) \nonumber \\
&& + 6\Delta'                       \\
\langle \Phi_{\Delta'},(1,1) |\Phi^{(h)}(1) |\Phi_{\Delta},(1,1) \rangle & = &  ( h + \Delta'-\Delta-1)   ( h + \Delta -\Delta' ) \nonumber\\
&& \times  [ (  h + \Delta -\Delta' -1 )   (  h - \Delta + \Delta') + 4\Delta' ] \nonumber \\
&& + 2(2\Delta+1)  [ (  h + \Delta' -\Delta -1 )   ( h - \Delta' + \Delta)\nonumber \\
&& + 2\Delta  ]
\end{eqnarray}

\section{$\mathbb{Z}_3$ Read-Rezayi state}
\label{RR}

In this appendix we show how to obtain the MPS representation of the Read-Rezayi state $\mathbb{Z}_3$.  This state  generalizes the Laughlin and Moore-Read states. The bosonic version of this state is the $(k,r)=(3,2)$ Jack polynomial, and its fermionic cousin is thought to describe states at filling $12/5$. Generically, the $\mathbb{Z}_k$ Read-Rezayi states have a rather involved MPS representation since the neutral CFT has a complicated chiral algebra ($\mathbb{Z}_k$ parafermions or $W_k$ theories). Due to its convolution, we will not present the general case here. However in the case of the $\mathbb{Z}_3$ state a simplification is possible, as the neutral CFT can be understood purely in terms of its Virasoro algebra, \emph{i.e.} without resorting to its extended $W_3$ (or $\mathbb{Z}_3$) structure.

\subsection{Hilbert Space}

The neutral CFT of the $\mathbb{Z}_3$ RR state is is referred to as $\mathbb{Z}_3$ parafermionic theory with central charge $c = 4/5$. This theory can be understood as the minimal model $M(5,6)$, but with some degeneracies in the field content (\emph{i.e.} a non-diagonal modular invariant theory). For instance it contains  two fields of dimension $2/3$ : $\Psi_1$  and $\Psi_{-1}$.  Fields in the $\mathbb{Z}_3$ parafermionic theory are classified according to an extra quantum number $\bar{n} \in \mathbb{Z}_3$ called the $\mathbb{Z}_3$ charge, and are given in Table[\ref{tableZ3RR}]

\begin{table*}[h]
 \begin{center}
\begin{tabular}{c|c|c|c|}
 &  $\bar{n}=0$     & $\bar{n}=1$ & $\bar{n}=-1$       \\
 \hline
 Vacuum &  $\mathbb{1}, \quad \Delta=0$   &   $\Psi_1, \quad  \Delta = \frac{2}{3}$  & $\Psi_{-1}, \quad \Delta = \frac{2}{3}$  \\
  sector &  $W, \quad \Delta=3$   &  &  \\
\hline
 Quasihole &  $\epsilon, \quad \Delta=\frac{2}{5}$   &   $\sigma_1, \quad  \Delta = \frac{1}{15}$  & $\sigma_{-1}, \quad \Delta = \frac{1}{15}$  \\
   sector &  $\varphi, \quad \Delta=\frac{7}{5}$   &  &  \\
\hline
\end{tabular} \label{tableZ3RR}
\end{center}
\caption{$\mathbb{Z}_3$ parafermionic theory : primary fields and their dimension according to their $\mathbb{Z}_3$ charge $\bar{n}$}
\label{tab:Z3 primaries}
\vspace{-5pt}
\end{table*}
\noindent The $\mathbb{Z}_3$ charge $\bar{n}$ is additive under fusion of two fields, as can be seen in the fusions of $\Psi_1$ with all the other fields: 
\begin{align}
\Psi_1 \times \Psi_1 & = \Psi_{-1} \\
\Psi_1 \times \Psi_{-1} & = 1+ W \\
\Psi_1 \times \mathbb{1} & = \Psi_1 \\
\Psi_1 \times W & = \Psi_1 
\end{align}
and
\begin{align}
\Psi_1 \times \sigma_{1} & = \sigma_{-1} \\
\Psi_1 \times \sigma_{-1} & = \epsilon+\varphi \\
\Psi_1 \times \epsilon & = \sigma_{1} \\
\Psi_1 \times\varphi & = \sigma_{1} 
\end{align}
The electron operator lives in the full CFT $M(5,6) \otimes U(1)_{3\sqrt{q}}$ and takes the form
\begin{align}
 V(z)=   \Psi_1(z) \, \otimes\, :e^{i\sqrt{q}\varphi (z)}: , \qquad q = 2/3 + m
\end{align}
The electron operator has conformal dimension $h= q/2+ 2/3 = 1 + m/2$.  As usual, the U$(1)$ charge of the various quasi-hole fields  has to be adjusted  in order to ensure trivial monodromies with electrons.  This leads to the following selection rule  in the tensor product  $M(5,6) \otimes U(1)_{3\sqrt{q}}$ :  a neutral field with a $\mathbb{Z}_3$ charge $\bar{n}$ must have a U$(1)$ charge $N = - \bar{n}$ mod $3$.

\subsection{Topological Sectors, Ground State Degeneracy and Thin-Torus Limit} 

The total Hilbert space can be decomposed into $6q$ topological sectors. Using the notations of \eqref{Heisenberg modulo module} and \eqref{Virasoro module}, the topological sectors can be written in compact form
\begin{itemize}
\item for the vacuum sector 
\begin{align}
\mathcal{H}_{(\mathbb{1},j)} =&  \left(  M_\mathbb{1}^{(0)}\otimes  h_{3j \text{ mod } 9q} \right) \,  & \\ 
& \oplus  \,\left( M_\mathbb{1}^{(1)} \otimes \ h_{3(j+q) \text{ mod }9q } \right)  \,     \oplus  \,\left( M_\mathbb{1}^{(2)} \otimes \ h_{3(j-q) \text{ mod }9q } \right)&\nonumber
\end{align}
where $M_{\mathbb{1}}^{(0)} =  \mathcal{V}_{\mathbb{1}} \oplus  \mathcal{V}_{W} $,  $M_{\mathbb{1}}^{(1)} =  \mathcal{V}_{\Psi_{ 1}}$ and $M_{\mathbb{1}}^{(2)} =  \mathcal{V}_{\Psi_{ -1}}$,
\item  and in the $\epsilon$ sector
\begin{align}
\mathcal{H}_{(\epsilon,j)} =&  \left(  M_\epsilon^{(0)}\otimes  h_{3j \text{ mod } 9q} \right) &\\
&     \oplus  \,\left( M_\epsilon^{(1)} \otimes \ h_{3(j+q) \text{ mod }9q } \right)  \,     \oplus  \,\left( M_\epsilon^{(2)} \otimes \ h_{3(j-q) \text{ mod }9q } \right)& \nonumber
\end{align}
with $M_{\epsilon}^{(0)} =  \mathcal{V}_{\epsilon} \oplus  \mathcal{V}_{\varphi} $,  $M_{\epsilon}^{(1)} =  \mathcal{V}_{\sigma_{ 1}}$ and $M_{\epsilon}^{(2)} =  \mathcal{V}_{\sigma_{ -1}}$,
\end{itemize}
where $j = 0 , \cdots, 3q-1$.  Each topological sector yields a unique ground state on the cylinder, with the corresponding topological charge at infinity. As for all Jack states, such ground states are characterized by a a root partition obeying a generalized Pauli principle for the $\mathbb{Z}_3$ RR state\cite{Bernevig-PhysRevB.77.184502,Bernevig-PhysRevLett.100.246802} :
\begin{itemize}
\item  for $\mathcal{H}_{(\mathbb{1},j)} $ we get
\begin{align}
\cdots0^{1+m-j} 10^{m-1}10^{m-1}10^j\cdots
\end{align}
\item  while for $\mathcal{H}_{(\epsilon,j)}$ we get
\begin{align}
\cdots0^{m-j} 10^{m}10^{m-1}10^j\cdots
\end{align}
\end{itemize}
where $m$ is related to the $q$ through $q = 2/3+m$.

\subsection{$\mathcal{W}_3$ vs Minimal Model Approach}

This CFT enjoys an extended $\mathcal{W}_3$ algebra, which is larger than Virasoro and which reduces the number of primary states. This algebra contains both the stress energy operator $T(z)$ and a spin $3$ current $W^{(3)}(z)$ whose modes $W_n$ can be used to build descendant states. The commutation relations of this algebra are too tedious to present here, but they can in principle be implemented.  In this language, the $W$ field above would not be a primary field, but rather a descendant of $1$ (to be more precise $|W\rangle=W_{-3}|0\rangle$ - for this reason the $W$ field only appears at ``momentum" $3$ even in the Virasoro formulation), and one recovers
\begin{align}
\Psi_1 \times \Psi_1 = \Psi_{-1} , \qquad \Psi_{1} \times \Psi_{-1} = \mathbb{1} \label{Z3_fusion_rules}
\end{align}
which implies working with $3$ sectors $|0\rangle$, $|\Psi_1\rangle$ and $|\Psi_{-1}\rangle$. The $B$ matrices are the same as before,
\begin{align}
\langle \Phi_{\Delta'}, \Theta' ; N', \mu' | B^0  | \Phi_{\Delta}, \Theta ; N, \mu \rangle = \langle \Phi_{\Delta'}, \Theta' | \Phi_{\Delta}, \Theta \rangle \times  \delta_{\mu,\mu'} \delta_{N',N-3} \label{B0 W}
\end{align}

\begin{eqnarray}
& & \langle \Phi_{\Delta'}, \Theta' ; Q', \mu' | B^1  | \Phi_{\Delta}, \Theta ; Q, \mu \rangle   \nonumber\\
&=&    \langle \Phi_{\Delta'}, \Theta' | \Psi_1(1) |\Phi_{\Delta},\Theta \rangle \label{B1 W}\\
&& \times  \delta_{\Delta'+|\Theta'|+|\mu'|+\frac{N}{3} + \frac{q-1}{2},\Delta +|\Theta|+|\mu|}   \delta_{N',N+3(q-1)} \,   A_{\mu',\mu}^{(\sqrt{q})}\nonumber
\end{eqnarray}

%%\begin{align}
%%  \addtolength{\fboxsep}{5pt}
%%   \begin{gathered}
%%     \langle \Phi_{\Delta'}, \Theta' ; Q', \mu' | B^1  | \Phi_{\Delta}, \Theta ; Q, \mu \rangle  =    \langle \Phi_{\Delta'}, \Theta' | \Psi_1(1) |\Phi_{\Delta},\Theta \rangle \ \times  \delta_{\Delta'+|\Theta'|+|\mu'|+N/3,\Delta +|\Theta|+|\mu|+ 2/3 + 1/2}   \delta_{N',N+3(q-1)} \,   A_{\mu',\mu}^{(\sqrt{q})}
%%  \end{gathered}
%%   \label{B1 W}
%%\end{align}
 $\langle \Phi_{\Delta'}, \Theta' | \Psi_1(1) |\Phi_{\Delta},\Theta \rangle$ and $\langle \Phi_{\Delta'}, \Theta' | \Phi_{\Delta}, \Theta  \rangle$ can be computed using standard CFT techniques \footnotemark \footnotetext{Although this is more involved than for Virasoro descendants and relies on the particular degeneracies of the field $\Psi$.} while  $A_{\mu',\mu}$ is given in \eqref{matrix A} for $\beta = \sqrt{q}$. $\Theta = (\theta,\rho)$ is a pair of (bosonic) partitions, and the descendant $| \Phi_{\Delta}, \Theta  \rangle$ is the descendant $\theta$ in the modes $W_\rho$ of the stress-energy operator and descendant $\rho$ in the modes of the spin field $W$:
\begin{align}
| \Phi_{\Delta}, \Theta  \rangle = L_{-\theta}W_{-\rho}| \Phi_{\Delta} \rangle \label{W_descendants}
\end{align}
Although this approach applies to all $\mathbb{Z}_3$ states, it has a huge disadvantage for the RR case. Since the RR CFT can be understood purely with Virasoro, it has much less independent descendants than a generic $W$ algebra. In working with descendants of the form \eqref{W_descendants}, a huge amount of null vectors appears. For instance in the sector of the identity the number of null vectors at level $14$ is $2600$. 
This makes the generic $W$ approach very inefficient. 

The way to improve this is to use the knowledge of the minimal model. In this approach, all the formalism is identical to the one previously described - we do not need to change the way we compute matrix elements and overlaps, as we do not need the commutation relations between the modes of $W$ with themselves and with other fields. In the Virasoro approach, the $\ket{W}= \sqrt{3/c} W_{-3}| 0\rangle$ state is a primary field (with conformal dimension $3$) with respect to the Virasoro algebra. The method then allows us to restrict  to descendants of the form
\begin{align}
L_{-\theta}| \Phi_{\Delta}  \rangle ,\qquad \text{in the sectors } \Psi_{\pm 1}   \\
L_{-\theta} | 1 \rangle,\text{ and } L_{-\theta} |W\rangle ,  \qquad \text{in the sector } 1,
\end{align}  As $|W\rangle = \sqrt{3/c} W_{-3}| 0 \rangle$ is a primary field with respect to the Virasoro algebra, we have that $L_n |W\rangle=0$ for $n>0$. In this basis the computation of both overlaps and matrix elements only involves commuting Virasoro modes just like in the $k=2$ case, and one does not need the W algebra,  except to compute one single coefficient, namely
\begin{align}
\frac{\langle \Psi_{1} | \Psi(1) | W \rangle}{\langle \Psi_{1} | \Psi(1) | 0\rangle} = \sqrt{3/c} \frac{ \langle \Psi_{1} | \Psi(1) W_{-3}|  0\rangle}{\langle \Psi_{1} | \Psi(1) | 0\rangle} = \frac{\sqrt{26}}{9} \label{sign choice W}
\end{align} Once this is known, the matrix elements can be computed in the exact way as for the $k=2$ states.

The minimal model description of the $\mathbb{Z}_3$ RR turns out to be numerically much more efficient than the $W_3$ approach. As mentioned previously, the reason for this is the proliferation of null vectors in the latter approach, which is to be expected since the Virasoro modes by themselves (i.e. without the $W$ modes) are enough to generate the whole Hilbert space form finitely many primary fields (note that this is no longer true for $k>3$ RR).

\newpage

\section{$\mathcal{N}=1$ Superconformal Theories}
\label{superconformal}

$\mathcal{N}=1$ superconformal field theories are the prototype of a CFT with an extended chiral algebra. Beyond the ubiquitous spin-$2$ stress energy tensor $T(z)$, the chiral algebra contains a fermonic field $\Psi$ with conformal weight $h_{\Psi}=3/2$ and OPE
\begin{align}
\Psi(z) \, \Psi(0) = \frac{1}{z^3} + \frac{1}{z}\frac{3T(0)}{c} + O(1)
\end{align}
where $c$ is the central charge and is a free parameter of the theory. The superconformal current $\Psi(z)$ can be used to build an electron operator in the spirit of \eqref{generic electron operator}. This construction yields a trial wave function with $(k=2,r=6)$ clustering properties, and inverse filling $q = \nu^{-1} = 3 + m$\cite{Estienne2010539}. 
In this section we describe the MPS representation of these trial wave functions.

\subsection{Structure of the Neutral Hilbert Space}

The field $\Psi$ is primary with respect to  the Virasoro algebra, but it is part of an extended chiral algebra : the ($\mathcal{N}=1$) super Virasoro algebra. It is generated by the modes of the stress-energy tensor $T(z) = \sum_n z^{n-2}L_{-n}$ and those of  $\Psi(z) = \sum_{n} z^{n-3/2} \Psi_{-n}$
\begin{align}
[ L_n,L_m] & = (n-m)L_{n+m} + \frac{c}{12}n(n^2-1)\delta_{n+m,0} \\
\{ \Psi_n ,\Psi_m\} &  = \frac{3}{c}L_{n+m} + \frac{1}{2}\left( n^2 -\frac{1}{4}\right)\delta_{n+m,0} \\
 [L_n,\Psi_m] & = \left( \frac{1}{2}n-m\right) \Psi_{n+m}
\end{align}
Super-primary states $\ket{\Phi}$ are the highest weight states of the super-Virasoro algebra:
\begin{align}
L_n \ket{\Phi} = 0, \qquad \Psi_m \ket{\Phi} =0, \qquad n,m >0  
\end{align} 
In particular the state $\ket{\Psi} = \Psi_{-3/2} \ket{0}$ is not a super-primary, although it is a primary in the sense of Virasoro.  This is because $\Psi_{3/2} \ket{\Psi} = | 0\rangle$ does not vanish. 

Descendants of a given super primary field $\ket{\Phi}$ are spanned by
\begin{align}
\ket{ \Phi, \theta, \kappa } = L_{-\theta_1} \cdots  L_{-\theta_n}  \Psi_{-\kappa_1} \cdots  \Psi_{-\kappa_m} \ket{\Phi} \label{SV descendants}
\end{align}
with $\theta_1 \geq \theta_2 \geq \cdots  \geq \theta_n >0$ and $\kappa_1 > \kappa_2 > \cdots > \kappa_m >0$. Note the strict inequalities for $\kappa$, due to the fermionic nature of $\Psi(z)$.  The entries of $\kappa$ must be either all integers or all half-integers. Super primary fields fall into two categories:
\begin{itemize}
\item Neveu-Schwarz (NS)sector, in which the modes of $\Psi$ are half-integer. 
\item Ramond (R) sector, in which the modes of $\Psi$ are integer. 
 \end{itemize}
Upon building the tensor product with the charged sector U$(1)_{2\sqrt{q}}$ - the U$(1)$ charge of the various quasi-holes must be constrained in order to ensure  trivial monodromy with the electron. The selection rule takes a very simple form: the U$(1)$ charge $N = 2\sqrt{q}Q$ must be odd for  Ramond quasi-holes, and even for NS quasi-holes. 
 
 A highest weight $\ket {\Phi}$ together with its  super-Virasoro descendants form an irreducible representation of the super-Virasoro algebra \footnotemark  \footnotetext{After removing the null vectors}. We denote this vector space by
 \begin{align}
 \mathcal{SV}_{\Phi} = \text{span} \{ L_{-\theta_1} \cdots  L_{-\theta_n}  \Psi_{-\kappa_1} \cdots  \Psi_{-\kappa_m} \ket{\Phi}  \} \label{supervirasoro module} 
 \end{align} 
 As in the Virasoro case there will be null vectors or states with negative norms on the neutral side, and these can be dealt with using the same method. Due to the paired nature (or $k=2$ clustering) of the corresponding trial wave function, it is convenient to split the module $\mathcal{SV}_{\Phi}$ according to the fermion parity $ \mathcal{SV}_{\Phi} =  \mathcal{SV}_{\Phi}^{(0)} \oplus  \mathcal{SV}_{\Phi}^{(1)}$, where $\mathcal{SV}_{\Phi}^{(0)}$  ($\mathcal{SV}_{\Phi}^{(1)}$) is spanned by descendants containing an even (odd) number of fermionic modes $\Psi_{-\kappa_i}$.

 As a benefit from working with a chiral algebra that contains $\Psi(z)$, matrix elements $\langle \alpha' | \Psi(1) | \alpha \rangle = \langle \alpha' | \Psi_{\Delta_{\alpha}-\Delta_{\alpha'}} | \alpha \rangle$ are relatively easy to compute. Taking $\ket{\alpha} = \ket{\Phi, \theta,\kappa}$ and $\ket{\alpha'} = \ket{\Phi',\theta',\kappa'}$ any two descendant states, the evaluation of the matrix element 
\begin{align}
\langle \alpha' | \Psi(1) | \alpha \rangle = \bra{\Phi',\theta',\kappa'}\Psi_{\Delta_{\alpha}-\Delta_{\alpha'}}L_{-\theta_1} \cdots  L_{-\theta_n}  \Psi_{-\kappa_1} \cdots  \Psi_{-\kappa_m} \ket{\Phi} 
\end{align}
 boils down to an overlap between two descendants.  Hence the overlap matrices between two descendants completely describe the MPS representation. In particular no OPE structure constant has to be computed separately, as opposed to the $k=2$ Jack state or the $\mathbb{Z}_3$ RR state.

 \subsection{Rational Theories and Neutral Field Content}
 
So far we have considered $\mathcal{N}=1$ superconformal theories with an arbitrary central charge $c$. As was the case for the Virasoro algebra - 
the corresponding theory is irrational: it contains an infinite number of super primary fields $\ket{\Phi}$. These cases are not relevant for the FQHE, because the corresponding FQH state would have a torus degeneracy that grows with system size, and ultimately becomes infinite in the thermodynamic limit. However for specific values of the central charge the CFT has a finite amount of super primary fields (i.e. it is rational). These are the (super) minimal models $SM(p,p')$, with  central charge 
\begin{align}
c(p,p') = \frac{3}{2}\left( 1 - \frac{2(p-p')^2}{pp'} \right)
\end{align} 
They are labeled by two integers $(p,p')$ of the form
\begin{itemize}
\item either $p,p'$ coprime and $p+p'$ even (type A)
\item or $p,p'$ even such that $p/2$ and $p'/2$ coprime, and $(p+p')/2$ odd (type B)
\end{itemize}

\noindent For the minimal model $SM(p,p')$ the list of primary fields is given by the Kac table $\Phi_{(n|m)}$, with conformal dimension
\begin{align}
\Delta_{(n|m)} = \frac{(np-mp')^2-(p-p')^2}{8pp'} + \frac{1}{16}\delta_{n-m,1 \text{ mod } 2}
\end{align}
They are labeled by two integers $(n,m)$ with $ 1 \leq n \leq p'-1$ and $ 1 \leq m \leq p-1$, with the identification $\Phi_{(n|m)} \equiv \Phi_{(p'-n|p-m)}$. The field $\Phi_{(n|m)}$ is in the NS sector when $n-m$ is even,  and in the $R$ sector for $n-m$ odd. In particular the last term $1/16$ in the conformal weight only appears in the Ramond sector.

\subsection{Topological Sectors}
  
As usual the total conformal field theory $SM(p,p') \otimes$ U$(1)_{2\sqrt{q}}$ can be decomposed into topological sectors $\mathcal{H}_{(n,m ; a)}$.
The ground-state degeneracy on the torus is given by the number of such sectors:

\begin{itemize}
\item $3q(p-1)(p'-1)/4$ such sectors for a type A theory ($p,p'$ odd), 
\item and  $3q[(p-1)(p'-1)+1]/4$ for type B ($p,p'$ even).
\end{itemize}
To be more explicit the topological sectors are 
\begin{itemize}
  \item for a given NS field $\Phi_{(n|m)}$ there are $2q$ topological sectors 
  \begin{align}
  \mathcal{H}_{(\Phi_{(n|m)}, j)} = \left(M_{(n|m)}^{(0)} \otimes h_{2j \text{ mod } 4q} \right) \oplus \left( M_{(n|m)}^{(1)} \otimes h_{2(j+q) \text{ mod } 4q} \right)
  \end{align}
with $j = 0,\cdots,2q-1$.
  \item for a generic R field $\Phi_{(n|m)}$ fermion parity is not well defined because $\Psi_0 \ket{\Phi_{(n|m)}} \propto \ket{\Phi_{(n|m)}}$. As a consequence there are only $q$ topological sectors
    \begin{align}
  \mathcal{H}_{\Phi_{(n|m)},j} = M_{(n|m)} \otimes h_{1+ 2j  \text{ mod } 2q}
  \end{align}
with $j= 0,\cdots,q-1$.
   \item however for type B theories (\emph{i.e.} $p,p'$ even) the Ramond field $\Phi_{(p'/2|p/2)}$ is annihilated by $\Psi_0$. As a consequence fermion parity is well defined, and there are $2q$ topological sectors
   \begin{align}
  \mathcal{H}_{(\Phi_{(\frac{p'}{2},\frac{p}{2})} ;j)} = \left( M_{(\frac{p'}{2}|\frac{p}{2}) }^{(0)} \otimes h_{1+2j \text{ mod } 4q} \right) \oplus \left( M_{(\frac{p'}{2}|\frac{p}{2})}^{(1)} \otimes h_{1+2(j+q) \text{ mod } 4q} \right) 
  \end{align}
with $j = 0,\cdots,2q-1$.
\end{itemize}
where we used the compact notation $M_{(n|m)} = \mathcal{SV}_{\Phi_{(n|m)}}$, $M_{(n|m)}^{(0)} = \mathcal{SV}^{(0)}_{\Phi_{(n|m)}}$ and $M_{(n|m)}^{(1)} = \mathcal{SV}^{(1)}_{\Phi_{(n|m)}}$ as defined in Eq[\ref{supervirasoro module}].

Each topological sector yields a different ground states on the cylinder. As opposed to the Jack cases treated so far, these ground states cannot be characterized uniquely by a root partition with a generalized Pauli principle. Many such ground states collapse to the same pattern in the thin cylinder limit.

%%%%%%%%%%%%%%%%%%%%%%%%%%%%%%%%%%%%%%%%%%%%%%%%%%%%%%%%%%%%%%%%%%%%%%%%%%%%%%%%%%%%%%%

\newpage 
\section{Thin Torus Limit for the $M(3,2+r)$ CFT and $\mathbb{Z}_3$ Read-Rezayi States and Other Matrix Elements}\label{AppendixThinTorus}

In this appendix, we show how to obtain the thin torus limit for the vacuum topological sector of the paired states and then Read-Rezayi $Z_3$ state. We also give, as an example, some of the matrix elements for the paired states at descendant level one.

\subsection{Paired States}

In the vacuum sector, paired states have the fusion $\Psi \times \Psi = 1$ and $\Psi \times 1 = \Psi$. We leave other sectors as exercise for the reader. This sector is relevant when there are no quasi holes in the system, either in the bulk or at infinity. As mentioned before and as the fusion rules clearly show, upon repeated fusion of $\Psi$ with itself, one can only end up in the module of the primary field $1$ (for an even number of $\Psi$), or in the module of $\Psi$ itself   (for an odd number of $\Psi$). If one starts from an in-vector $\alpha$ in \eqref{conformal block wf} which is either the primary field $1$ or $\Psi$, there are only two neutral sectors for $k=2$ Jack states which correspond to $n=0$ in \eqref{topologicalsectorspairedstates1}  and which boil down to electron parity. We now perform a truncation on the absolute value of $|\theta| + |\mu| =P$ in \eqref{B0 Virasoro} and \eqref{B1 Virasoro}. The $P$ is the ``momentum" of the entanglement spectrum, or the level of the descendant fields in the CFT. It is one of the possible truncations (the other truncation being in the total conformal dimension $\Delta+ P+ Q^2/2$).

\subsubsection{The $P=0$ Truncation}

For generic  $r$ we have $q= \nu^{-1}=r/2+m$. From \eqref{B0 Virasoro} and \eqref{B1 Virasoro},  $P=0$ means the only possible $\theta= \theta'= \mu = \mu'=0$, which we will suppress as indices in the state $\ket{\Phi_{\Delta}, \theta; N, \mu}$ and we have ($k=2$ for the paired states):
\begin{align}
B^0_{x',N' ; x, N} & \propto \delta_{x',x}\,  \delta_{N',N-2}, \\
B^1_{x',N' ; x, N} & \propto \delta_{x'+x,1}\, \delta_{N + r + m,r x +1}\, \delta_{N',N+r + 2m -2}
\end{align} where $x \in \{0,1\}$ encodes in the neutral sector :  $x=0$ corresponds to the primary field $\mathbb{1}$ and $x=1$ to the primary field $\Psi$. In this case, there is only one occupation configuration for which the MPS matrices do not vanish\footnotemark \footnotetext{For a given number of electrons in the droplet. Outside the droplet it is always possible to put an infinite string of $B^0$ matrices.}. To see this, first observe that only two states are not annihilated by $B^1$ : 
\begin{align}
|x=0;N=1-r-m \rangle, \qquad |x=1;N=1-m \rangle 
\end{align}
Acting with $B^1$ on the first one, namely $|x=0;N=1-r-m \rangle$, we get to the state $\ket{x=1; N=  m-1}$. Since both $N$ and $x$ are integers, acting with $B^1$ again will make the state vanish. The only thing we can do is act with $(B^{0})^{m-1}$ to get to the state $\ket{x=0; N= 1-m}$. Acting on this with $B^1$ we obtain $\ket{x=0; N= r+m-1}$ which now requires the application of $(B^0)^{r+m-1}$ to get us back to $\ket{x=0;N=1-r-m}$ state from which the cycle repeats. We then obtain the Slater determinant of occupation numbers:
\begin{align}
(\cdots 0^{r+m-1}10^{m-1}1 0^{r+m-1} 1 0^{m-1}1)
\end{align} of the $(2,r)$ Jack states multiplied by $m$ Jastrow factors. This is known as the ``root partition" or thin torus limit.

\subsubsection{The First Few Matrix Elements : Truncation at $P=1$ for Paired States}

In this case we have generically three states in each sector $|\Phi_{\Delta} ; N \rangle$:
\begin{itemize}
\item the primary state at level $0$: $|\Phi,0 ; N,0 \rangle = | \Phi_{\Delta} \rangle \otimes | N \rangle$
\item two descendant states at level 1 : $|\Phi_\Delta, (0) ; N,(1) \rangle = | \Phi_{\Delta} \rangle \otimes a_{-1} | N \rangle$ and $|\Phi_\Delta,(1) ; N \rangle =  L_{-1}| \Phi_{\Delta} \rangle \otimes  | N \rangle$
\end{itemize}
Note that $|0,(1);N \rangle = L_{-1}| 0 \rangle \otimes |N \rangle= 0$ is a null vector, and all matrix elements involving this vector do vanish. For the neutral side, the only neutral matrix elements when truncating at $P=1$ are
\begin{align}
\langle \Psi | \Psi(1) |0 \rangle & =  C \\
\langle \Psi, (1) | \Psi(1) |0 \rangle & =  \frac{r}{2}C \\
\langle \Psi | \Psi(1) |0,(1) \rangle & =0
\end{align}
where $C$ is a constant that we need not compute since it will only contribute to the wave function through an overall factor $C^N$ where $N$ is the number of electrons. This basis is almost orthonormal. One simply has to normalize $| \Psi, (1) \rangle$, whose norm squared is $\langle \Psi, (1) | \Psi, (1) \rangle = r/2$.

For the $\text{U(1)}$ side, the vertex operator has a charge $\sqrt{q} = \sqrt{r/2+m}$. In a given sector  $N$, we only have to consider two states : the primary state $|N \rangle$, and its descendant $|N,(1)\rangle = a_{-1} |N\rangle$. The matrix elements we need are
\begin{align}
\langle N' | :e^{i\sqrt{q} \varphi (1)}: |N \rangle & =  \delta_{N',N+q}\\
\langle N',(1) | :e^{i\sqrt{q} \varphi (1)}: |N \rangle & =  \delta_{N',N+q}  \sqrt{q} \\
\langle N' | :e^{i\sqrt{q} \varphi (1)}: |N,(1) \rangle & = - \delta_{N',N+q}  \sqrt{q} \\
\langle N',(1) | :e^{i\sqrt{q} \varphi (1)}: |N ,(1) \rangle & =  \delta_{N',N+q}\,  (1 - q)
\end{align}

\subsection{Thin Torus Limit of the $\mathbb{Z}_3$ RR State from MPS}\label{Appendix6}

We again show how obtain the root partition in the vacuum sector, leaving those for the quasihole sectors as an exercise for the reader. 
In \eqref{B0 Virasoro} and \eqref{B1 Virasoro} we particularize at $\theta=\theta'= \mu =\mu'=\emptyset$ and $k=3, r=2$ for the RR state. From the $W$ approach of Appendix[\ref{RR}], we know $|W\rangle$ is a level three descendant of $|0\rangle$. Therefore it will play no role for $P \leq 2$, and there is a unique primary field per sector (for $P >2$ in the Virasoro approach there are two primary fields in the sector $1$ : $|0 \rangle$  and $|W \rangle$). At level $P=0$, and with $q = 2/3+m$,  we find
\begin{align}
B^0 \propto  \delta_{x',x} \delta_{N',N-3},\qquad B^1 \propto \delta_{x',x+1} \delta_{N',N+3m-1}\delta_{ 2N + 3m-1, 4(|x|- |x+1|)}
\end{align}
 where $x\in \{-1,0,1\}$ (defined $mod$ $3$) corresponds to $\Psi_{-1}, \mathbb{1}, \Psi_1$ fields respectively of conformal dimension $\frac{2}{3}|x|$. Very much like it was the case for the $k=2$ Jack states, very few states are not annihilated by $B^1$. These are
 \begin{align}
 | x = 0; N = -3 (m+1)/2 \rangle,\\
  | x = 1; N = (1-3m)/2 \rangle, \\
 | x = -1; N =  (5-3m)/2 \rangle
 \end{align}
 The presence of the Kronecker delta's now severely constraints the possible configurations. We observe that at this level, acting on the state $|x=0;N=-3 (m+1)/2\rangle$ we obtain only one configuration. First
 \begin{align}
 B^1 |x=0;N= -3(m+1)/2\rangle \propto \ket{x=1;N= (3m-5)/2},
 \end{align}  
which means we need $(B^0)^{m-1}$ to reach the next state not annihilated by $B^1$
 \begin{align}
  (B^0)^{m-1} B^1 |0 ; N= -3(m+1)/2 \rangle \propto \ket{ 1; N=(1-3m)/2}.
\end{align}  
 This now allows for another application of $B^1$: 
  \begin{align}
 B^1  (B^0)^{m-1} B^1 |0 ; N= -3(m+1)/2 \rangle \propto \ket{ -1; N=(3m-1)/2}.
\end{align}  
 Continuing, we need again $m-1$ matrices $B^0$ before we can act with the next $B^1$ 
 \begin{align}
B^1(B^0)^{m-1}  B^1  (B^0)^{m-1} B^1 |0 ;  -3(m+1)/2 \rangle \propto \ket{0; 3(m+1)/2}.
\end{align}  
 and finally we need $m+1$ matrices $B^0$ to go back to the initial state. Then the cycle repeats (until we reach the edge in the case of a finite droplet). Hence the thin torus limit or the root partition is 
\begin{align}
\cdots 0^{m+1}1 0^{m-1} 1 0 ^{m-1} 1 0^{m+1} 1 0^{m-1} 1 0^{m-1}1
\end{align}
In particular for $m=1$ the root partition is $\cdots 1^3 0^2 1^3 0^2 1^3$.

%%%%%%%%%%%%%%%%%%%%%%%%%%%%%%%%%%%%%%%%%%%%%%%%%%%%%%%%%%%%%%%%%%%%%%%%%%%%%%%%%%%%%%%

\newpage 
\section{Some Structure Constants for the 3 State Potts Model}

We can use the underlying $W_3$ symmetry of the 3 state Potts model to compute some otherwise more difficult to access structure constants. In particular we need to compute
\begin{align}
 \langle \Psi | \Psi(1) | W \rangle
\end{align}
where $W$, which was a primary field in the minimal model $M(5,6)$ description, is now a  descendant field in the $W_3$ model:
\begin{align}
| W \rangle = \sqrt{\frac{15}{4}} W_{-3} | 1 \rangle
\end{align}
The matrix elements are not easy to find as they involve complicated commutations relations with $W$ modes. These relations, available in Ref.\cite{Fateev-1988IJMPA...3..507F}, together with the method presented in Ref.\cite{Estienne-2009JPhA...42R5209E}, give:
\begin{eqnarray}
 \langle \Psi | \Psi(1) | W \rangle = -  \sqrt{\frac{15}{4}}\langle \Psi | [W_{-3} ,\Psi(1)] | 0 \rangle =\mp   \sqrt{\frac{15}{4}}\frac{4}{9}\sqrt{\frac{13}{30}} \langle \Psi | \Psi(1) | 0 \rangle
\end{eqnarray}
where the sign ambiguity comes from the possibility to change $W \to -W$.  Choosing one of the two conventions,  we get
\begin{align}
 \langle \Psi | \Psi(1) | W \rangle =  \frac{\sqrt{26}}{9}. \qquad  \label{W check}
\end{align}  Then the other sign appears in $
 \langle W| \Psi(1) |  \Psi_{-1} \rangle =  - \frac{\sqrt{26}}{9}$

%%%%%%%%%%%%%%%%%%%%%%%%%%%%%%%%%%%%%%%%%%%%%%%%%%%%%%%%%%%%%%%%%%%%%%%%%%%%%%%%%%%%%%%

\newpage 
\section{Squeezing}\label{AppendixSqueezing}

All model FQH states, when expanded in many-body non-interacting states (Slater determinants or monomials) have a property called squeezing from a nontrivial root configuration (partition).  This means only configurations obtained by the repeated application, starting with the root configuration of two body operators which brings particles inward \cite{Bernevig-PhysRevB.77.184502,Bernevig-PhysRevLett.100.246802} are allowed. Squeezing was shown to be a consequence of writing model wave functions as correlators  of vertex operators in CFT \cite{Thomale-PhysRevB.84.045127}. The squeezing property is better expressed  in angular momentum basis in which a string of occupation numbers $\{m_k\}$ for $N_e$ electrons in the orbital of angular momentum $k$   is re-written as   $\{k_1, k_2, \ldots k_{N_e}\}$ (with $0\le k_1\le k_2 \ldots \le k_{N_e}$ - notice the inverse ordering of partition convention adopted here for convenience purposes to be linked with MPS results) where $m_{k_i} >0$ (no momentum of an unoccupied orbitals are written down). If the root partition has the orbital occupation $\ell=\{l_1,l_2, \ldots, l_{N_e}\}$ then the squeezing property says that any configuration $k=\{k_1, k_2, \ldots k_{N_e}\}$ with nonzero weight in a Quantum Hall State of root partition $\ell$ must satisfy:

\beq
\sum_{i=1}^j (k_i- l_i) \ge 0 , \qquad  \forall j 
\eneq 
The root partition is the same as the charge density wave of the thin torus limit. In this limit there is no charge fluctuation and the state is a single Slater determinant with electrons pinned at a repeating pattern of certain orbitals. In MPS, this limit can be obtained in the case when no descendants are included in the CFT spectrum ($P=0$). Only in this MPS limit we obtain a single state, and the electron fluctuations are  frozen.  The descendant level  \eqref{Vir x U(1) basis} of the MPS matrix after  $j$ orbitals (the maximum orbital momentum is $j-1$, as we include the zeroth orbital) is denoted to be $P_j$ and is related to the conformal dimension $\Delta_j$ in \eqref{conformaldimensionsolution1} by
\beq
P_j +\frac{1}{2} Q_j^2 + h_j = \Delta_j
\eneq
where $h_j$ is the conformal dimension of the neutral CFT primary field  on the MPS bond after $j$ orbitals. It is a finite number. We then have
\beq
P_j + h_j = \Delta_0- \frac{1}{2} Q_0^2 + \left(\frac{1}{2}   -\frac{Q_0}{\sqrt{\nu}}  \right) \sum_{k=0}^{j-1} k m_k  - \frac{1}{2\nu} \left( \sum_{k=0}^{j-1} m_k \right)
\eneq
which can be rewritten
\beq
P_j+ h_j  = h_0 + \left(\frac{1}{2}   -\frac{Q_0}{\sqrt{\nu}}  \right) N_{j} -\frac{1}{2 \nu} N_j^2 + L_{j} \label{momentumlinktosqueezing}
\eneq
where $N_j = \sum_{k=0}^{j-1} m_k, L_j=\sum_{k=0}^{j-1} k  m_k$ are the number of particles in the first $j$ orbitals (remember we have orbital $0$). For a partition $k$ we can write $L_j = \sum_{i=1}^{N_{j}} k_{i}$. The root partition $\ell$ is then by definition the one for $P_j=0 \forall j$  in \eqref{momentumlinktosqueezing}and we have 
\beq
P_j = \sum_{i=1}^{N_{j}} k_{i} - \sum_{i=1}^{N_{j}} \ell_{i}
\eneq Since $P_j \ge 0$, we see that the MPS implies squeezing.

%%%%%%%%%%%%%%%%%%%%%%%%%%%%%%%%%%%%%%%%%%%%%%%%%%%%%%%%%%%%%%%%%%%%%%%%%%%%%%%%%%%%%%%

\newpage

\section{Trimming Examples for Laughlin and Moore-Read States}\label{AppendixTrimming}

\subsubsection{Laughlin States}

The naive way to build the $P, Q$ Hilbert space for a $U(1)$ boson at truncation level $P_{\text{max}}$ is: for every momentum sector $P=P_{\text{max}}, P_{\text{max}}-1, P_{\text{max}}-2, \ldots 2,1,0$ build the partitions of $P$. The charge sectors  $N= \sqrt{q} Q$ run from $-P_{\text{max}} \ldots P_{\text{max}}$ due to the $\delta_{P', P-N}$ term in $B^1$. Then the $B^0$ term has two more charge sectors because it can add (or subtract $1$) from whatever maximum charge sector one reached. Therefore, in the naive way, we are find at $2 P_{\text{max}} + 3$ charge sectors for every $P_{\text{max}}$. This means that the number of matrix elements in a $B$ matrix  is $(2 P_{\text{max}} + 3) \sum_{i=0}^{P_{\text{max}}} N_{\text{partitions} (i)}$ where $N_{\text{partitions} (i)}$ is the number of partitions of  $i$ (the momentum sectors of the $U(1)$ boson. This is a huge number inhibiting large scale computations but it is severely redundant, as it trivially considers that the number of matrix elements $\{N, P\}$ is just a product of the number in these sectors, $\{N\} \otimes \{P\}$. However we cannot access most of the $N, P$ sectors built this way. For example, for $P= P_{\text{max}}$ we can only access the sectors $N=0,\ldots q-1$, which leads to a huge decrease, as we do not need to keep the matrix elements with $P=P_{\text{max}}$ and $N<0$ or $N>q-1$. In fact, for a given $N$, we can only reach levels of $P$ given by

\beq
P+ \frac{1}{2} \left[\frac{N}{q-1} \right] ( 2 N - (q-1)  \left[\frac{N}{q-1} \right] - (q-1) )  \le |P_{\text{max}}|
\eneq
where $ \left[\frac{N}{q-1} \right]$ is the floor function for $N$. Any $P, N$ that does not satisfy the above relation can be thrown out as it will not be used in any calculation at truncation level $P_{\text{max}}$. This provides a huge decrease in Hilbert space size.

The formula above can be justified per the method explained in the text: the root partition of the ground-state of the filling $1/q$ Laughlin state is
\beq
\ldots 1 0^{q-1} 1 0^{q-1} 10^{q-1} 1 0^{q-1} 10^{q-1} 10^{q-1} 10^{q-1} 1\ldots
\eneq
where $\ldots$ means that this is an infinite state. Assume the truncation level is $P_{\text{max}}$. We want to determine  the possible indices $N, P$ on some auxiliary bond (cut).  We first determine the $N,P=0$ root partition. We decompose $N= q N_p +k$, $0\le k \le q-1$ which means $N_p$ particles moved to the right of a cut made $k$ orbitals to the right of a $1$ in the root partition. We  ask what is the root configuration with those properties? To the left of the cut, it is just the  ground-state root followed by a bunch of zeros. The right of the cut, every partition configuration must be squeezed from a configuration of the type:

\beq
_{N/\sqrt{q}, P=0}| \underbrace{111\ldots111}_{2N_p \; \text{or} \; 2 N_p-1} 0\ldots 01\underbrace{ 0 \ldots 0}_{\ge q-1}1 0^{q-1} 10^{q-1} 10^{q-1} 10^{q-1} 1\ldots
\eneq   where the number of consecutive  $1$'s to the right of the cut is either $2 N_p$ or $2 N_p-1$. The number of zeroes immediately after this string of integers is determined by squeezing,  can be obtained as an analytic formula, but is not important for the present discussion. For example, there can be no number of zeroes, as in the $q=3$ state. The next number of zeroes, after the lonely $1$ which follows the string of $1$'s is $\ge q-1$, and then we have the root partition.   For example, $q=3, N=4$ has the root:
\beq
\ldots 10010010010000 | \boxed{11100001001001001} \ldots
\eneq
It has $P=0$ at the boundary. However, moving away from the boundary to the right into the bulk of the string of $2N_p$ occupied sites, the momentum of each auxiliary bond first increases then decreases again, as the root partition to the very far right of the cut has momentum $0$.  The string of $2 N_p$ $B^1$ matrices on the right of the boundary  has the highest momentum on an auxiliary bond easily computable:
\beq
P_1 =N' N - \frac{1}{2} N' (N'+1) (q-1)
\eneq where $N'$ is the number of particles to the right of the cut that need to be transversed such that the momentum doesn't increase anymore. This happens when the $U(1)$ index after those particles becomes negative $N- N'(q-1) <0$ - then the momentum starts decreasing. So $N'= Floor[N/(q-1)] = [N/(q-1)]$ giving the formula
\beq  P_{1}= \frac{1}{2} \left[\frac{N}{q-1} \right] ( 2 N - (q-1)  \left[\frac{N}{q-1} \right] - (q-1) ) \eneq

Now assume that, at the cut, we want to look for a configuration of momentum $P$, and not zero as above. Once we want to increase the momentum at the cut side, we must squeeze particles across the cut, and since the root partition to the right of the cut already has a large momentum inside of it for $N>0$, we will not be able to get every value of the momentum because we have a cutoff $P_{\text{max}}$. It is clear that we will only be able to get momenta $P$ at the cut such that $P + P_{1} \le P_{\text{max}}$. Notice that in the thermodynamic limit, this condition becomes
\beq
P+\frac{1}{2 (q-1)} N^2< |P_{\text{max}}|
\eneq which is the $L_0$ of a $1/(q-1)$ filling system, and not of a $1/q$-filling system.  This seems at first counter-intuitive, but we remind the reader that the $q=1$ ground-state state is a single Slater determinant and should be computable with $P=0, N=0$, which the formula gives.

\subsubsection{Moore-Read and Read-Rezayi}

The prior procedure obviously be generalized to Moore-Read and Read Rezayi, with certain small modifications. Lets first generalize it to the $1/2$ Moore-Read state.  Analytic formulas can be found for any states once an MPS form is known, but its quite inefficient to repeat that for every state, especially when the generic prescription given in the text can be applied numerically to all states.  For Moore-Read at $\nu=1/2$ in the $\Delta= I, \psi$ sector we find the following truncation 

\begin{eqnarray}
\Delta=I \rightarrow P &+&  N(N-  (N \mod 2))  \\
&-& \text{Floor}\left[ \frac{(N+1 -  (N \mod 2))^2}{2}  \right] \le P_{\text{max}}\nonumber
\end{eqnarray}

\begin{eqnarray}
\Delta = \psi  \rightarrow P&+& N(N-  ( (1+(N \mod 2) ) \mod 2)) \\
&-& \text{Ceiling}\left[ \frac{(N-  ( (1+(N \mod 2) ) \mod 2))^2}{2}  \right] \le P_{\text{max}}\nonumber
\end{eqnarray}

Note that for $\Delta=1, N=0$ the above function gives: $- \text{Ceiling}[1/2]=-1$, obviously not true as it would say that we can reach the level $P_{\text{max}}+1$ although we have kept only the truncation $P_{\text{max}}$. This is because for $N=0$ (root partition), there cannot be \emph{any} $\Delta=1$ as the root partition for this state is made out of even number of electrons, and the quantity above should be replaced to with $0$.

\bibliography{fqhempslong}
\bibliographystyle{model1-num-names}

\end{document}